\documentclass[aps,prd,twocolumn,nofootinbib,superscriptaddress]{revtex4-1}

\usepackage{graphicx}
\usepackage{amsmath,amssymb,amsfonts,mathtools,bm}
\usepackage{physics}
\usepackage{placeins}
\usepackage{float}
\usepackage{comment}
\usepackage{csquotes}
\usepackage{epstopdf}
\usepackage{subcaption}
\usepackage{capt-of}
\usepackage{hyperref}

\allowdisplaybreaks

\hypersetup{
     colorlinks = true,
     citecolor  = red,
     linkcolor  = blue,
     urlcolor   = blue,
}

\def\As{A_{\rm s}}
\def\HI{H_{\rm end}}
\def\mH{\mathcal{H}}
\def\ee{\eta_{\rm end}}
\def\ae{a_{\rm end}}
\def\are{a_{\rm re}}

\def\ns{n_s}
\def\ere{\eta_{\rm re}}

\def\Tre{T_{\rm re}}

\def\wre{w_{\rm re}}
\def\kpv{k_*}
\def\Mp{\mathrm{M}_{\rm P}}
\def\Gev{\mathrm{GeV}}

\def\l{\left}
\def\r{\right}
\def\d{\mathrm{d}}
\def\nn{\nonumber}
\def\Mpc{\mathrm{Mpc}}
\def\vk{\mathbf{k}}
\def\vq{\mathbf{q}}
\def\vx{\mathbf{x}}

\def\nb{n_{\rm b}}

\def\mPb{\mathcal{P}_{\rm B}}
\def\tmPb{\tilde{\mathcal{P}}_{\rm B}}
\def\rhoem{\rho_{\rm EM}}
\def\mF_{\mathcal{F}}

\def\rhob{\rho_{\rm B}}

\def\mPc{\mathcal{P}_\zeta}

\def\ke{k_{\rm end}}
\def\kre{k_{\rm re}}
\def\ksb{k_{\rm SB1}}
\def\ksbs{k_{\rm SB2}}
\def\xre{x_{\rm re}}
\def\xe{x_{\rm e}}

\def\mF{\mathcal{F}}
\def\mI{\mathcal{I}}

\def\hij{h_{ij}}
\def\hk{h_{\vk}}

\def\Pt{\mathcal{P}_{\rm T}}

\def\ogw{\Omega_{\rm GW}}
\def\ogwh{\Omega_{\rm GW}h^2}
\def\ogws{\Omega_{\rm GW}^{\rm sec}h^2}
\def\ogwp{\Omega_{\rm GW}^{\rm pri}h^2}

\def\rhogw{\rho_{\rm GW}}

\def\mGk{\mathcal{G}_{k}}

\def\mIrds{\mathcal{I}_{\rm RD,s}}

\def\f{\frac}

\def\teta{\tilde{\eta}}

\def\nw{n_w}
\def\nbs{n_{\rm b,2}}
\def\kp{k_{\rm p}}
\def\fp{f_{\rm peak}}
\def\ogws{\Omega_{\rm GW,S}}
\def\ogwm{\Omega_{\rm GW,M}}
\def\ogwhs{\Omega_{\rm GW}^{\rm scalar}h^2}
\def\ogwhm{\Omega_{\rm GW}^{\rm mag}h^2}
\def\hz{\text{Hz}}

\def\ve{\mathbf{e}}

\def\gsp{g_{*,0}}
\def\gseq{g_{*,\rm eq}}
\def\fre{f_{\rm re}}

\def\mPcv{\mathcal{P}_{\zeta}^{\rm vac}}

\def\mSem{\mathcal{S}_{\rm em}}
\def\Ni{N_{\rm inf}}
\def\Nre{N_{\rm re}}
\def\nc{n_{\zeta}}

\def\mAk{\mathcal{A}_{\vk}^{\lambda}}
\def\Ak{A_{\vk}^{\lambda}}
\def\tmPb{\tilde{\mathcal{P}}_{\rm B}}
\def\tmPe{\tilde{\mathcal{P}}_{\rm E}}

\def\xe{x_{\rm e}}

\def\mAre{\mathcal{A}^{\rm re}_{\vk}}
\def\qnb{\mathrm{q}_{\nb}}
\def\gsre{g^{*}_{\rm s,re}}
\def\GeV{\text{GeV}}

\def\Pijmn{\mathrm{P}_{ij}^{mn}}
\def\mSmn{\mathcal{S}_{mn}}
\def\mSs{\mathcal{S}_{\rm scalar}}
\def\mPh{P_{\rm T}}
\def\mPhv{P_{\rm T}^{\rm vac}}
\def\mPhs{P_{\rm T}^{\rm scalar}}
\def\mPhm{\mPh^{\rm mag}}
\def\mrP{\mathrm{P}}

\def\mPci{\mathcal{P}_{\zeta}^{\rm ind}}

\begin{document}

\title{Gravitational Waves from Post-Inflationary Magnetism: Direct and Scalar-Induced Contributions}

\author{Subhasis Maiti}
\email{E-mail: subhashish@iitg.ac.in}
\affiliation{Department of Physics, Indian Institute of Technology, Guwahati, 
Assam, India}
\date{\today}

\begin{abstract}
We study stochastic gravitational waves generated in a post-inflationary magnetogenesis scenario with time-dependent gauge couplings during inflation and reheating. In this setup, magnetic anisotropic stress directly sources gravitational waves, while the induced curvature perturbations generate an additional scalar-induced GW component.
We compare the spectral behavior of the two contributions and find that the magnetic-originated GW dominates the peak amplitude, whereas the scalar-induced contribution becomes important on larger scales. For $n_{\rm b}\ge 3/2$, both contributions exhibit the universal infrared scaling $\Omega_{\rm GW}(f<f_{\rm peak})\propto f^3$, while their ultraviolet slopes differ, leading to distinct spectral signatures.
For suitable reheating and magnetogenesis parameters, the resulting GW signal naturally extends into the nano-Hz range relevant for pulsar timing array observations, and it remains consistent with other bounds. The distinct spectral features of the two components may provide a useful probe of reheating dynamics and primordial magnetogenesis.
\end{abstract}

\maketitle


\section{Introduction}

The direct detection of gravitational waves (GWs) has opened a new observational window into the physics of the early Universe~\cite{LIGOScientific:2016aoc, LIGOScientific:2016vlm, LIGOScientific:2016emj, LIGOScientific:2016vbw, LIGOScientific:2017bnn, LIGOScientific:2016jlg}. Unlike electromagnetic radiation, GWs propagate almost freely after their production and retain information about processes occurring at epochs far earlier than recombination. The stochastic gravitational-wave backgrounds generated in the primordial Universe can serve as unique messengers of inflation, reheating, phase transitions, topological defects, and primordial density fluctuations~\cite{Planck:2018jri, BICEP:2021xfz, BICEP2:2018kqh, Clarke:2020bil, Watanabe:2006qe, Caprini:2018mtu,Acquaviva:2002ud, Planck:2018vyg, PhysRevD.85.123002, NANOGrav:2023hvm, Sorbo:2011rz, Caprini:2014mja, Ito:2016fqp, Sharma:2019jtb, Starobinsky:1979ty, Grishchuk:1974ny, Guzzetti:2016mkm, Haque:2021dha, PhysRevD.64.123514, Di:2017ndc,Fu:2019vqc,PhysRevD.103.083510, Bhaumik:2020dor, Solbi:2021wbo, Figueroa:2021zah, Domenech:2020kqm, Cai:2018dig, Maiti:2025rkn,Maiti:2025awl, Maiti:2025cbi, Chakraborty:2024rgl, Chakraborty:2023ocr, Hoory:2025qgm,Inui:2023qsd, Abe:2022xur}.The increasing sensitivity of current and future GW observatories motivates studying possible cosmological sources of stochastic backgrounds.

Among these possibilities, primordial magnetic fields (PMFs) have been widely studied~\cite{Sorbo:2011rz,Caprini:2014mja,Ito:2016fqp,Sharma:2019jtb,Okano:2020uyr}. Magnetic fields are observed on a wide range of astrophysical scales, from galaxies to galaxy clusters, with strengths of order $\mu{\rm G}$~\cite{fletcher2011magnetic,beck2016magnetic,Haverkorn:2008tb,kronberg2001magnetic}. Their origin remains one of the longstanding open questions in cosmology. While astrophysical dynamo mechanisms may amplify seed fields, they still require an initial magnetic component whose origin is not yet fully understood~\cite{brandenburg:2023}. This has motivated extensive studies of primordial magnetogenesis scenarios, where magnetic fields are generated in the early Universe during inflation, reheating, or subsequent cosmological epochs~\cite{PhysRevD.37.2743, PhysRevD.52.6694, PhysRevD.70.063502, Maiti:2025awl, Maiti:2025cbi, Maiti:2025rkn, 2023arXiv231207938Y, Sharma:2019jtb, Ito:2016fqp,PhysRevD.65.023517, Sorbo:2011rz}.

If primordial magnetic fields were generated in the early Universe, they can produce gravitational waves through their anisotropic stress~\cite{Sorbo:2011rz,Caprini:2014mja,Ito:2016fqp,Sharma:2019jtb,Okano:2020uyr, Maiti:2025rkn,Maiti:2025awl, Maiti:2025cbi, Maiti:2025ijr}. In this case, the transverse-traceless part of the magnetic energy-momentum tensor sources tensor perturbations and generates a secondary GW background. The spectral shape of this signal depends on the amplitude, coherence scale, and evolution of the primordial magnetic fields~\cite{Maiti:2025cbi, Maiti:2025awl}.

Primordial magnetic fields can also generate scalar curvature perturbations through their energy-density fluctuations~\cite{Fujita:2013qxa, PhysRevD.94.043523, Maiti:2025ijr}. These scalar perturbations later source gravitational waves at second order when the corresponding modes re-enter the Hubble horizon during the post-inflationary era. This produces another stochastic GW background, which we refer to as magnetically scalar induced gravitational waves (MSIGWs). Since these two GW contributions arise from different physical processes, they can show different spectral features.

In this paper, we study both of these GW contributions generated by primordial magnetic fields. We analyze the direct tensor contribution sourced by magnetic anisotropic stress as well as the scalar-induced GW contribution generated from magnetic-field-induced curvature perturbations. With the improving sensitivities of pulsar timing arrays, space-based interferometers such as LISA~\cite{Amaro-Seoane:2012aqc,Barausse:2020rsu,Caprini:2015zlo}, and future terrestrial GW detectors~\cite{Crowder:2005nr,Corbin:2005ny,Baker:2019pnp,Seto:2001qf,Kawamura:2011zz,Suemasa:2017ppd,Janssen:2014dka,NANOGrav:2023gor,2023arXiv230616224A,Reardon:2023gzh,Zic:2023gta,Xu:2023wog,TianQin:2015yph}, a wide range of frequencies relevant for cosmological magnetic fields is becoming observable. Motivated by recent indications of a stochastic GW background in the nanohertz regime, we study the spectral behaviour, relative amplitudes, and reheating dependence of these two contributions. We find that their different evolution after horizon re-entry leads to distinct features in the GW spectrum, which may help to probe primordial magnetogenesis and reheating dynamics with future GW observations.

\section{Magnetogenesis Model}\label{sec_magnetogenesis}
 We consider a phenomenological magnetogenesis scenario with a sawtooth-type coupling function. The conformal invariance of the gauge field is broken through a time-dependent coupling $I(\eta)$, which allows the amplification of vacuum fluctuations. During inflation, the coupling function increases with time and enhances gauge field production, while during reheating it decreases, leading to a second stage of amplification even on super-horizon scales.

After reheating, the Universe quickly becomes highly conducting due to the production of charged particles. As a result, the electric field decays rapidly and the magnetic field effectively freezes on large scales, preventing any further amplification of the gauge field. For simplicity, we assume that conformal invariance is effectively restored after reheating so that no additional gauge field production occurs during the standard radiation-dominated era.

The action for the gauge field sector in a flat FLRW background is given by~\cite{Ferreira:2013sqa,Kobayashi:2014sga,PhysRevD.100.023524,Haque:2020bip,Maity:2021qps,Tripathy:2021sfb,Li:2022yqb,Benevides:2018mwx,Kobayashi:2019uqs,PhysRevD.94.043523,Hortua:2014wna,Campanelli:2008kh,Jain:2012jy,Caprini:2014mja,Sharma:2018kgs,Bamba:2021wyx,PhysRevD.37.2743,PhysRevD.94.043523}
\begin{align}
    \mSem= \frac{1}{4}\int d^4 x\sqrt{g}I^2(\eta)F_{\mu\nu}F^{\mu\nu}.
\end{align}
Here, $I(\eta)$ determines the deviation from conformal invariance and controls the efficiency of gauge field production.

To proceed, one must specify the evolution of $I(\eta)$. Motivated by earlier studies~\cite{Sharma:2017eps, Papanikolaou:2024cwr, Maiti:2025awl, Maiti:2025ijr}, we adopt a sawtooth-type parametrization,
\begin{align}\label{eq:f_coupling}
    I(\eta)=
    \begin{cases}
        (a/a_i)^{n}, & a_i < a \leq a_e, \\[6pt]
        \left(\dfrac{a_e}{a_i}\right)^{n} \left(\dfrac{a_e}{a}\right)^{n\beta}, & a_e \leq a \leq a_{\rm re}, \\[6pt]
        1, & a \geq a_{\rm re},
    \end{cases}
\end{align}
where $n$ is the magnetogenesis parameter that determines the time evolution of the coupling function during inflation. We further define $ \beta \equiv \Ni/\Nre$,
where $\Ni=\ln(\ae/a_i)$ denotes the number of inflationary $e$-folds after the pivot scale $\kpv$ exits the horizon, and $\Nre=\ln(\are/\ae)$ is the reheating duration in $e$-folds. Here, the scale factors $a_i$, $\ae$, and $\are$ correspond to pivot-scale horizon crossing, the end of inflation, and the end of reheating, respectively.

{\bf Strong coupling and Backreaction issue:}
In standard electromagnetism, the gauge coupling is constant. By contrast, inflationary and post-inflationary magnetogenesis requires a time-dependent coupling function, $I(\eta)$, to break conformal invariance and amplify gauge-field fluctuations. In such scenarios, the effective electric charge becomes $e_{\rm eff}=e/I$. If $I\ll1$, one obtains $e_{\rm eff}\gg1$, signaling a strong-coupling regime in which perturbative calculations are no longer reliable.

To avoid this issue, we restrict the parameter space such that $I(\eta)\geq 1$ throughout the evolution. For the coupling function defined in Eq.~\eqref{eq:f_coupling}, this condition is satisfied by taking $n>0$.
We also impose the backreaction constraint by requiring the generated electromagnetic energy density to remain subdominant to the background energy density, $ \rho_{\rm em}<\rho_c$,
where $\rho_c=3H^2(\eta)\Mp^2$ denotes the total background energy density at conformal time $\eta$.

\textbf{Relating inflationary and reheating parameters:}

Inflation provides the initial conditions for magnetogenesis by generating fluctuations at horizon crossing. The subsequent reheating phase determines how these modes evolve before entering the standard radiation-dominated era.
Instead of specifying a particular inflationary model, we parametrize the background using the reheating equation of state $\wre$ and reheating temperature $\Tre$, assuming perturbative reheating~\cite{Kamionkowski-2014}. 

The inflationary Hubble scale is related to observables by
$\HI = \pi \Mp \sqrt{{r_{k_*}\,\As}/{2}}$. We use $r_{k_* = 0.05} \simeq 0.036$ for the tensor-to-scalar ratio and $\As \simeq 2.1 \times 10^{-9}$ for the amplitude of scalar perturbations at CMB scales~\cite{Planck:2018jri, Planck:2018vyg}. Here, $\Mp$ is the reduced Planck mass.

The comoving wavenumber exiting the horizon at the end of inflation is $\ke = \kpv e^{\Ni}$, which can be expressed as~\cite{Chakraborty:2024rgl}
\begin{align}\label{eq:ke}
    \ke = \left( \frac{43\,\gsre}{11} \right)^{1/3}
    \left( \frac{\pi^2 \gsre}{90} \right)^{\sigma}
    \frac{\HI^{1 - 2\sigma}\, \Tre^{4\sigma - 1} \, T_0}{\Mp^{2\sigma}}.
\end{align}
Here $\gsre$ is the relativistic degree of freedom defined at the end of reheating and where \( \sigma = {1}/{3(1 + \wre)} \), and \( T_0 = 2.725 \, \text{K} \) is the present-day CMB temperature.

Similarly, the comoving wavenumber re-entering the horizon at the end of reheating is~\cite{Chakraborty:2024rgl}
\begin{align}\label{eq:kre}
    \kre = 3.9 \times 10^6 \left( \frac{\Tre}{10^{-2}} \right) \, \Mpc^{-1}.
\end{align}
These relations demonstrate that $\wre$ and $\Tre$ control both the duration of reheating and the inflationary era. Throughout the reheating periods, we consider matter like evolution with the average equation of state $\wre=\langle w_\phi\rangle=0$, where $w_\phi$ is the equation of state of the inflation field.


\paragraph{\bf Production during inflation:}

In Fourier space, the equation of motion for the canonically normalized mode function$\mAk=I(\eta)\Ak$ is\cite{Subramanian:2015lua, Tripathy:2021sfb, Maiti:2025awl,Sharma:2018kgs}
\begin{align}\label{eq:gauge_field}
    {\mAk}''+\l(k^2-\frac{I''}{I}\r)\mAk=0.
\end{align}
The amplification of the gauge field is controlled by the effective mass term $I''/I$, which becomes significant when conformal invariance is broken. We suppress the polarization index $\lambda$, since both polarization states evolve identically.

During inflation, the rapidly evolving background leads to efficient gauge field production, particularly for modes exiting the horizon. Assuming a de Sitter background, the solution is~\cite{Subramanian:2015lua, Tripathy:2021sfb, Maiti:2025awl,Sharma:2018kgs}
\begin{align}
    \mAk(\eta)=\sqrt{-\frac{\pi\eta}{4}}e^{i(n+1/2)\pi/2}\mathrm{H}^{(1)}_{n+1/2}(-k,\eta) ,
\end{align}
where $\mathrm{H}^{(1)}_{n+1/2}(x)$ is the Hankel function of the first kind. This solution follows from imposing the Bunch-Davies vacuum condition in the sub-horizon limit $(-k\eta\rightarrow\infty $), where $\mAk \simeq {e^{-ik\eta}}/{\sqrt{2k}}$.

As modes cross the horizon, $k \sim aH$, their oscillatory behavior stops and the modes become amplified. The time-dependent coupling function plays an important role here, as it continuously transfers energy into the system and helps sustain the growth of large-scale magnetic fields.

At the end of inflation, the comoving magnetic and electric energy spectra are~\cite{Maiti:2025awl, Maiti:2025cbi}
\begin{subequations} \label{6a}
   \begin{align}
    \tmPb(k,\ee)&=\frac{\ke^4}{8\pi}\frac{2^{2n+1}\Gamma^2(n+1/2)}{\pi^2}\l(\frac{k}{\ke}\r)^{\nb}\\
    \label{6b}
    \tmPe(k,\ee)&=\frac{\ke^4}{8\pi}\frac{2^{2n-1}\Gamma^2(n-1/2)}{\pi^2}\l(\frac{k}{\ke}\r)^{\nb+2} 
\end{align} 
\end{subequations}
where $\nb = 4 - 2|n|$ is the magnetic spectral index. These expressions indicate that, during inflation, both spectra exhibit a single power-law behavior determined by the parameter $n$.

\textbf{Production during reheating:}
The time dependence of $I(\eta)$ during reheating leads to a second phase of gauge field amplification. This arises because the effective mass term $I''/I$ remains non-negligible even after inflation ends.

We solve Eq.~\eqref{eq:gauge_field} in this phase, assuming negligible electrical conductivity so that the gauge field evolves freely. So the general solution of the gauge field is~\cite{Maiti:2025awl}
\begin{align}
    \mAre(k,\eta>\ee) = \sqrt{k\tilde{\eta}} \left\{ d_1 J_{\alpha+1/2}(k\tilde{\eta}) + d_2 J_{-\alpha-1/2}(k\tilde{\eta}) \right\}.
\end{align}  
where $\tilde{\eta}=\eta-\ee+\frac{\delta}{\HI}$ with $\delta=2/(1+3\wre)$. Here, $J_{\alpha+1/2}$ denotes the Bessel function of the first kind, with $\alpha =n\beta \delta$. The coefficients $d_1$ and $d_2$ are fixed by matching the solution to the inflationary mode functions at $\eta=\ee$, thereby encoding the information associated with the production of the gauge field during inflation.

Focusing on modes with $k < \ke$ (i.e., $\xe \ll 1$), one finds that the reheating phase modifies the spectral shape. For super-horizon modes at the end of reheating, i.e., ($k <\kre$),
\begin{subequations}\label{eq:mag_spec}
    \begin{align}
        \tmPb(k,\ere) &=\frac{\ke^4}{8\pi}\frac{\qnb^1(n)}{(\alpha+1/2)^2}\l(\frac{\ke}{\kre}\r)^{2(\alpha+1)}\l(\frac{k}{\ke}\r)^{\nb} ,\\
        \tmPe(k,\ere) &=\frac{\ke^4}{8\pi}\frac{\qnb^1(n)}{4}\l(\frac{\ke}{\kre}\r)^{2\alpha}\l(\frac{k}{\ke}\r)^{\nb-2} .
    \end{align}
\end{subequations}
Whereas for the sub-horizon modes $(k>\kre)$, the magnetic and electric spectra behave as~\cite{Maiti:2025awl}
\begin{align}
    \tmPb(k>\kre,\ere)\simeq \frac{\ke^4}{4\pi}\frac{\qnb^2(n)}{\pi}\l(\frac{k}{\ke}\r)^{\nb-2(\alpha+1)} ,\\
    \tmPe(k>\kre,\ere)\simeq \frac{\ke^4}{4\pi}\qnb^2(n)\l(\frac{k}{\ke}\r)^{\nb-2(\alpha+1)} .
\end{align}
Where $\qnb^1(n)$ and $\qnb^2(n)$ is defined as~\cite{Maiti:2025awl}
\begin{subequations}
    \begin{align}\label{eq:fnb1}
    \qnb^1(n)&=\frac{2^{2n+1}(\alpha-n)^2\Gamma^2(n+1/2)}{\cos^2(\pi\alpha)\Gamma^2(1/2-\alpha)\Gamma^2(\alpha+1/2)} ,\\
    \qnb^2(n)&=\frac{2^{2(n+\alpha)}(\alpha-n)^2\Gamma^2(n+1/2)}{\cos^2(\pi\alpha)\Gamma^2(1/2-\alpha)} .
\end{align}
\end{subequations}
We find that, although the electric and magnetic spectra follow simple power laws during inflation, reheating changes their scale dependence and turns them into broken power-law spectra. Both spectra peak around $k=\kre$, which means that most of the electromagnetic energy is carried by modes with $k\sim\kre$. In addition, post-inflationary production introduces an extra enhancement factor, $(\ke/\kre)^{2(\alpha+1)}$, in both the magnetic and electric energy spectra. The amount of this enhancement depends on the reheating dynamics $(\wre,\Tre)$ and the magnetogenesis parameters $(n)$.

\begin{figure}
    \centering
    \includegraphics[width=0.99\linewidth]{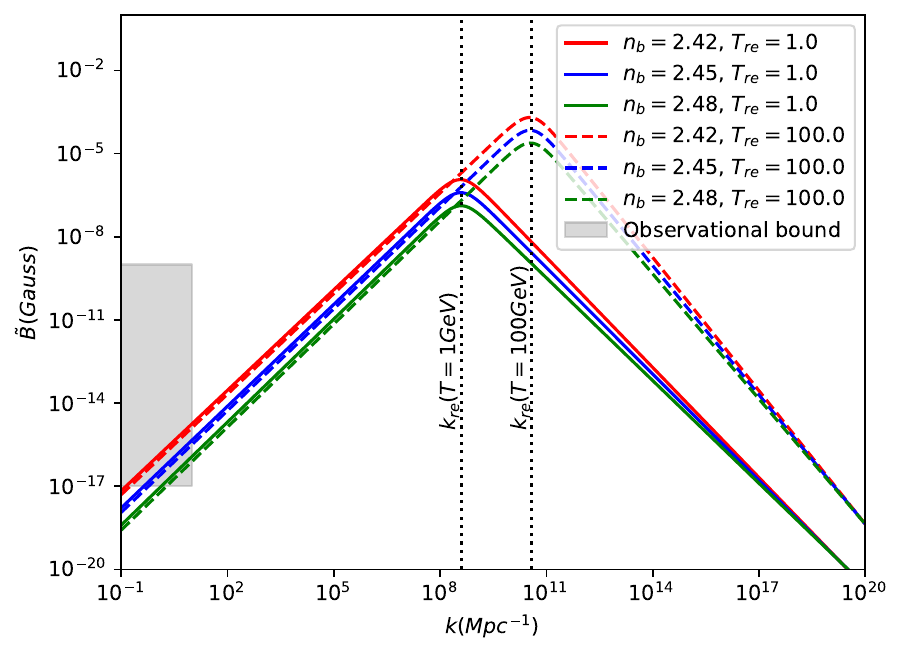}
    \caption{Comoving magnetic field strength as a function of comoving wavenumber $k\,(\Mpc^{-1})$ for three values of the magnetic spectral index, $\nb=2.42$, $2.45$, and $2.48$. Solid and dashed lines correspond to reheating temperatures $\Tre=1\,\GeV$ and $\Tre=10^2\,\GeV$, respectively. The gray shaded region denotes the observational limit on large-scale magnetic fields.}
    \label{fig:comoving_mag}
\end{figure}
In Fig.~\ref{fig:comoving_mag}, we show the comoving magnetic field strength. i.e., $\tilde{B}=a^2B$, as a function of comoving wavenumber for two reheating temperatures, $\Tre=1\,\GeV$ (solid lines) and $\Tre=10^2\,\GeV$ (dashed lines). The different colors correspond to $\nb=2.42$, $2.45$, and $2.48$. The gray shaded regions represent the observational bounds on large-scale magnetic fields from CMB~\cite{Paoletti:2008ck, Planck:2015zrl, Zucca:2016iur, paoletti2022constraints, BICEP2:2017lpa} and $\gamma$-ray observations~\cite{MAGIC:2022piy,takahashi2011lower,Arlen_2014,Arlen:2012iy}.

We find that the magnetic-field spectral tilt for $k<\kre$ is set by the magnetogenesis parameter $n$, namely $\tilde{B}^2(k)\propto k^{4-2|n|}$. On the other hand, modes that reenter the horizon during reheating, $k>\kre$, follow a red-tilted spectrum with spectral index $\nbs=\nb-2(\alpha+1)$. Therefore, both the amplitude and the shape of the spectrum depend on the reheating dynamics as well as the magnetogenesis parameters.
As the reheating temperature increases, the peak of the spectrum shifts to larger wavenumbers since the comoving horizon scale at the end of reheating becomes smaller. While the magnetic-field strength is only weakly affected by the reheating temperature, the total magnetic energy density depends strongly on the duration of reheating.
We also find that smaller values of $\nb$ lead to a larger magnetic-field amplitude on large scales. This enhancement originates from the factor $(\ke/\kre)^{2(\alpha+1)}$, which carries the dependence on the magnetogenesis parameters. As shown in the figure, suitable choices of these parameters can generate large-scale magnetic fields consistent with current observational bounds~\cite {Paoletti:2008ck, Planck:2015zrl, Zucca:2016iur,paoletti2022constraints, BICEP2:2017lpa,2011A&A...529A.144T,2010Sci...328...73N, PhysRevLett.116.191302}.

Recently, it was pointed out that sufficiently strong primordial magnetic fields present before the electroweak symmetry breaking (EWSB) can generate baryon isocurvature perturbations through electroweak sphaleron processes, leading to stringent constraints on inflationary magnetogenesis scenarios~\cite{Kamada:2016cnb,Kamada:2016eeb,Fujita:2016igl,Kamada:2020bmb}. However, the analysis of Ref.~\cite{Kamada:2020bmb} assumes that the Universe enters the standard radiation-dominated era immediately after inflation and that significant magnetic fields are already present before the electroweak crossover. In contrast, our scenario includes a prolonged matter-like reheating phase ($w_{\rm re}=0$), during which the dominant amplification of the gauge field occurs. Moreover, throughout this work we consider a low reheating temperature, $T_{\rm re}<T_{\rm EW}$, such that the dominant modes, which peak around $k\simeq\kre$, re-enter the horizon only after the electroweak crossover. Therefore, the baryon isocurvature constraint derived in Ref.~\cite{Kamada:2020bmb} is not directly applicable to our setup. A dedicated analysis of baryon isocurvature perturbations in the presence of a non-standard reheating history is beyond the scope of the present work and is left for future investigation.

\section{Magnetically induced curvature perturbations}
 In this process, magnetic fields are continuously generated during reheating, leading to a broken power-law spectrum and the energy density reaches its maximum at the end of reheating, i.e., $\eta\simeq\ere$. If the electric or magnetic field strength is sufficiently large, then it will automatically induce enhanced secondary curvature perturbations that can easily overtake the primary curvature perturbations near $k\sim\kre$. The comoving curvature perturbations on uniform-density hypersurfaces is defined as~\cite{Bassett:2005xm, Wands:2000dp, PhysRevD.94.043523, Baumann:2009ds, Malik:2008im}
\begin{align}
    \zeta=-\Phi-\mH\frac{\delta\rho}{\rho'}\,,
\end{align}
where $\Phi$ is the Bardeen potential, $\delta\rho$ is the density fluctuation, $\rho'$ denotes the background energy-density evolution, and $\mH=a'/a$ is the conformal Hubble parameter~\cite{Bassett:2005xm, Wands:2000dp, PhysRevD.94.043523, Baumann:2009ds, Maiti:2025ijr}.

Magnetic fields generate curvature perturbations both through energy-density fluctuations, $\mH\delta\rhoem/\rho'$, and indirectly through the Bardeen potential $\Phi$~\cite{PhysRevD.88.083515, PhysRevD.81.043517, Saga_2020}.
 Since the magnetic field peaks at the end of reheating, after which standard radiation domination begins, the induced curvature power spectrum can be written as~\cite{Maiti:2025ijr}
\begin{align}\label{eq:P_zeta_deltab}
    \mPc^{\rm ind}(k)\simeq \frac{1}{16}\frac{P_{\delta\rhob}(k)}{\rho_{\rm ra}^2}\,.
\end{align}
where $P_{\delta\rhob}$ is the power spectrum associated with the density fluctuations induced by the magnetic field, which can be written as~\cite{Fujita:2013qxa, PhysRevD.94.043523, Maiti:2025ijr}
\begin{align}\label{eq:P_deltab}
    P_{\delta\rhob}(k,\eta)=\int \frac{\d q}{q}\int_{-1}^1\d \gamma\,\frac{(1-\gamma^2)\mPb(q,\eta)\mPb(|\vk-\vq|,\eta)}{\l(1+\frac{q^2}{k^2}-2\gamma\frac{q}{k}\r)^{5/2}},
\end{align}
where $\gamma=\hat{k}\cdot\hat{q}=\cos(\theta)$ where $\theta$ is the angle between two vector $\vk$ and $\vq$.
Here, $\mPb(k)=\tmPb(k,\eta)a^{-4}$ denotes the magnetic power spectrum defined in Eq.~\eqref{eq:mag_spec}. During the radiation-dominated era, both the magnetic and background radiation energy densities scale as $a^{-4}$, so their relative energy density remains nearly constant.

We numerically evaluate Eq.~\eqref{eq:P_deltab} to compute the curvature power spectrum induced by the magnetic field. The resulting spectrum follows a broken power-law form. On large scales, the inflationary contribution dominates, giving $\mPcv(k\ll\kre)\propto k^{\ns-1}$. For modes close to the reheating scale, the spectrum behaves as $\mPci(k\leq\kre)\propto k^{2\nb}$ for $\nb\leq3/2$, while for $\nb>3/2$ it scales as $\mPci(k<\kre)\propto k^3$. For modes re-entering the horizon around the end of reheating, $k\geq\kre$, the spectrum scales as
$\mPci(k\geq\kre)\propto k^{\nb-2(\alpha+1)}$. At much smaller scales, the magnetic contribution becomes negligible and the spectrum approaches the standard inflationary behavior, $\mPcv(k\gg\kre)\propto k^{\ns-1}$.

\begin{figure}
    \centering
    \includegraphics[width=0.97\linewidth]{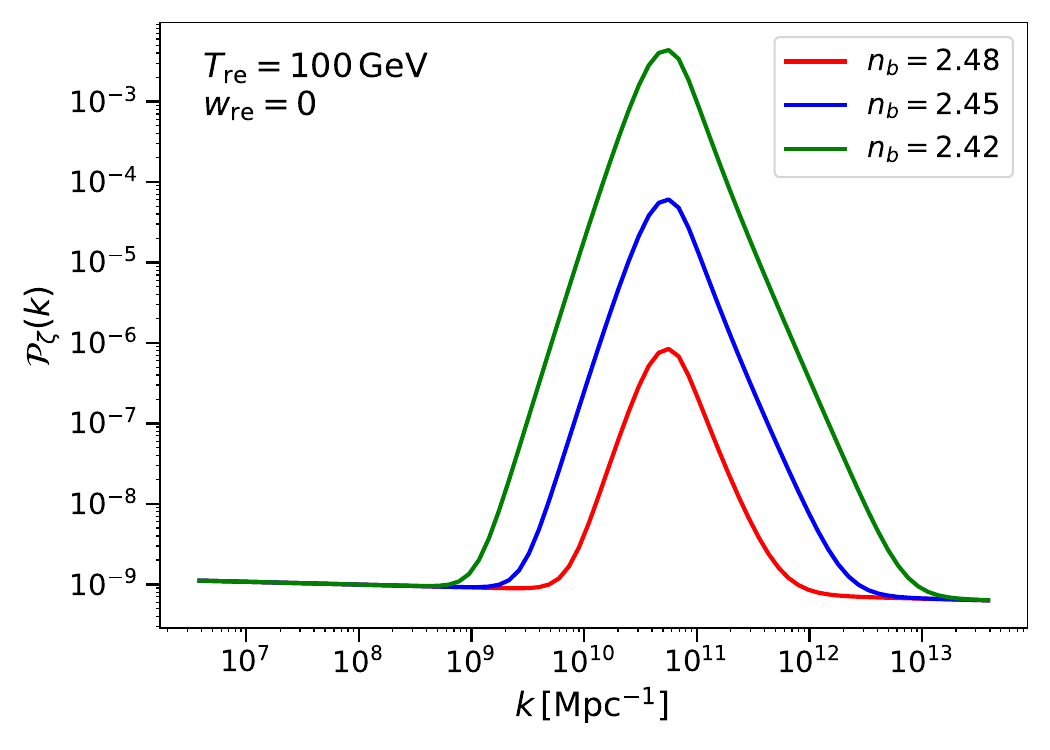}
    \caption{Total curvature power spectrum (primary + secondary) as a function of wavenumber $k\,(\Mpc^{-1})$ for a fixed reheating scenario with $\wre=0$ and $\Tre=100\,\GeV$. The three curves correspond to magnetic spectral indices $\nb=2.42$, $2.45$, and $2.48$.}
    \label{fig:curvature_power}
\end{figure}

Combining these behaviors, the curvature power spectrum can be approximated as
\begin{align}\label{eq:curvature}
    \mPc(k)\simeq 
    \l\{
    \begin{matrix}
        \As (k/k_{*})^{\ns-1} & k_{*}\leq k<\ksb,\\
        A_\zeta (k/\kre)^{\nc} & \ksb <k\leq\kre,\\
        A_\zeta (k/\kre)^{\nb-2(\alpha+1)} & \kre<k<\ksbs,\\
        \As (k/k_{*})^{\ns-1} &\ksbs<k<\ke,
    \end{matrix}
    \r.
\end{align}
where $\nc$ is defined as
\begin{align}
    \nc=\l\{ 
    \begin{matrix}
        2\nb & \text{for}~\nb\leq 3/2\\
        3 & \text{for}~\nb>3/2
   \end{matrix}
   \r.
\end{align}
Here $\As\simeq 2.1\times 10^{-9}$ denotes the amplitude of the curvature power spectrum at CMB scales~\cite{Planck:2018jri}. Here, $A_\zeta$ is the amplitude of the peak of the curvature power spectrum derived from the magnetic field, and it can be defined as~\cite{Maiti:2025ijr}
\begin{align}\label{eq:curv_peak}
  A_\zeta \simeq \frac{1}{24}\l(\frac{\qnb^1}{24\pi}\r)^2\l(\frac{\HI}{\Mp}\r)^4\l(\frac{\are}{\ae}\r)^{2\alpha-\nb}.
    \end{align}
In the above,  $\ksb$ and $\ksbs$ correspond to the mode associated with the first and third Spectral Breaks (SB), respectively. For $\ksb<k<\ksbs$, the secondary (magnetically induced) contribution dominates, while for $k<\ksb$ and $k>\ksbs$ the inflationary curvature power spectrum becomes the dominant component (see in Fig.\ref{fig:curvature_power}).

In Fig.~\ref{fig:curvature_power}, we plot the total curvature power spectrum, $\mPc(k)=\mPcv(k)+\mPci(k)$, as a function of the comoving wavenumber $k$ in $\Mpc^{-1}$. We fix the reheating parameters to $\wre=0$ and $\Tre=10^2\,\GeV$, and consider three values of the magnetic spectral index, $\nb=2.42$, $2.45$, and $2.48$.
Since the reheating temperature is fixed, the peak of the spectrum appears at the same position for all values of $\nb$. For $\nb>3/2$, the induced curvature power spectrum has a similar shape in the intermediate region, while the slope for $k>\kre$ shows a dependence on $\nb$. Because the variation in $\nb$ is small, the overall spectral shape remains almost unchanged.

However, the amplitude of the induced curvature power spectrum changes noticeably with $\nb$, reflecting the sensitivity of the magnetic-field energy density to this parameter. As a result, this class of magnetogenesis models can produce a significant enhancement of curvature perturbations at intermediate scales. Such an enhancement may lead to the formation of primordial black holes (PBHs), which can provide a viable dark matter candidate, as discussed in~\cite{Maiti:2025ijr}.

\section{Secondary Gravitational Waves}
Gravitational waves provide a direct probe of the early Universe. Since they decouple at very high energies, they can carry information about inflation and the physical processes active during the post-inflationary era.
In the present scenario, a significant amount of magnetic fields are generated after inflation. The corresponding electromagnetic energy density can become a few percent of the background energy density, i.e., $\rhoem/\rho_c\sim \mathcal{O}(0.1)$, which naturally leads to the production of secondary gravitational waves~\cite{Maiti:2026lvx,Maiti:2025awl}.

The magnetic field contributes to GW production in two ways. First, it directly sources tensor perturbations through its anisotropic stress~\cite{Sorbo:2011rz,Caprini:2014mja,Ito:2016fqp,Sharma:2019jtb,Okano:2020uyr,Maiti:2025rkn,Maiti:2025awl,Maiti:2025cbi,Maiti:2025ijr}. Second, it can generate enhanced curvature perturbations~\cite{Fujita:2013qxa,PhysRevD.94.043523,Maiti:2025ijr}, which later source tensor modes through quadratic scalar. This second contribution is known as scalar-induced gravitational waves~\cite{PhysRevD.64.123514,Di:2017ndc,Fu:2019vqc,PhysRevD.103.083510,Bhaumik:2020dor,Solbi:2021wbo,Figueroa:2021zah,Domenech:2020kqm,Cai:2018dig,Domenech:2025ffb,Domenech:2024rks,Domenech:2021ztg,Domenech:2020xin,Domenech:2020kqm,Domenech:2019quo,Balaji:2023ehk,Balaji:2022dbi, Inui:2024fgk}. Since the magnetic and scalar perturbations evolve differently, they produce distinct features in the GW spectrum.
Below, we derive the evolution equations for tensor modes and compute the GW spectra from both the magnetic and scalar-induced contributions.

The equation of motion for the tensor perturbations $h_{ij}(\vx,\eta)$ is given by~\cite{Acquaviva:2002ud, Baumann:2007zm, Caprini:2018mtu}
\begin{align}\label{eq:hk_master}
    \hij''+2\mH\hij'-\nabla^2\hij=\Pijmn\l[ \mSmn+\frac{2}{\Mp^2}T_{mn} \r],
\end{align}
where $\mSmn$ arises from quadratic combinations of scalar perturbations~\cite{PhysRevD.64.123514, Di:2017ndc, Fu:2019vqc, PhysRevD.103.083510, Bhaumik:2020dor, Solbi:2021wbo, Figueroa:2021zah, Ragavendra:2020sop, Ragavendra:2023ret, Domenech:2025ffb, Domenech:2024rks, Balaji:2023ehk, Balaji:2022dbi, Domenech:2021ztg, Domenech:2020xin, Domenech:2020kqm, Domenech:2019quo, Dimastrogiovanni:2022eir, Cai:2018dig, Cai:2019cdl}, and $T_{mn}$ denotes the anisotropic part of the energy-momentum tensor, sourced here by the electromagnetic fields~\cite{Sorbo:2011rz,Caprini:2014mja,Ito:2016fqp,Sharma:2019jtb,Okano:2020uyr, Maiti:2025rkn,Maiti:2025awl, Maiti:2025cbi, Maiti:2025ijr}. 

The operator $\Pijmn$ projects onto the transverse-traceless (TT) subspace, and in Fourier space it is defined as
\begin{align}
{\mrP}^{mn}_{ij}(k)=\mrP^m_i(k)\mrP^n_j(k)-\frac{1}{2}\mrP^{mn}(k)\mrP_{ij}(k),
\end{align}
with $\mrP_{ij}=\delta_{ij}-\partial_i\partial_j/\Delta$~\cite{Acquaviva:2002ud, Baumann:2007zm}.

It is convenient to work in Fourier space, where the tensor perturbations is decomposed as
\begin{align}
    h_{ij}(\vx,\eta)=\int \frac{d^3\vk}{(2\pi)^{3/2}} e^{i\vk\cdot\vx}\Big[ h_{\vk}^{+}(\eta)e^{+}_{ij}(\vk)+h^{\times}_{\vk}(\eta)e^{\times}_{ij}(\vk)\Big].
\end{align}

The polarization tensors are constructed from two orthonormal basis vectors orthogonal to $\vk$:
\begin{subequations}
\begin{align}
    e_{ij}^{+}(\vk)=\frac{1}{\sqrt{2}}\l( e_i^{+}(\vk)e^{+}_j(\vk)-e^{\times}_i(\vk)e^{\times}_j(\vk)\r),\\
    e_{ij}^{\times}(\vk)=\frac{1}{\sqrt{2}}\l( e_i^{+}(\vk)e^{\times}_j(\vk)+e^{\times}_i(\vk)e^{+}_j(\vk)\r).
\end{align}
\end{subequations}
where \(e_i^{+}(\vk)\) and \(e_i^{\times}(\vk)\) are two orthonormal polarization vectors transverse to the wavevector \(\vk\). They satisfy
\begin{align}
    \vk \cdot \ve^{+}(\vk) = 0, \qquad
    \vk \cdot \ve^{\times}(\vk) = 0,
\end{align}
along with the orthonormality conditions
\begin{align}
    \ve^{+}(\vk)\cdot \ve^{\times}(\vk) = 0, \qquad
    |\ve^{+}(\vk)| = |\ve^{\times}(\vk)| = 1.
\end{align}
The corresponding polarization tensors \(e_{ij}^{+}\) and \(e_{ij}^{\times}\) describe the two independent tensor modes, namely the plus and cross polarizations.
For each polarization mode $\lambda$, the equation of motion reduces to~\cite{Sorbo:2011rz, Caprini:2014mja,Sharma:2019jtb}
\begin{align}\label{eq:hk_em}
    h_{\vk}^{''(\lambda)}+2\mH h_{\vk}^{'(\lambda)}+k^2 h_{\vk}^{(\lambda)}=\mSs(\vk,\eta)+\mSem(\vk,\eta).
\end{align}

The scalar-induced source term is given by~\cite{PhysRevD.64.123514, Di:2017ndc, Fu:2019vqc, PhysRevD.103.083510, Bhaumik:2020dor, Solbi:2021wbo, Figueroa:2021zah, Ragavendra:2020sop, Ragavendra:2023ret, Domenech:2025ffb,Domenech:2024rks,Balaji:2023ehk,Balaji:2022dbi,Domenech:2021ztg,Domenech:2020xin,Domenech:2020kqm,Domenech:2019quo,Dimastrogiovanni:2022eir,Papanikolaou:2022chm, Maity:2024odg, Kohri:2018awv,Cai:2018dig, Cai:2019cdl,Inomata:2016rbd}
\begin{align}\label{eq:tensor_scalr_source}
    \mSs &(\vk,\eta)=4\,e^{ij}_\lambda(\vk)\int \frac{d^3\vq}{(2\pi)^{3/2}}q_iq_j\Big(2\Phi_{\vq}\Phi_{\vk-\vq}+\nn\\
   & \frac{4}{3(1+w)}(\mH^{-1}\Phi'_{\vq}+\Phi_\vq)(\mH^{-1}\Phi_{\vk-\vq}'+\Phi_{\vk-\vq})\Big),
\end{align}
which arises from second-order scalar perturbations.

The electromagnetic contribution is~\cite{Sorbo:2011rz, Caprini:2014mja, Ito:2016fqp, Sharma:2019jtb, Okano:2020uyr, Maiti:2025cbi, Maiti:2025rkn, Maiti:2024nhv, Bhaumik:2025kuj,Ragavendra:2026fgs, Maiti:2026lvx}
\begin{align}
    \mSem &=-\f{2}{\Mp^2}e^{ij}_{\lambda}({\bm k})
\int \f{\d^3\vq}{(2\pi)^{3/2}} [E_i(\vq,\eta)E_j(\vk-\vq,\eta)\nn\\
&\quad +B_i(\vq,\eta)B_j(\vk-\vq,\eta)],
\end{align}
which explicitly shows that both electric and magnetic fields contribute through their anisotropic stresses. In the present setup, the magnetic component typically dominates after inflation, although the electric contribution can be relevant depending on the magnetogenesis mechanism.

We solve Eq.~\eqref{eq:hk_em} using the Green's function method. The full solution can be written as
\begin{align}\label{eq:hk_sol_def}
    \hk^{(\lambda)}(\eta)= \hk^{\rm{vac}(\lambda)}(\eta)+\int_{\eta_i}^{\eta}\d \teta\, \mGk(\eta,\teta)(\mSs+\mSem),
\end{align}
where $\hk^{\rm{vac}(\lambda)}(\eta)$ denotes the homogeneous (vacuum) solution~\cite{Starobinsky:1979ty,Grishchuk:1974ny,Guzzetti:2016mkm,Haque:2021dha,Boyle:2005se,Hoory:2025qgm}.

The retarded Green's function $\mathcal{G}_k(\eta,\eta_1)$ satisfies
\begin{align}
    \left[
    \frac{\partial^2}{\partial \eta^2}
    + 2\mathcal{H}\frac{\partial}{\partial \eta}
    + k^2
    \right]
    \mathcal{G}_k(\eta,\eta_1)
    =
    \delta^{(3)}(\eta-\eta_1).
\end{align}

During the radiation-dominated era, the Green's function takes the form~\cite{Maiti:2024nhv}
\begin{align}\label{eq:green_function}
    \mGk(\eta,\eta_1)
    =
    \Theta(\eta-\eta_1)
    \frac{\eta_1\,\sin(k(\eta_1-\eta))}{k\eta}.
\end{align}
Assuming statistical independence of the two polarization states, the two-point correlation function is given by~\cite{Acquaviva:2002ud, Baumann:2007zm, Sorbo:2011rz, Caprini:2014mja, Sharma:2019jtb}
\begin{align}\label{eq:Ph_def}
    \langle \hk^{(\lambda)}(\eta)\,
    h^{*(\lambda')}_{\vk'}(\eta)\rangle
    =\delta^{(3)}(\mathbf{k}-\mathbf{k}')\,
    \delta_{\lambda\lambda'}\,
    \frac{2\pi^2}{k^3}\, \mPh^{\lambda}(k,\eta).
\end{align}
In the present non-helical magnetogenesis scenario, both polarization modes are equally enhanced. So, for the rest of the calculations, we dropped this index.

The total tensor power spectrum receives three contributions:
\begin{align}
    \mPh(k,\eta)=\mPhv(k,\eta)+\mPhm(k,\eta)+\mPhs(k,\eta),
\end{align}
where $\mPhv$ corresponds to the primary (vacuum) tensor modes generated during inflation, $\mPhm$ arises from direct magnetic field sourcing, and $\mPhs$ denotes the scalar-induced contribution.

The behavior of the primary tensor spectrum during reheating, particularly in a matter-like background, has been extensively studied~\cite{Haque:2021dha, Maiti:2025cbi}. Since our primary interest lies in the effects of post-inflationary magnetogenesis, we focus in the following on the secondary contributions $\mPhm$ and $\mPhs$, and analyze their relative amplitudes, spectral shapes, and time evolution.

\paragraph{\bf Present-Day Gravitational Wave Spectrum:}
Once produced, gravitational waves propagate essentially freely due to their weak coupling to matter. Neglecting possible GW self-interactions, sub-Hubble tensor modes evolve as free radiation after production. Their energy density therefore scales as $\rhogw\propto a^{-4}$, while the physical momentum redshifts as $k/a$.

During the radiation-dominated era, the GW energy density normalized to the critical density is given by
\begin{align}\label{eq92}
\ogw(k,\eta)=\frac{\rhogw(k,\eta)}{\rho_c(\eta)}=\frac{1}{12}\frac{k^2\Pt(k,\eta)}{a^2(\eta)H^2(\eta)}.
\end{align}
Here, $\rho_c(\eta)=3H^2(\eta)\Mp^2$, with $\Mp\simeq 2.43\times10^{18}\Gev$, and $\Pt(k,\eta)$ denotes the tensor power spectrum during radiation domination.

Since the GW energy density redshifts like radiation after production, the present-day spectrum can be related to its value during the radiation era through
\begin{align}
\ogw(k)h^2 
&\simeq \left(\frac{\gsp}{\gseq}\right)^{1/3}\Omega_Rh^2\ogw(k,\eta).
\end{align}
Here, $\Omega_Rh^2\simeq 4.3\times 10^{-5}$ is the present-day radiation density parameter, while $\gseq\simeq\gsp\simeq 3.35$ denote the effective relativistic degrees of freedom at matter-radiation equality and at present, respectively.

\subsection{Tensor Power Spectrum Sourced by Magnetic Field Anisotropies}

We first consider tensor perturbations sourced directly by post-inflationary magnetic fields. 
After generation during inflation and reheating, the magnetic field evolves adiabatically. Although magnetohydrodynamic (MHD) effects such as turbulence can modify the evolution of magnetic fields, these effects mainly influence modes well inside the horizon. Since in this work we are interested in modes near the horizon scale at the end of reheating, we neglect such effects~\cite{Maiti:2025rkn, Maiti:2025awl, Maiti:2025cbi}.

During radiation domination, both the background and magnetic energy densities scale as $a^{-4}$. Under these assumptions, the tensor power spectrum sourced by magnetic fields is given by~\cite{Maiti:2024nhv,Atkins:2025pvg}
\begin{align}\label{eq:ph_mag_s}
\mPhm(k,\eta) &=\frac{2}{\Mp^4}
\l( \int_{\eta_i}^{\eta}\d \eta_1\,\frac{\mGk(\eta,\eta_1)}{a^2(\eta_1)}\r)^2\times\nn\\
&\times\int_0^{\infty} \frac{\d q}{q}\int_{-1}^1 \d \mu \,\frac{\mathcal{F}(k, q, \mu){\tmPb}(q){\tmPb}(|\vk-\vq|)}{[1+(q/k)^2-2\mu(q/k)]^{3/2}}.
\end{align}
Here, $\tmPb(k)=a^4\mPb(k,\eta)$ denotes the comoving magnetic energy spectrum. The kernel $\mF(k,q,\mu)$ for a non-helical magnetic field is given by $\mF(k,q,\gamma)=2(1+\gamma^2)(1+\mu^2)$, where $\gamma=\hat{k}\cdot\hat{q}=\cos(\theta)$ and $\mu=\widehat{\vk-\vq}\cdot\hat{k}$~\cite{Maiti:2025rkn,Maiti:2025awl, Maiti:2025cbi, Ragavendra:2026fgs, Atkins:2025pvg}.

For computational convenience and to facilitate comparison with scalar-induced gravitational waves, we introduce the dimensionless variables
\begin{align}\label{eq:v_u}
    v=\frac{q}{k}, \qquad u=\frac{|\vk-\vq|}{k}.
\end{align}
In terms of these variables, Eq.~\eqref{eq:ph_mag_s} becomes
\begin{align}\label{eq:ph_mag}
    \mPhm(k,\eta)=\frac{18}{\rho_c^2(\eta)}\l( \int_{\xre}^{x}\d x_1 \frac{\tilde{\mGk}(x,x_1)}{x_1^2} \r)^2\nn\\
    \times \int_{0}^{\infty}\d v\int_{|1-v|}^{1+v}\d u \frac{\mF(u,v)}{u^2v^2}\mPb(ku)\mPb(kv),
\end{align}
where $\tilde{\mGk}=k\mGk$. Here we have used the following relation $\xre^4/\Mp^4k^4\are^4=9/\are^8\rho^2_c(\ere)$, which is valid for a radiation-dominated background.

In this parametrization, the kernel takes the form~\cite{Ragavendra:2026fgs, Atkins:2025pvg}
\begin{align}
    \mF(u,v)=\frac{[4v^2+(1+v^2-u^2)^2][4u^2+(1-v^2+u^2)^2]}{8v^2u^2}.
\end{align}
 In the limit $v \gg 1$, one has $u \rightarrow v$, and the kernel $\mF(u,v)$ becomes approximately independent of $u$ and $v$. In contrast, in the limit $v \ll 1$, one finds $u \simeq 1$, leading to $\mF(u,v\ll1)\sim 1/v^2$. This indicates that the integral receives contributions from the infrared region, which depends on the shape of the magnetic power spectrum.

During the radiation-dominated era, the retarded Green's function is governed by Eq.~\eqref{eq:green_function}. Performing the time integral, we define
\begin{align}
    \mI_{\rm RD}(x,\xre)=\int_{\xre}^{x}\d x_1 \frac{\tilde{\mGk}(x,x_1)}{x_1^2}=\frac{1}{x}\int_{\xre}^x \frac{dx_1}{x_1}\sin(x-x_1).
\end{align}
For super-horizon limit $x \ll 1$, we obtain
\begin{align}
    \lim_{x<<1}\overline{\mI^2_{\rm RD}(x,\xre)}\simeq (\ln(x/\xre)-1)^2
\end{align}
which implies a logarithmic growth of the tensor power spectrum during the radiation-dominated era. Correspondingly, the spectral energy density of gravitational waves grows as $\ogw(x\ll1)\propto x^2 (\ln(x/\xre)-1)^2$,
reflecting the continuous sourcing of tensor modes while they remain outside the horizon.

For sub-horizon limit, i.e., $x \gg 1$, the integral behaves as
\begin{align}
    \lim_{x>>1}\overline{\mI_{\rm RD}^2(x,\xre)}\simeq \frac{\pi^2}{8x^2}.
\end{align}
Although this shows a decay of the tensor amplitude, the gravitational-wave energy density $\ogw\sim (k^2/a^2H^2)\overline{h^2}$ grows logarithmically with time, since $x\equiv k\eta \sim k/(aH)$ increases as the Universe expands.

To determine the precise spectral behavior of the tensor power spectrum and the resulting gravitational wave spectrum, we numerically evaluate the above integral. The detailed results will be presented in a later section.

\subsection{Tensor power spectrum sourced by the magnetically induced curvature perturbations:}

Scalar perturbations sourced by the magnetic field generate secondary tensor modes through quadratic terms in the gravitational potential $\Phi$, as given in Eq.~\eqref{eq:tensor_scalr_source}. In a radiation-dominated era, the scalar-induced tensor power spectrum can be written as~\cite{PhysRevD.64.123514, Di:2017ndc, Fu:2019vqc, PhysRevD.103.083510, Bhaumik:2020dor, Solbi:2021wbo, Figueroa:2021zah, Ragavendra:2020sop, Ragavendra:2023ret, Domenech:2025ffb,Domenech:2024rks,Balaji:2023ehk,Balaji:2022dbi,Domenech:2021ztg,Domenech:2020xin,Domenech:2020kqm,Domenech:2019quo,Dimastrogiovanni:2022eir,Papanikolaou:2020qtd, Inomata:2016rbd}.
\begin{align}\label{eq:ph_scalar}
    \mPhs(k,\eta) &=4\int_{0}^{\infty}\d v\int_{|1-v|}^{1+v}\d u \left(\frac{4v^2-(1+v^2-u^2)^2}{4u v}\right)^2\nn\\
    &\times \mIrds^2(v,u,x)\mPc(vk)\mPc(uk)\,.
\end{align}

Here, $v$ and $u$ are the same dimensionless variables introduced in Eq.\eqref{eq:v_u}. The kernel $\mIrds(v,u,x)$ encodes the time evolution of the tensor modes and is defined as~\cite{Papanikolaou:2022chm}
\begin{align}\label{eq:Ird_scalar}
    \mIrds(v,u,x)=\int_{\xre}^{x}\d x_1 \tilde{\mGk}(x,x_1) f(v,u,x_1)\,,
\end{align}
where the function $f(v,u,x_1)$ contains the information about the scalar source term, and it is given by~\cite{PhysRevD.64.123514, Di:2017ndc, Fu:2019vqc, PhysRevD.103.083510, Bhaumik:2020dor, Solbi:2021wbo, Figueroa:2021zah, Ragavendra:2020sop, Ragavendra:2023ret, Domenech:2025ffb,Domenech:2024rks,Balaji:2023ehk,Balaji:2022dbi,Domenech:2021ztg,Domenech:2020xin,Domenech:2020kqm,Domenech:2019quo,Dimastrogiovanni:2022eir,Papanikolaou:2022chm, Maity:2024odg, Kohri:2018awv,Cai:2018dig, Cai:2019cdl}
\begin{align}\label{eq:fvux}
    &f(v,u,x_1) =\frac{4}{3}\Phi_{\vq}(vx_1)\Phi_{\vk-\vq}(ux_1)\nn\\
   & +\frac{4 x_1}{9}\Big(\partial_{\eta_1}\Phi_{\vq}(vx_1)\Phi_{\vk-\vq}(ux_1)+\Phi_{\vq}(vx_1)
    \partial_{\eta_1}\Phi_{\vk-\vq}(ux_1)\Big)\nn\\
   & +\frac{4 x_1^2}{9}\partial_{\eta_1}\Phi_{\vq}(vx_1)\partial_{\eta_1}\Phi_{\vk-\vq}(ux_1)\,.
\end{align}
In the above  Eq.\eqref{eq:fvux}, we consider a radiation-dominant universe, so we put $w=1/3$ and simplify the $f(v,ux_1)$.

For the super-horizon limit $x \ll 1$, the gravitational potential remains approximately constant in time. Consequently, $f(v,u,x_1) \sim \Phi(vx_1)\Phi(ux_1)$ becomes effectively independent of $x_1$, since $\Phi(x_1 \ll 1) \sim \text{const}$. Substituting this behavior into the integral expression for $\mIrds$, one finds that $\mIrds(v,u,x_1 \ll 1)\propto\l(\ln(x/\xre)-1\r)$, which implies
$\ogw(x\ll1)\propto x^2 \l(\ln(x/\xre)-1\r)^2$. Therefore, the GW energy density grows with time while the corresponding mode remains outside the horizon.

For the sub-horizon limit, i.e., $x \gg 1$, the gravitational potential decays after horizon entry. In this regime, one can approximate $f(v,u,x_1) \sim 1/x_1^2$. Using this asymptotic behavior in Eq.\eqref{eq:Ird_scalar}, the kernel admits the following late-time form~\cite{PhysRevD.64.123514, Di:2017ndc, Fu:2019vqc, PhysRevD.103.083510, Bhaumik:2020dor, Solbi:2021wbo, Figueroa:2021zah, Ragavendra:2020sop, Ragavendra:2023ret, Domenech:2025ffb,Domenech:2024rks,Balaji:2023ehk,Balaji:2022dbi,Domenech:2021ztg,Domenech:2020xin,Domenech:2020kqm,Domenech:2019quo,Dimastrogiovanni:2022eir,Papanikolaou:2022chm, Maity:2024odg, Kohri:2018awv,Cai:2018dig, Cai:2019cdl,Inomata:2016rbd}
\begin{align}
    \overline{\mIrds^2(v,u,x\rightarrow\infty)}\simeq \frac{1}{2x^2}\left( \frac{3(v^2+u^2-3)}{4v^3u^3}\right)^2\nn\\
    \times \Bigg( \left( -4vu+(v^2+u^2-3)\log\left| \frac{3-(v+u)^2}{3-(v-u)^2}\right|\right)^2\nn\\
    +\pi^2(v^2+u^2-3)^2\Theta(v+u-\sqrt{3})\Bigg)\,.
\end{align}
From the above analysis, it is clear that both the magnetic and scalar-induced sources show similar time dependence. The scalar-induced gravitational waves, however, include an additional contribution in the kernel arising from the structure of the source term, which appears in the momentum integral.
To determine the spectral shape of the resulting gravitational waves, including both magnetic-induced and scalar-induced contributions, we solve the above integral numerically.

In this work, we focus on the production of secondary gravitational waves after the end of reheating. Although the gauge field is amplified during both inflation and reheating, the inflationary contribution is too small to produce an observable gravitational-wave signal for the parameter space considered here. Instead, the dominant amplification occurs during reheating due to the time-dependent gauge-field coupling, causing the magnetic field to reach its maximum strength at the end of reheating. Consequently, the magnetic anisotropic stress, which sources both tensor and scalar perturbations, is also maximized at this epoch. As shown in Ref.\cite{Maiti:2025awl}, the resulting secondary gravitational-wave signal is generated predominantly during the subsequent radiation-dominated era. Since both the magnetic and radiation energy densities scale as $a^{-4}$ after reheating, the magnetic anisotropic stress remains an efficient source of gravitational waves throughout radiation domination. Furthermore, after horizon re-entry, the amplification of UV modes ceases once the condition $k^2 \geq I''/I$ is satisfied [see Eq.\eqref{eq:gauge_field} and Ref.~\cite{Maiti:2025awl} for details].

\subsection{Spectral energy density of the GW at present}
\paragraph{\underline{Primary GWs:}}
Primordial gravitational waves (PGWs) arise from vacuum tensor fluctuations generated during inflation and acquire a nearly scale-invariant spectrum in a de Sitter background. Their present-day amplitude and spectral shape depend on the subsequent expansion history of the Universe. The non-standard reheating can significantly modify modes that re-enter the horizon before reheating ends. The resulting present-day PGW spectrum is given by~\cite{Maiti:2025cbi, Maiti:2024nhv, Chakraborty:2024rgl}
\begin{align}
\ogwp(k) h^2\simeq  
\frac{\Omega_{\rm r}h^2}{6}\frac{\HI^2}{\Mp^2}  
\times\left\{\begin{array}{ll}
1, & k<\kre,\\
\mathcal{D}_1\l(\frac{k}{\kre}\r)^{-\nw}, & k>\kre,
\end{array}\right.
\end{align}
where $\mathcal{D}_1\simeq 2^{1-2l}\Gamma^2(1-l)/2\pi\simeq \mathcal{O}(1)$ and
$\nw=2(1-3\wre)/(1+3\wre)$~\cite{Maiti:2025cbi, Maiti:2024nhv}.
\paragraph{\underline{Secondary GWs}:}
In our scenario, the magnetic field follows a broken power-law spectrum with a peak at $k=\kp$. For $k<\kp$, the spectrum has a blue tilt with index $\nb$, while for $k>\kp$, the slope changes to $\nbs=\nb-2(\alpha+1)$, determined by both the magnetogenesis parameter $n$ and the reheating dynamics.

The gravitational-wave spectrum does not directly trace the magnetic spectrum. Since the magnetic energy is concentrated around the peak scale $\kp(\sim\kre)$, the GW signal at $k<\kp$ receives its dominant contribution from modes near the peak rather than from the local spectral slope. This leads to a universal infrared behavior for $\nb>3/2$, with $\ogw(f<\fp)\sim f^3$ as required by causality.

At higher frequencies, $f>\fp$, the spectrum is dominated by local contributions from the source and follows the magnetic-field scaling more closely, giving $\ogw(f>\fp)\propto f^{\nbs}$.

Combining the two regimes, the spectral energy density (SED) of gravitational waves can be approximated as
\begin{align}
    \ogwhm(f)\simeq \ogwm^{\rm peak}\times \l\{ 
    \begin{matrix}
          (f/\fre)^{\nc} & \text{for}~ f\ll\fre\\
          (f/\fre)^{\nbs} & \text{for}~ f\gg\fre
    \end{matrix}
    \r.
\end{align}
where $\ogwm^{\rm peak}$ is the amplitude at the peak frequency.

\begin{figure}
    \centering
    \includegraphics[width=0.95\linewidth]{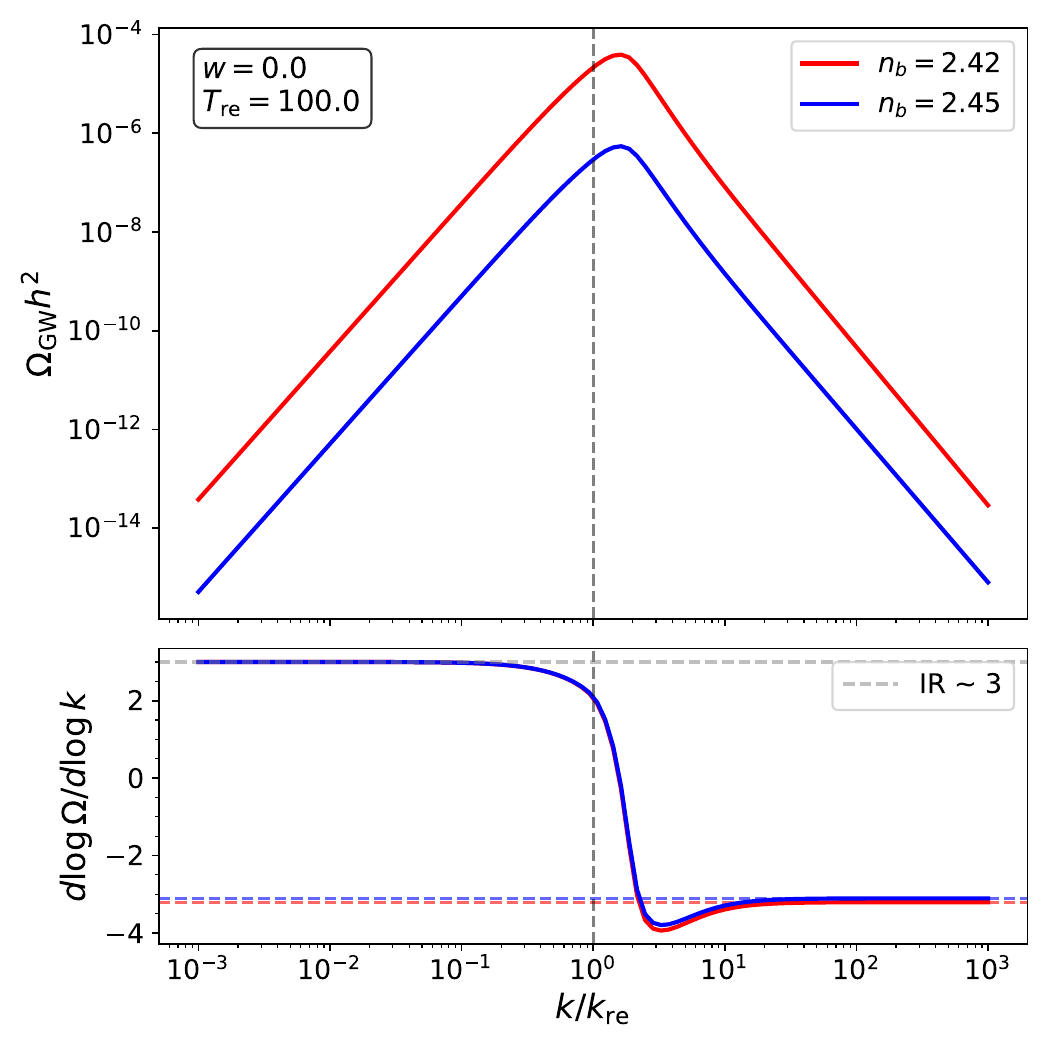}
    \caption{Top panel: Secondary GW spectral energy density induced by the magnetic field as a function of the wavenumber normalised by $\kre$. Bottom panel: GW spectral index near the peak as a function of the wavenumber. Here we consider a fixed reheating scenario with $\wre=0$ and $\Tre=100\,\GeV$. The two different colours correspond to two different values of the magnetic spectral index, $\nb=2.42$ (red) and $\nb=2.45$ (blue).}
    \label{fig:sgw_mag}
\end{figure}

Similarly, we can characterize the secondary GW spectrum sourced by scalar perturbations. As follows from Eq.~\eqref{eq:curvature}, the induced curvature power spectrum exhibits a broken power-law behaviour. For $k<\kp$, it scales as $\mPci(k<\kp)\propto k^{3}$, whereas for $k>\kp$, it follows $\mPci(k>\kp)\propto k^{\nbs}$.
At large scales, the induced curvature power spectrum is steeper than the magnetic-field spectrum. In contrast, at smaller scales, $k>\kp$, both the magnetic energy spectrum and the induced curvature power spectrum follow the same scaling behaviour.

\begin{figure}
    \centering
    \includegraphics[width=0.95\linewidth]{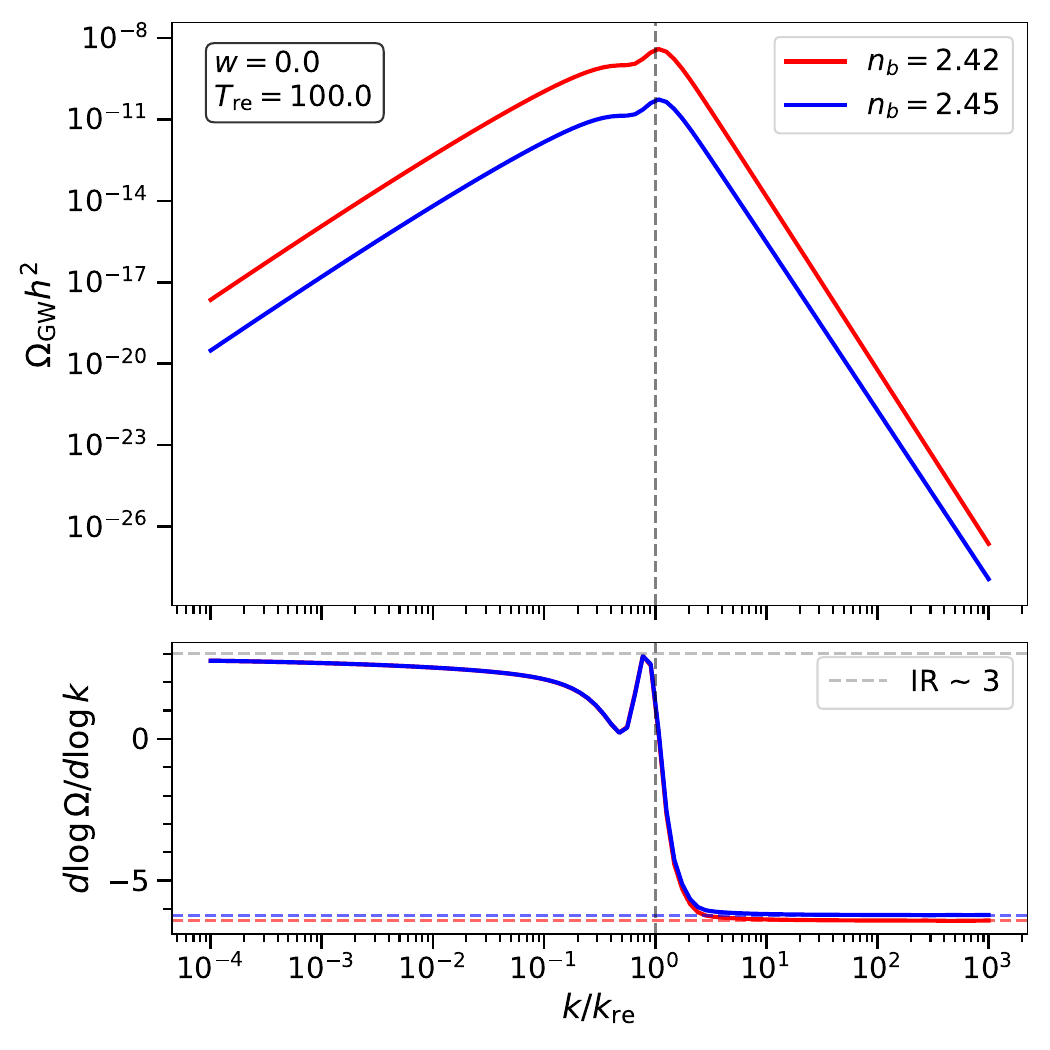}
    \caption{Top panel: GW spectral energy density induced by curvature perturbations as a function of the wavenumber $(k)$ normalised by $\kre$, for $\nb=2.42$ (red) and $\nb=2.45$ (blue). Bottom panel: running spectral index as a function of the wavenumber $(k)$. Here we consider a fixed reheating scenario with $\wre=0$ and $\Tre=100\,\GeV$. }
    \label{fig:sgws_scalar}
\end{figure}
\begin{figure*}
    \centering
    \includegraphics[width=0.48\linewidth]{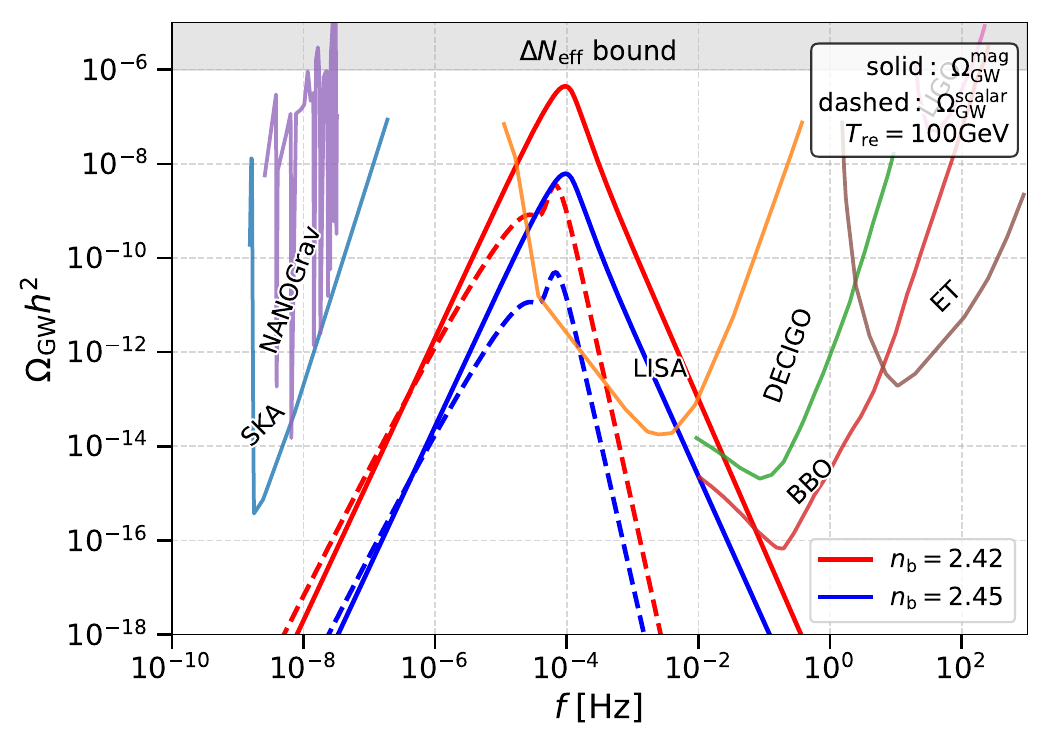}
    \includegraphics[width=0.48\linewidth]{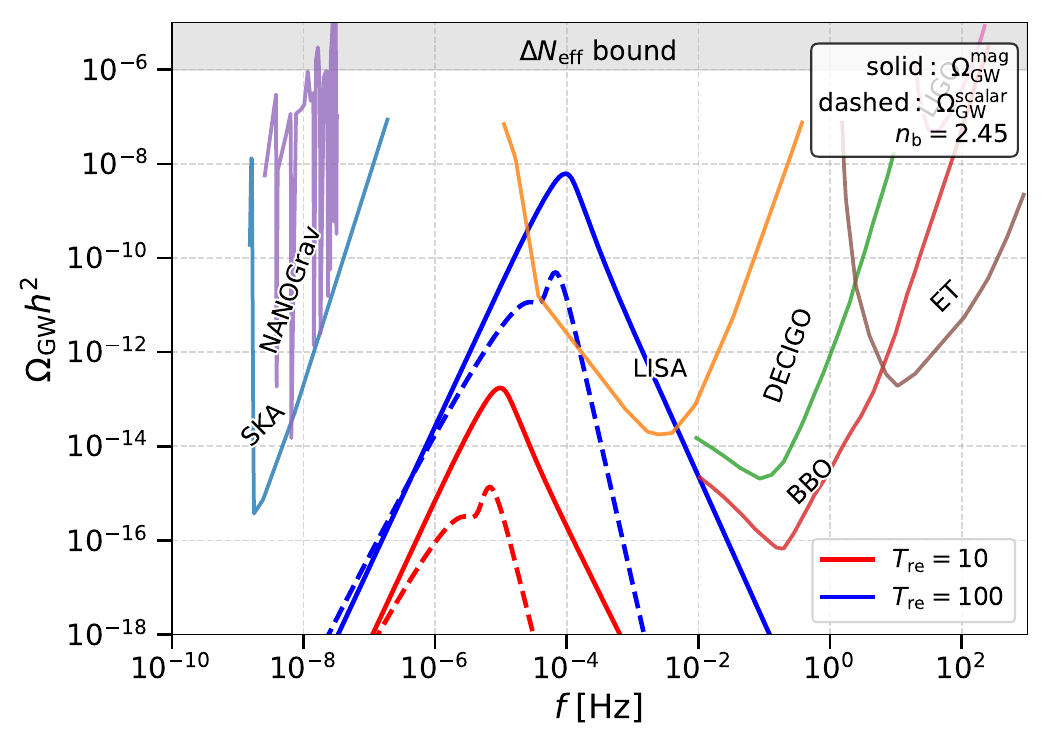}
    \caption{Present-day GW spectral energy density $\ogwh$ as a function of observable frequency $f(\hz)$ for two different scenarios. In the left panel, the reheating temperature is fixed, while the magnetic spectral index is varied with $\nb=2.42$ (red) and $\nb=2.45$ (blue). In the right panel, the magnetic spectral index is fixed at $\nb=2.45$, and two different reheating temperatures $\Tre=10\,\GeV$ (red) and $\Tre=100\,\GeV$(blue) are denoted by two different colors. Solid and dashed lines represent the magnetic and scalar-induced contributions, respectively. Both panels correspond to $\wre=0$ during reheating.}
    \label{fig:sgws_vs_f_nb}
\end{figure*}
The scalar-induced GW spectrum also exhibits a broken power-law structure. In the infrared regime, it follows the universal scaling $\ogw(f)\propto f^3$, while in the ultraviolet regime it traces the spectral index of the source. The spectrum can be approximated as
\begin{align}
    \ogwhs(f)\simeq \ogws^{\rm peak}\times \l\{ 
    \begin{matrix}
          (f/\fre)^3 & \text{for}~ f\ll\fre\\
        (f/\fre)^{2\nbs} & \text{for}~f\gg\fre
    \end{matrix}
    \r.
\end{align}
where $\ogws^{\rm peak}$ denotes the peak amplitude.
The peak amplitudes satisfy $\ogwm^{\rm peak}/\ogws^{\rm peak}\sim \mathcal{O}(100)$, with the magnetic contribution being dominant.
For $f\ll\fre$, both magnetic and scalar-induced spectra show the same infrared scaling, $\ogwh(f)\propto f^3$. In the ultraviolet regime, they differ: the magnetic component scales as $\ogwhm(f)\propto f^{\nbs}$, while the scalar-induced one scales as $\ogwhs(f)\propto f^{2\nbs}$, where $\nbs=\nb-2(\alpha+1)$.

\subsection{Results}

In this section, we present the numerical results obtained from Eq.~\eqref{eq:ph_mag} and Eq.~\eqref{eq:ph_scalar}, focusing on the spectral features of the GW spectra and their physical implications.

In Fig.~\ref{fig:sgw_mag}, we show the present-day GW spectral energy density $\ogwh$ generated by the anisotropic stress of the magnetic field as a function of the normalized wavenumber $k/\kre$. We fix the reheating parameters to $\Tre=100\,\GeV$ and $\wre=0$, and consider two representative values of the magnetic spectral index, $\nb=2.42$ and $\nb=2.45$.
For $k<\kre$, both cases follow the same infrared scaling, $\ogwhm(k<\kre)\propto k^3$. For $k>\kre$, the spectrum shows a mild dependence on $\nbs$, leading to a change in the spectral slope. This behavior is clearly reflected in the running spectral index shown in the lower panel, which agrees with the analytic estimates obtained in the previous subsection.
We also find that increasing $\nb$ suppresses the amplitude of the induced GW signal, since the magnetic-field strength itself decreases for larger values of $\nb$ (see Fig.~\ref{fig:comoving_mag}).

In Fig.~\ref{fig:sgws_scalar}, we show the scalar-induced contribution. It peaks at the same scale, $k\sim\kp$, since the curvature perturbations inherit the scale dependence of the magnetic source. At large scales, the spectrum again approaches $\Omega_{\rm GW}\propto k^3$, indicating that the infrared behavior is insensitive to the details of the source.

The main difference appears at high $k$, where the scalar-induced spectrum has a different slope compared to the magnetic contribution. This arises from the convolution structure in Eq.~(\ref{eq:ph_scalar}) and the scalar transfer kernel. Although the scalar-induced signal is about two orders of magnitude smaller at the peak, its high-$k$ fall-off differs from the magnetically induced GW spectrum.

The lower panel of Fig.~\ref{fig:sgws_scalar} shows the running of the GW spectral index induced by curvature perturbations near the peak. We find that the spectral index deviates from the infrared scaling and differs significantly from that of the magnetically induced GW spectrum (see Fig.~\ref{fig:sgw_mag}). As a result, the scalar-induced contribution can become relatively more relevant at large scales, despite its suppressed peak amplitude.

In Fig.~\ref{fig:sgws_vs_f_nb}, we plot the GW spectral energy density $\ogwh$ as a function of the observed frequency in Hz. In the left panel, we consider a fixed reheating scenario with $\wre=0$ and $\Tre=10^2\,\GeV$. The two colors correspond to $\nb=2.42$ (red) and $\nb=2.45$ (blue). For clarity, we separately show the magnetic and scalar-induced contributions: the solid lines denote magnetically induced GWs, $\ogwhm$, while the dashed lines represent scalar-induced GWs, $\ogwhs$.

For $f>\fp$, the two sources exhibit different spectral tilts, which is clearly visible in both panels of Fig.~\ref{fig:sgws_vs_f_nb}. Since the reheating temperature is fixed in the left panel, the peak position remains nearly the same for both values of $\nb$. However, the scalar-induced contribution is suppressed near the peak and in the ultraviolet regime. At large scales, the scalar-induced signal becomes comparable to or larger than the magnetic contribution and can dominate the spectrum.

In the right panel, we fix $\nb=2.45$ and vary the reheating temperature, considering $\Tre=10\,\GeV$ (blue) and $\Tre=10^2\,\GeV$ (red). Changing $\Tre$ shifts both the peak position and the overall amplitude. A higher reheating temperature moves the peak to higher frequencies and enhances the amplitude of the GW spectrum. This behavior is directly related to the strength of the magnetic field.

Finally, we compare the predicted spectra with sensitivity curves of current and future GW detectors, including SKA~\cite{Janssen:2014dka}, LISA~\cite{Amaro-Seoane:2012aqc,Barausse:2020rsu}, DECIGO~\cite{Seto:2001qf,Kawamura:2011zz,Suemasa:2017ppd}, BBO~\cite{Crowder:2005nr,Corbin:2005ny,Baker:2019pnp}, ET~\cite{Punturo:2010zz,Sathyaprakash:2012jk}, as well as recent PTA results~\cite{NANOGrav:2023gor,Reardon:2023gzh,Xu:2023wog}. The predicted spectra fall within the sensitivity ranges of several experiments, with the nanohertz regime making this scenario particularly relevant for PTA observations.
\paragraph{\textbf{$\Delta N_{\rm eff}$ bound:}}

The energy density stored in the stochastic gravitational-wave background contributes to the total radiation content of the Universe. This additional radiation component is commonly expressed in terms of the effective number of relativistic species, $\Delta N_{\rm eff} \equiv N_{\rm eff}-N_{\rm eff}^{\rm SM}$,
which measures the deviation from the Standard Model prediction $N_{\rm eff}^{\rm SM}$.

Current constraints from Planck, combined with Big Bang Nucleosynthesis (BBN) and Cosmic Microwave Background (CMB) measurements, impose a stringent upper bound $\Delta N_{\rm eff} \lesssim 0.3$ at $95\%$ confidence level. This bound directly translates into a constraint on the total energy density stored in primordial GWs. So, it implies an approximate upper limit on the GW spectrum amplitude, $\Omega_{\rm GW} h^2(f) \lesssim 1.6 \times 10^{-6}$,
as discussed in~\cite{Watanabe:2006qe, Caprini:2018mtu, Planck:2018vyg, clarke2020constraints, PhysRevD.85.123002}.

More generally, the contribution of a stochastic GW background to $\Delta N_{\rm eff}$ is obtained by integrating over all frequencies~\cite{Watanabe:2006qe, Caprini:2018mtu, Planck:2018vyg, clarke2020constraints, PhysRevD.85.123002},
\begin{align}
    \Delta N_{\rm eff} = \frac{1}{\Omega_{\gamma} h^2} \frac{8}{7} \left( \frac{11}{4} \right)^{4/3}
    \int_{f_*}^{f_{\rm end}} \mathrm{d} \ln f \, \Omega_{\rm GW} h^2(f),
\end{align}
where $\Omega_{\gamma} h^2 \simeq 2.4 \times 10^{-5}$ denotes the present photon energy density.

 Since the GW spectrum is strongly localized around the peak $(\fp\sim\fre)$, the integral is dominated by this region. So if the peak amplitude satisfies $\Omega_{\rm GW}^{\rm peak} \lesssim 1.6 \times 10^{-6}$, it
is sufficient to ensure consistency with the $\Delta N_{\rm eff}$ bound.
 This implies that the maximum allowed GW amplitude in this framework can naturally reach values as large as $\mathcal{O}(10^{-6})$ without violating cosmological bounds.

\vspace{0.3cm}
\paragraph{\textbf{PTA signals:}}

Recent observations from Pulsar Timing Array (PTA) collaborations, including NANOGrav, EPTA, PPTA, and CPTA, have provided compelling evidence for a stochastic gravitational wave background (GWB) in the nanohertz frequency range~\cite{NANOGrav:2023gor, Reardon:2023gzh, Xu:2023wog}. While the leading astrophysical interpretation attributes this signal to supermassive black hole binaries (SMBHBs)~\cite{1995ApJ...446..543R, Jaffe:2002rt, Wyithe:2002ep, Enoki:2006kj, Sesana:2008mz, Kelley:2017lek}, primordial origins remain viable and are actively being explored~\cite{NANOGrav:2023hvm, Benetti:2021uea, Vagnozzi:2020gtf,Vagnozzi:2023lwo, Zhu:2023gmx}.

In particular, the NANOGrav 15-year dataset (NG15) has motivated investigation of new-physics scenarios that can generate a primordial GWB consistent with the observed signal~\cite{NANOGrav:2023gor}. We test our model against PTA data using the \texttt{PTArcade} framework~\cite{mitridate2023ptarcade}, which uses the \textit{ENTERPRISE}~\cite{2019ascl.soft12015E} and \textit{ceffyl}~\cite{lamb2023need}.

Given a PTA dataset $\{\vec{d t}\}$ and model parameters $\{\vec{\theta}\}$, the posterior distribution is defined as
\begin{align}
    p(\vec{\theta}|\vec{d t}) = \frac{p(\vec{d t}|\vec{\theta})\, p(\vec{\theta})}{p(\vec{d t})},
\end{align}
where $p(\vec{d t}|\vec{\theta})$ is the likelihood and $p(\vec{\theta})$ denotes the prior. The Bayesian evidence for a model $X$ is given by
\begin{align}
    \mathcal{Z}_X = p(\vec{d t}|X) = \int d\vec{\theta} \, p(\vec{d t}|\vec{\theta})\, p(\vec{\theta}),
\end{align}
which allows for model comparison via the Bayes factor, $ \mathcal{B}_{X,Y} = {\mathcal{Z}_X}/{\mathcal{Z}_Y}$.
We compare our model ($X$) with the standard SMBHB scenario ($Y$) by computing the corresponding Bayes factor. The analysis is performed using \texttt{PTArcade} in \textit{ENTERPRISE} mode for evidence evaluation and in \textit{ceffyl} mode for parameter estimation.

To determine the best-fit model parameters $(r, T_{\rm re}, n_B)$, we perform a Markov Chain Monte Carlo (MCMC) analysis using the NG15 dataset. We impose two physically motivated constraints:
(i) the backreaction condition, requiring the electromagnetic energy density to remain subdominant, $ \delta_{\rm em}(\ere) = \frac{\rho_{\rm em}}{\rho_{\rm rad}} < 1,$
and (ii) the $\Delta N_{\rm eff}$ bound discussed above.
\begin{figure*}
    \centering
    \includegraphics[width=0.48\linewidth]{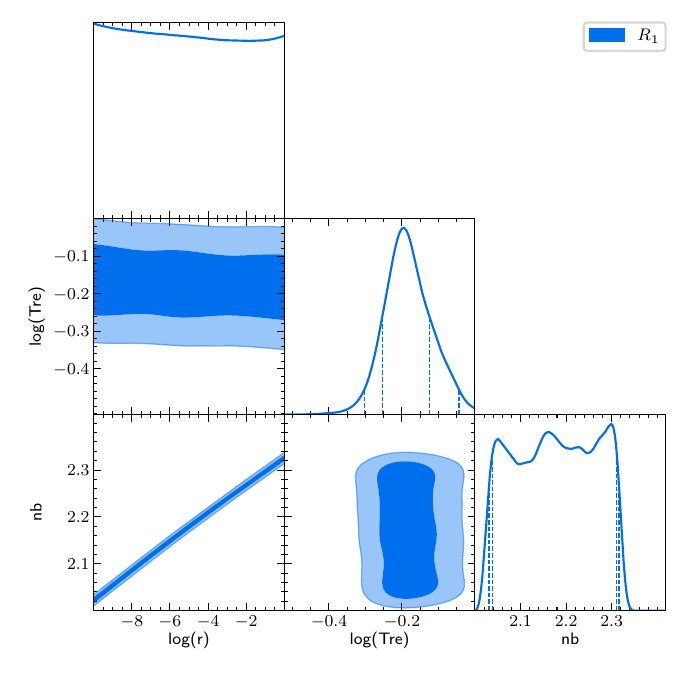}
    \includegraphics[width=0.48\linewidth]{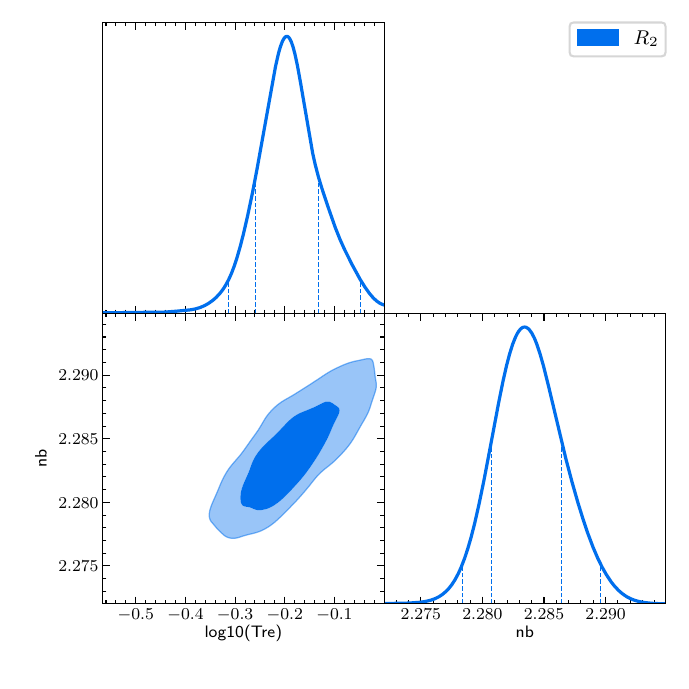}
    \caption{We have plotted the posterior distributions of the parameters. In the left panel, we have considered that $r,\Tre$ and $\nb$ are the free parameters, and in the right panel, we fixed the tensor-to-scalar ratio as $r=0.036$ and consider the same uniform prior for the other two parameters.}
    \label{fig:posterior}
\end{figure*}

\begin{center}
\begin{table}[t]
\centering
    \begin{tabular}{ c c c c }
    \hline
    \hline
    Model & Prior $(\mathcal{U})$ & Mean values & Bayes-Factor\\
      \hline
       & $\log_{10}(r)(-10.0,0.0)$ & $-5.02^{+3.44}_{-3.40}$ & \\
      $R_1$ & $\log_{10}(\Tre)~(-0.5,0.5)$ & $-0.19^{+0.07}_{-0.06}$ &  $23.36\pm 4.17$  \\
                     & $\nb~(2.0,3.0)$              & $2.18^{+0.10}_{-0.11}$  &  \\
      \hline
       & $r=0.036$ & $0.036$ & \\
      $R_2$ & $\log_{10}(\Tre)(-2.0,0.0)$ & $-0.19^{+0.07}_{-0.06}$ & $13.36\pm 1.34$ \\
      & $\nb(2.0,3.0)$ & $2.28^{+0.005}_{-0.004}$ & \\
      \hline
    \end{tabular}
         \caption{Here we listed the best-fitted values of the parameters using the MCMC analysis for different scenarios. We also compute it corresponding Bayesian-Factor $\mathcal{B}_{\rm X, Y}$ that invokes
primordial physics as the source of the stochastic GW background observed by the NANOGrav 15-year
data, when compared to the astrophysical scenario of SMBHBs.}
    \label{tab:param_est}
 \end{table}
\end{center}

Since the GW spectrum in our model peaks at $f\sim\fre$, the signal naturally falls within the PTA frequency band for reheating temperatures in the range $0.1\,\GeV\lesssim\Tre\lesssim1\,\GeV$. All viable parameter combinations satisfy the backreaction condition and remain consistent with the $\Delta N_{\rm eff}$ constraint, typically yielding $\Delta N_{\rm eff}\lesssim0.28$.

In Table~\ref{tab:param_est}, we present the parameter constraints obtained from two representative runs. In $R_1$, the tensor-to-scalar ratio is treated as a free parameter with prior $\log_{10}r\in[-10,0]$, while in both runs we assume $\log_{10}(\Tre)\in[-1,1]$ and $n_B\in[2,3]$. The posterior mean values are reported as the best-fit estimates.

For $R_1$, we obtain $\log_{10}(r_{0.05})\simeq -5.02^{+3.44}_{-3.40}$, $\log_{10}(\Tre)\simeq -0.19^{+0.07}_{-0.06}$, and $\nb\simeq 2.18^{+0.10}_{-0.11}$ at the $2\sigma$ confidence level, with a Bayes factor $\mathcal{B}_{\rm X,Y}\simeq 23.36\pm4.17$. These results show that PTA observations place strong constraints on the reheating temperature and magnetic spectral index, while the tensor-to-scalar ratio remains weakly constrained.

For $R_2$, we fix the tensor-to-scalar ratio to $r_{0.05}=0.036$ and vary only $\Tre$ and $n_B$. In this case, we find $\log_{10}(\Tre)\simeq -0.19^{+0.07}_{-0.06}$ and $\nb\simeq 2.28^{+0.008}_{-0.007}$, with a Bayes factor $\mathcal{B}_{\rm X,Y}\simeq 13.36\pm1.34$.

In Fig.~\ref{fig:posterior}, we show the posterior distributions for the two runs listed in Tab.~\ref{tab:param_est}, fixing the reheating equation of state to $\wre=0$.
In run $R_1$, we vary $r_{k_*}$, $\Tre$, and $\nb$. The results show that $\Tre$ and $\nb$ are tightly constrained, while $r_{k_*}$ remains weakly constrained.
In run $R_2$, we fix $r_{k_*}$ and vary only $\Tre$ and $\nb$, which are again well constrained by the data.
Overall, $\Tre$ and $\nb$ are robustly constrained, whereas $r_{k_*}$ is not. This indicates that the model can generate GW signals compatible with PTA observations for suitable choices of the magnetogenesis parameter $\nb$ and reheating parameters $\wre$ and $\Tre$.

\section{Conclusions}

The origin of large-scale magnetic fields remains an important open problem in cosmology. Among the proposed mechanisms, post-inflationary magnetogenesis provides a simple framework in which time-dependent couplings during inflation and reheating can generate magnetic fields consistent with current observational bounds~\cite{Paoletti:2008ck, Planck:2015zrl, Zucca:2016iur, paoletti2022constraints, BICEP2:2017lpa, 2011A&A...529A.144T, 2010Sci...328...73N, 2011A&A...529A.144T,
Planck:2015zrl, Paoletti:2008ck, PhysRevLett.116.191302}. In the sawtooth-type coupling scenario, the magnetic field develops a broken power-law spectrum, where the peak scale is determined by the horizon size at the end of reheating~\cite{Maiti:2025ijr, Maiti:2025awl}.
Besides generating large-scale magnetic fields, this scenario also produces curvature perturbations through electromagnetic energy density fluctuations. For suitable parameter choices, these perturbations can become large at intermediate scales and may lead to the formation of primordial black holes (PBHs), which can act as dark matter candidates~\cite{Maiti:2025ijr}.

Large magnetic anisotropic stress also sources the stochastic gravitational wave (GW) background~\cite{Maiti:2025ijr, Maiti:2025rkn, Maiti:2025awl, Maiti:2025cbi, Maiti:2024nhv, Sorbo:2011rz, Caprini:2014mja, Ito:2016fqp, Sharma:2019jtb, Okano:2020uyr, Bhaumik:2025kuj, Ragavendra:2026fgs, Maiti:2026lvx}. The resulting magnetically-induced GW spectrum follows a broken power-law form. In addition, the enhanced curvature perturbations generate scalar-induced GWs. Although the scalar-induced contribution is formally higher order and generally smaller, its spectral behavior is qualitatively different from the magnetic contribution.

After doing the comparison between the two contributions, we find that the peak amplitude of the magnetically-induced GW spectrum is typically about two orders of magnitude larger than the scalar-induced contribution, ${\ogwhm(\fp)}/{\ogwhs(\fp)}
\sim \mathcal{O}(10^2)$.
However, the two spectra evolve differently away from the peak. In the infrared regime, both contributions scale as $\ogwh(f\ll\fp)\propto f^3$~\cite{Cai:2019cdl}. In the ultraviolet regime, their behaviors differ:$
\ogwhm(f\gg\fp)\propto f^{\nbs},
\qquad
\ogwhs(f\gg\fp)\propto f^{2\nbs}$.
The difference in spectral slopes provides a characteristic feature of this scenario.

Although the magnetically-induced GW spectrum has a larger peak amplitude, the scalar-induced contribution becomes comparable and eventually dominates at sufficiently low frequencies ($f\ll\fp$). This is because the scalar-induced spectrum is broader and decreases more slowly below the peak frequency. The different spectral shapes originate from the distinct transfer functions associated with the two source terms after horizon re-entry. In particular, the scalar-induced spectrum shows a more pronounced deviation from the asymptotic infrared scaling $(n_{\rm IR}\simeq 3)$ near the peak frequency, which we have verified is a physical feature rather than a numerical artifact. Although the difference in the spectral slope may provide a useful signature for distinguishing the two GW sources with future high-precision gravitational-wave observations, the amplitude difference in the deep infrared is very small, making this regime challenging to probe observationally.

Depending on the magnetogenesis and reheating parameters, these models can generate GW signals over a wide range of frequencies. In particular, the spectrum can naturally peak in the nano-Hz range relevant for recent PTA observations. This does not require an additional ultra-slow-roll phase, making magnetogenesis a viable alternative explanation of the PTA signal. Future GW observations may also help constrain both the magnetogenesis parameters and the reheating history through spectral reconstruction.

The combined analysis of magnetically-induced and scalar-induced GWs provides complementary information about reheating and primordial magnetogenesis. The distinct spectral features discussed here may help test magnetogenesis scenarios and shed light on the origin of primordial magnetic fields and stochastic GW backgrounds.

\acknowledgments

SM acknowledges financial support from the Council of Scientific and Industrial Research (CSIR), Ministry of Science and Technology, Government of India. SM is grateful to Prof. Debaprasad Maity and Prof. L. Sriramkumar for many useful discussions.


\bibliographystyle{apsrev4-1}
\bibliography{references}

\begin{thebibliography}{152}%
\makeatletter
\providecommand \@ifxundefined [1]{%
 \@ifx{#1\undefined}
}%
\providecommand \@ifnum [1]{%
 \ifnum #1\expandafter \@firstoftwo
 \else \expandafter \@secondoftwo
 \fi
}%
\providecommand \@ifx [1]{%
 \ifx #1\expandafter \@firstoftwo
 \else \expandafter \@secondoftwo
 \fi
}%
\providecommand \natexlab [1]{#1}%
\providecommand \enquote  [1]{``#1''}%
\providecommand \bibnamefont  [1]{#1}%
\providecommand \bibfnamefont [1]{#1}%
\providecommand \citenamefont [1]{#1}%
\providecommand \href@noop [0]{\@secondoftwo}%
\providecommand \href [0]{\begingroup \@sanitize@url \@href}%
\providecommand \@href[1]{\@@startlink{#1}\@@href}%
\providecommand \@@href[1]{\endgroup#1\@@endlink}%
\providecommand \@sanitize@url [0]{\catcode `\\12\catcode `\$12\catcode `\&12\catcode `\#12\catcode `\^12\catcode `\_12\catcode `\%12\relax}%
\providecommand \@@startlink[1]{}%
\providecommand \@@endlink[0]{}%
\providecommand \url  [0]{\begingroup\@sanitize@url \@url }%
\providecommand \@url [1]{\endgroup\@href {#1}{\urlprefix }}%
\providecommand \urlprefix  [0]{URL }%
\providecommand \Eprint [0]{\href }%
\providecommand \doibase [0]{http://dx.doi.org/}%
\providecommand \selectlanguage [0]{\@gobble}%
\providecommand \bibinfo  [0]{\@secondoftwo}%
\providecommand \bibfield  [0]{\@secondoftwo}%
\providecommand \translation [1]{[#1]}%
\providecommand \BibitemOpen [0]{}%
\providecommand \bibitemStop [0]{}%
\providecommand \bibitemNoStop [0]{.\EOS\space}%
\providecommand \EOS [0]{\spacefactor3000\relax}%
\providecommand \BibitemShut  [1]{\csname bibitem#1\endcsname}%
\let\auto@bib@innerbib\@empty
\bibitem [{\citenamefont {Abbott}\ \emph {et~al.}(2016{\natexlab{a}})\citenamefont {Abbott} \emph {et~al.}}]{LIGOScientific:2016aoc}%
  \BibitemOpen
  \bibfield  {author} {\bibinfo {author} {\bibfnamefont {B.~P.}\ \bibnamefont {Abbott}} \emph {et~al.} (\bibinfo {collaboration} {LIGO Scientific, Virgo}),\ }\href {\doibase 10.1103/PhysRevLett.116.061102} {\bibfield  {journal} {\bibinfo  {journal} {Phys. Rev. Lett.}\ }\textbf {\bibinfo {volume} {116}},\ \bibinfo {pages} {061102} (\bibinfo {year} {2016}{\natexlab{a}})},\ \Eprint {http://arxiv.org/abs/1602.03837} {arXiv:1602.03837 [gr-qc]} \BibitemShut {NoStop}%
\bibitem [{\citenamefont {Abbott}\ \emph {et~al.}(2016{\natexlab{b}})\citenamefont {Abbott} \emph {et~al.}}]{LIGOScientific:2016vlm}%
  \BibitemOpen
  \bibfield  {author} {\bibinfo {author} {\bibfnamefont {B.~P.}\ \bibnamefont {Abbott}} \emph {et~al.} (\bibinfo {collaboration} {LIGO Scientific, Virgo}),\ }\href {\doibase 10.1103/PhysRevLett.116.241102} {\bibfield  {journal} {\bibinfo  {journal} {Phys. Rev. Lett.}\ }\textbf {\bibinfo {volume} {116}},\ \bibinfo {pages} {241102} (\bibinfo {year} {2016}{\natexlab{b}})},\ \Eprint {http://arxiv.org/abs/1602.03840} {arXiv:1602.03840 [gr-qc]} \BibitemShut {NoStop}%
\bibitem [{\citenamefont {Abbott}\ \emph {et~al.}(2016{\natexlab{c}})\citenamefont {Abbott} \emph {et~al.}}]{LIGOScientific:2016emj}%
  \BibitemOpen
  \bibfield  {author} {\bibinfo {author} {\bibfnamefont {B.~P.}\ \bibnamefont {Abbott}} \emph {et~al.} (\bibinfo {collaboration} {LIGO Scientific, Virgo}),\ }\href {\doibase 10.1103/PhysRevLett.116.131103} {\bibfield  {journal} {\bibinfo  {journal} {Phys. Rev. Lett.}\ }\textbf {\bibinfo {volume} {116}},\ \bibinfo {pages} {131103} (\bibinfo {year} {2016}{\natexlab{c}})},\ \Eprint {http://arxiv.org/abs/1602.03838} {arXiv:1602.03838 [gr-qc]} \BibitemShut {NoStop}%
\bibitem [{\citenamefont {Abbott}\ \emph {et~al.}(2016{\natexlab{d}})\citenamefont {Abbott} \emph {et~al.}}]{LIGOScientific:2016vbw}%
  \BibitemOpen
  \bibfield  {author} {\bibinfo {author} {\bibfnamefont {B.~P.}\ \bibnamefont {Abbott}} \emph {et~al.} (\bibinfo {collaboration} {LIGO Scientific, Virgo}),\ }\href {\doibase 10.1103/PhysRevD.93.122003} {\bibfield  {journal} {\bibinfo  {journal} {Phys. Rev. D}\ }\textbf {\bibinfo {volume} {93}},\ \bibinfo {pages} {122003} (\bibinfo {year} {2016}{\natexlab{d}})},\ \Eprint {http://arxiv.org/abs/1602.03839} {arXiv:1602.03839 [gr-qc]} \BibitemShut {NoStop}%
\bibitem [{\citenamefont {Abbott}\ \emph {et~al.}(2017{\natexlab{a}})\citenamefont {Abbott} \emph {et~al.}}]{LIGOScientific:2017bnn}%
  \BibitemOpen
  \bibfield  {author} {\bibinfo {author} {\bibfnamefont {B.~P.}\ \bibnamefont {Abbott}} \emph {et~al.} (\bibinfo {collaboration} {LIGO Scientific, VIRGO}),\ }\href {\doibase 10.1103/PhysRevLett.118.221101} {\bibfield  {journal} {\bibinfo  {journal} {Phys. Rev. Lett.}\ }\textbf {\bibinfo {volume} {118}},\ \bibinfo {pages} {221101} (\bibinfo {year} {2017}{\natexlab{a}})},\ \bibinfo {note} {[Erratum: Phys.Rev.Lett. 121, 129901 (2018)]},\ \Eprint {http://arxiv.org/abs/1706.01812} {arXiv:1706.01812 [gr-qc]} \BibitemShut {NoStop}%
\bibitem [{\citenamefont {Abbott}\ \emph {et~al.}(2017{\natexlab{b}})\citenamefont {Abbott} \emph {et~al.}}]{LIGOScientific:2016jlg}%
  \BibitemOpen
  \bibfield  {author} {\bibinfo {author} {\bibfnamefont {B.~P.}\ \bibnamefont {Abbott}} \emph {et~al.} (\bibinfo {collaboration} {LIGO Scientific, Virgo}),\ }\href {\doibase 10.1103/PhysRevLett.118.121101} {\bibfield  {journal} {\bibinfo  {journal} {Phys. Rev. Lett.}\ }\textbf {\bibinfo {volume} {118}},\ \bibinfo {pages} {121101} (\bibinfo {year} {2017}{\natexlab{b}})},\ \bibinfo {note} {[Erratum: Phys.Rev.Lett. 119, 029901 (2017)]},\ \Eprint {http://arxiv.org/abs/1612.02029} {arXiv:1612.02029 [gr-qc]} \BibitemShut {NoStop}%
\bibitem [{\citenamefont {Akrami}\ \emph {et~al.}(2020)\citenamefont {Akrami} \emph {et~al.}}]{Planck:2018jri}%
  \BibitemOpen
  \bibfield  {author} {\bibinfo {author} {\bibfnamefont {Y.}~\bibnamefont {Akrami}} \emph {et~al.} (\bibinfo {collaboration} {Planck}),\ }\href {\doibase 10.1051/0004-6361/201833887} {\bibfield  {journal} {\bibinfo  {journal} {Astron. Astrophys.}\ }\textbf {\bibinfo {volume} {641}},\ \bibinfo {pages} {A10} (\bibinfo {year} {2020})},\ \Eprint {http://arxiv.org/abs/1807.06211} {arXiv:1807.06211 [astro-ph.CO]} \BibitemShut {NoStop}%
\bibitem [{\citenamefont {Ade}\ \emph {et~al.}(2021)\citenamefont {Ade} \emph {et~al.}}]{BICEP:2021xfz}%
  \BibitemOpen
  \bibfield  {author} {\bibinfo {author} {\bibfnamefont {P.~A.~R.}\ \bibnamefont {Ade}} \emph {et~al.} (\bibinfo {collaboration} {BICEP, Keck}),\ }\href {\doibase 10.1103/PhysRevLett.127.151301} {\bibfield  {journal} {\bibinfo  {journal} {Phys. Rev. Lett.}\ }\textbf {\bibinfo {volume} {127}},\ \bibinfo {pages} {151301} (\bibinfo {year} {2021})},\ \Eprint {http://arxiv.org/abs/2110.00483} {arXiv:2110.00483 [astro-ph.CO]} \BibitemShut {NoStop}%
\bibitem [{\citenamefont {Ade}\ \emph {et~al.}(2018)\citenamefont {Ade} \emph {et~al.}}]{BICEP2:2018kqh}%
  \BibitemOpen
  \bibfield  {author} {\bibinfo {author} {\bibfnamefont {P.~A.~R.}\ \bibnamefont {Ade}} \emph {et~al.} (\bibinfo {collaboration} {BICEP2, Keck Array}),\ }\href {\doibase 10.1103/PhysRevLett.121.221301} {\bibfield  {journal} {\bibinfo  {journal} {Phys. Rev. Lett.}\ }\textbf {\bibinfo {volume} {121}},\ \bibinfo {pages} {221301} (\bibinfo {year} {2018})},\ \Eprint {http://arxiv.org/abs/1810.05216} {arXiv:1810.05216 [astro-ph.CO]} \BibitemShut {NoStop}%
\bibitem [{\citenamefont {Clarke}\ \emph {et~al.}(2020{\natexlab{a}})\citenamefont {Clarke}, \citenamefont {Copeland},\ and\ \citenamefont {Moss}}]{Clarke:2020bil}%
  \BibitemOpen
  \bibfield  {author} {\bibinfo {author} {\bibfnamefont {T.~J.}\ \bibnamefont {Clarke}}, \bibinfo {author} {\bibfnamefont {E.~J.}\ \bibnamefont {Copeland}}, \ and\ \bibinfo {author} {\bibfnamefont {A.}~\bibnamefont {Moss}},\ }\href {\doibase 10.1088/1475-7516/2020/10/002} {\bibfield  {journal} {\bibinfo  {journal} {JCAP}\ }\textbf {\bibinfo {volume} {10}},\ \bibinfo {pages} {002} (\bibinfo {year} {2020}{\natexlab{a}})},\ \Eprint {http://arxiv.org/abs/2004.11396} {arXiv:2004.11396 [astro-ph.CO]} \BibitemShut {NoStop}%
\bibitem [{\citenamefont {Watanabe}\ and\ \citenamefont {Komatsu}(2006)}]{Watanabe:2006qe}%
  \BibitemOpen
  \bibfield  {author} {\bibinfo {author} {\bibfnamefont {Y.}~\bibnamefont {Watanabe}}\ and\ \bibinfo {author} {\bibfnamefont {E.}~\bibnamefont {Komatsu}},\ }\href {\doibase 10.1103/PhysRevD.73.123515} {\bibfield  {journal} {\bibinfo  {journal} {Phys. Rev. D}\ }\textbf {\bibinfo {volume} {73}},\ \bibinfo {pages} {123515} (\bibinfo {year} {2006})},\ \Eprint {http://arxiv.org/abs/astro-ph/0604176} {arXiv:astro-ph/0604176} \BibitemShut {NoStop}%
\bibitem [{\citenamefont {Caprini}\ and\ \citenamefont {Figueroa}(2018)}]{Caprini:2018mtu}%
  \BibitemOpen
  \bibfield  {author} {\bibinfo {author} {\bibfnamefont {C.}~\bibnamefont {Caprini}}\ and\ \bibinfo {author} {\bibfnamefont {D.~G.}\ \bibnamefont {Figueroa}},\ }\href {\doibase 10.1088/1361-6382/aac608} {\bibfield  {journal} {\bibinfo  {journal} {Class. Quant. Grav.}\ }\textbf {\bibinfo {volume} {35}},\ \bibinfo {pages} {163001} (\bibinfo {year} {2018})},\ \Eprint {http://arxiv.org/abs/1801.04268} {arXiv:1801.04268 [astro-ph.CO]} \BibitemShut {NoStop}%
\bibitem [{\citenamefont {Acquaviva}\ \emph {et~al.}(2003)\citenamefont {Acquaviva}, \citenamefont {Bartolo}, \citenamefont {Matarrese},\ and\ \citenamefont {Riotto}}]{Acquaviva:2002ud}%
  \BibitemOpen
  \bibfield  {author} {\bibinfo {author} {\bibfnamefont {V.}~\bibnamefont {Acquaviva}}, \bibinfo {author} {\bibfnamefont {N.}~\bibnamefont {Bartolo}}, \bibinfo {author} {\bibfnamefont {S.}~\bibnamefont {Matarrese}}, \ and\ \bibinfo {author} {\bibfnamefont {A.}~\bibnamefont {Riotto}},\ }\href {\doibase 10.1016/S0550-3213(03)00550-9} {\bibfield  {journal} {\bibinfo  {journal} {Nucl. Phys. B}\ }\textbf {\bibinfo {volume} {667}},\ \bibinfo {pages} {119} (\bibinfo {year} {2003})},\ \Eprint {http://arxiv.org/abs/astro-ph/0209156} {arXiv:astro-ph/0209156} \BibitemShut {NoStop}%
\bibitem [{\citenamefont {Aghanim}\ \emph {et~al.}(2020)\citenamefont {Aghanim} \emph {et~al.}}]{Planck:2018vyg}%
  \BibitemOpen
  \bibfield  {author} {\bibinfo {author} {\bibfnamefont {N.}~\bibnamefont {Aghanim}} \emph {et~al.} (\bibinfo {collaboration} {Planck}),\ }\href {\doibase 10.1051/0004-6361/201833910} {\bibfield  {journal} {\bibinfo  {journal} {Astron. Astrophys.}\ }\textbf {\bibinfo {volume} {641}},\ \bibinfo {pages} {A6} (\bibinfo {year} {2020})},\ \bibinfo {note} {[Erratum: Astron.Astrophys. 652, C4 (2021)]},\ \Eprint {http://arxiv.org/abs/1807.06209} {arXiv:1807.06209 [astro-ph.CO]} \BibitemShut {NoStop}%
\bibitem [{\citenamefont {Sendra}\ and\ \citenamefont {Smith}(2012)}]{PhysRevD.85.123002}%
  \BibitemOpen
  \bibfield  {author} {\bibinfo {author} {\bibfnamefont {I.}~\bibnamefont {Sendra}}\ and\ \bibinfo {author} {\bibfnamefont {T.~L.}\ \bibnamefont {Smith}},\ }\href {\doibase 10.1103/PhysRevD.85.123002} {\bibfield  {journal} {\bibinfo  {journal} {Phys. Rev. D}\ }\textbf {\bibinfo {volume} {85}},\ \bibinfo {pages} {123002} (\bibinfo {year} {2012})}\BibitemShut {NoStop}%
\bibitem [{\citenamefont {Afzal}\ \emph {et~al.}(2023)\citenamefont {Afzal} \emph {et~al.}}]{NANOGrav:2023hvm}%
  \BibitemOpen
  \bibfield  {author} {\bibinfo {author} {\bibfnamefont {A.}~\bibnamefont {Afzal}} \emph {et~al.} (\bibinfo {collaboration} {NANOGrav}),\ }\href {\doibase 10.3847/2041-8213/acdc91} {\bibfield  {journal} {\bibinfo  {journal} {Astrophys. J. Lett.}\ }\textbf {\bibinfo {volume} {951}},\ \bibinfo {pages} {L11} (\bibinfo {year} {2023})},\ \bibinfo {note} {[Erratum: Astrophys.J.Lett. 971, L27 (2024), Erratum: Astrophys.J. 971, L27 (2024)]},\ \Eprint {http://arxiv.org/abs/2306.16219} {arXiv:2306.16219 [astro-ph.HE]} \BibitemShut {NoStop}%
\bibitem [{\citenamefont {Sorbo}(2011)}]{Sorbo:2011rz}%
  \BibitemOpen
  \bibfield  {author} {\bibinfo {author} {\bibfnamefont {L.}~\bibnamefont {Sorbo}},\ }\href {\doibase 10.1088/1475-7516/2011/06/003} {\bibfield  {journal} {\bibinfo  {journal} {JCAP}\ }\textbf {\bibinfo {volume} {06}},\ \bibinfo {pages} {003} (\bibinfo {year} {2011})},\ \Eprint {http://arxiv.org/abs/1101.1525} {arXiv:1101.1525 [astro-ph.CO]} \BibitemShut {NoStop}%
\bibitem [{\citenamefont {Caprini}\ and\ \citenamefont {Sorbo}(2014)}]{Caprini:2014mja}%
  \BibitemOpen
  \bibfield  {author} {\bibinfo {author} {\bibfnamefont {C.}~\bibnamefont {Caprini}}\ and\ \bibinfo {author} {\bibfnamefont {L.}~\bibnamefont {Sorbo}},\ }\href {\doibase 10.1088/1475-7516/2014/10/056} {\bibfield  {journal} {\bibinfo  {journal} {JCAP}\ }\textbf {\bibinfo {volume} {10}},\ \bibinfo {pages} {056} (\bibinfo {year} {2014})},\ \Eprint {http://arxiv.org/abs/1407.2809} {arXiv:1407.2809 [astro-ph.CO]} \BibitemShut {NoStop}%
\bibitem [{\citenamefont {Ito}\ and\ \citenamefont {Soda}(2017)}]{Ito:2016fqp}%
  \BibitemOpen
  \bibfield  {author} {\bibinfo {author} {\bibfnamefont {A.}~\bibnamefont {Ito}}\ and\ \bibinfo {author} {\bibfnamefont {J.}~\bibnamefont {Soda}},\ }\href {\doibase 10.1016/j.physletb.2017.05.017} {\bibfield  {journal} {\bibinfo  {journal} {Phys. Lett. B}\ }\textbf {\bibinfo {volume} {771}},\ \bibinfo {pages} {415} (\bibinfo {year} {2017})},\ \Eprint {http://arxiv.org/abs/1607.07062} {arXiv:1607.07062 [hep-th]} \BibitemShut {NoStop}%
\bibitem [{\citenamefont {Sharma}\ \emph {et~al.}(2020)\citenamefont {Sharma}, \citenamefont {Subramanian},\ and\ \citenamefont {Seshadri}}]{Sharma:2019jtb}%
  \BibitemOpen
  \bibfield  {author} {\bibinfo {author} {\bibfnamefont {R.}~\bibnamefont {Sharma}}, \bibinfo {author} {\bibfnamefont {K.}~\bibnamefont {Subramanian}}, \ and\ \bibinfo {author} {\bibfnamefont {T.~R.}\ \bibnamefont {Seshadri}},\ }\href {\doibase 10.1103/PhysRevD.101.103526} {\bibfield  {journal} {\bibinfo  {journal} {Phys. Rev. D}\ }\textbf {\bibinfo {volume} {101}},\ \bibinfo {pages} {103526} (\bibinfo {year} {2020})},\ \Eprint {http://arxiv.org/abs/1912.12089} {arXiv:1912.12089 [astro-ph.CO]} \BibitemShut {NoStop}%
\bibitem [{\citenamefont {Starobinsky}(1979)}]{Starobinsky:1979ty}%
  \BibitemOpen
  \bibfield  {author} {\bibinfo {author} {\bibfnamefont {A.~A.}\ \bibnamefont {Starobinsky}},\ }\href@noop {} {\bibfield  {journal} {\bibinfo  {journal} {JETP Lett.}\ }\textbf {\bibinfo {volume} {30}},\ \bibinfo {pages} {682} (\bibinfo {year} {1979})}\BibitemShut {NoStop}%
\bibitem [{\citenamefont {Grishchuk}(1974)}]{Grishchuk:1974ny}%
  \BibitemOpen
  \bibfield  {author} {\bibinfo {author} {\bibfnamefont {L.~P.}\ \bibnamefont {Grishchuk}},\ }\href@noop {} {\bibfield  {journal} {\bibinfo  {journal} {Zh. Eksp. Teor. Fiz.}\ }\textbf {\bibinfo {volume} {67}},\ \bibinfo {pages} {825} (\bibinfo {year} {1974})}\BibitemShut {NoStop}%
\bibitem [{\citenamefont {Guzzetti}\ \emph {et~al.}(2016)\citenamefont {Guzzetti}, \citenamefont {Bartolo}, \citenamefont {Liguori},\ and\ \citenamefont {Matarrese}}]{Guzzetti:2016mkm}%
  \BibitemOpen
  \bibfield  {author} {\bibinfo {author} {\bibfnamefont {M.~C.}\ \bibnamefont {Guzzetti}}, \bibinfo {author} {\bibfnamefont {N.}~\bibnamefont {Bartolo}}, \bibinfo {author} {\bibfnamefont {M.}~\bibnamefont {Liguori}}, \ and\ \bibinfo {author} {\bibfnamefont {S.}~\bibnamefont {Matarrese}},\ }\href {\doibase 10.1393/ncr/i2016-10127-1} {\bibfield  {journal} {\bibinfo  {journal} {Riv. Nuovo Cim.}\ }\textbf {\bibinfo {volume} {39}},\ \bibinfo {pages} {399} (\bibinfo {year} {2016})},\ \Eprint {http://arxiv.org/abs/1605.01615} {arXiv:1605.01615 [astro-ph.CO]} \BibitemShut {NoStop}%
\bibitem [{\citenamefont {Haque}\ \emph {et~al.}(2021{\natexlab{a}})\citenamefont {Haque}, \citenamefont {Maity}, \citenamefont {Paul},\ and\ \citenamefont {Sriramkumar}}]{Haque:2021dha}%
  \BibitemOpen
  \bibfield  {author} {\bibinfo {author} {\bibfnamefont {M.~R.}\ \bibnamefont {Haque}}, \bibinfo {author} {\bibfnamefont {D.}~\bibnamefont {Maity}}, \bibinfo {author} {\bibfnamefont {T.}~\bibnamefont {Paul}}, \ and\ \bibinfo {author} {\bibfnamefont {L.}~\bibnamefont {Sriramkumar}},\ }\href {\doibase 10.1103/PhysRevD.104.063513} {\bibfield  {journal} {\bibinfo  {journal} {Phys. Rev. D}\ }\textbf {\bibinfo {volume} {104}},\ \bibinfo {pages} {063513} (\bibinfo {year} {2021}{\natexlab{a}})},\ \Eprint {http://arxiv.org/abs/2105.09242} {arXiv:2105.09242 [astro-ph.CO]} \BibitemShut {NoStop}%
\bibitem [{\citenamefont {Adams}\ \emph {et~al.}(2001)\citenamefont {Adams}, \citenamefont {Cresswell},\ and\ \citenamefont {Easther}}]{PhysRevD.64.123514}%
  \BibitemOpen
  \bibfield  {author} {\bibinfo {author} {\bibfnamefont {J.}~\bibnamefont {Adams}}, \bibinfo {author} {\bibfnamefont {B.}~\bibnamefont {Cresswell}}, \ and\ \bibinfo {author} {\bibfnamefont {R.}~\bibnamefont {Easther}},\ }\href {\doibase 10.1103/PhysRevD.64.123514} {\bibfield  {journal} {\bibinfo  {journal} {Phys. Rev. D}\ }\textbf {\bibinfo {volume} {64}},\ \bibinfo {pages} {123514} (\bibinfo {year} {2001})}\BibitemShut {NoStop}%
\bibitem [{\citenamefont {Di}\ and\ \citenamefont {Gong}(2018)}]{Di:2017ndc}%
  \BibitemOpen
  \bibfield  {author} {\bibinfo {author} {\bibfnamefont {H.}~\bibnamefont {Di}}\ and\ \bibinfo {author} {\bibfnamefont {Y.}~\bibnamefont {Gong}},\ }\href {\doibase 10.1088/1475-7516/2018/07/007} {\bibfield  {journal} {\bibinfo  {journal} {JCAP}\ }\textbf {\bibinfo {volume} {07}},\ \bibinfo {pages} {007} (\bibinfo {year} {2018})},\ \Eprint {http://arxiv.org/abs/1707.09578} {arXiv:1707.09578 [astro-ph.CO]} \BibitemShut {NoStop}%
\bibitem [{\citenamefont {Fu}\ \emph {et~al.}(2020)\citenamefont {Fu}, \citenamefont {Wu},\ and\ \citenamefont {Yu}}]{Fu:2019vqc}%
  \BibitemOpen
  \bibfield  {author} {\bibinfo {author} {\bibfnamefont {C.}~\bibnamefont {Fu}}, \bibinfo {author} {\bibfnamefont {P.}~\bibnamefont {Wu}}, \ and\ \bibinfo {author} {\bibfnamefont {H.}~\bibnamefont {Yu}},\ }\href {\doibase 10.1103/PhysRevD.101.023529} {\bibfield  {journal} {\bibinfo  {journal} {Phys. Rev. D}\ }\textbf {\bibinfo {volume} {101}},\ \bibinfo {pages} {023529} (\bibinfo {year} {2020})},\ \Eprint {http://arxiv.org/abs/1912.05927} {arXiv:1912.05927 [astro-ph.CO]} \BibitemShut {NoStop}%
\bibitem [{\citenamefont {Ragavendra}\ \emph {et~al.}(2021{\natexlab{a}})\citenamefont {Ragavendra}, \citenamefont {Saha}, \citenamefont {Sriramkumar},\ and\ \citenamefont {Silk}}]{PhysRevD.103.083510}%
  \BibitemOpen
  \bibfield  {author} {\bibinfo {author} {\bibfnamefont {H.~V.}\ \bibnamefont {Ragavendra}}, \bibinfo {author} {\bibfnamefont {P.}~\bibnamefont {Saha}}, \bibinfo {author} {\bibfnamefont {L.}~\bibnamefont {Sriramkumar}}, \ and\ \bibinfo {author} {\bibfnamefont {J.}~\bibnamefont {Silk}},\ }\href {\doibase 10.1103/PhysRevD.103.083510} {\bibfield  {journal} {\bibinfo  {journal} {Phys. Rev. D}\ }\textbf {\bibinfo {volume} {103}},\ \bibinfo {pages} {083510} (\bibinfo {year} {2021}{\natexlab{a}})}\BibitemShut {NoStop}%
\bibitem [{\citenamefont {Bhaumik}\ and\ \citenamefont {Jain}(2021)}]{Bhaumik:2020dor}%
  \BibitemOpen
  \bibfield  {author} {\bibinfo {author} {\bibfnamefont {N.}~\bibnamefont {Bhaumik}}\ and\ \bibinfo {author} {\bibfnamefont {R.~K.}\ \bibnamefont {Jain}},\ }\href {\doibase 10.1103/PhysRevD.104.023531} {\bibfield  {journal} {\bibinfo  {journal} {Phys. Rev. D}\ }\textbf {\bibinfo {volume} {104}},\ \bibinfo {pages} {023531} (\bibinfo {year} {2021})},\ \Eprint {http://arxiv.org/abs/2009.10424} {arXiv:2009.10424 [astro-ph.CO]} \BibitemShut {NoStop}%
\bibitem [{\citenamefont {Solbi}\ and\ \citenamefont {Karami}(2021)}]{Solbi:2021wbo}%
  \BibitemOpen
  \bibfield  {author} {\bibinfo {author} {\bibfnamefont {M.}~\bibnamefont {Solbi}}\ and\ \bibinfo {author} {\bibfnamefont {K.}~\bibnamefont {Karami}},\ }\href {\doibase 10.1088/1475-7516/2021/08/056} {\bibfield  {journal} {\bibinfo  {journal} {JCAP}\ }\textbf {\bibinfo {volume} {08}},\ \bibinfo {pages} {056} (\bibinfo {year} {2021})},\ \Eprint {http://arxiv.org/abs/2102.05651} {arXiv:2102.05651 [astro-ph.CO]} \BibitemShut {NoStop}%
\bibitem [{\citenamefont {Figueroa}\ \emph {et~al.}(2022)\citenamefont {Figueroa}, \citenamefont {Raatikainen}, \citenamefont {Rasanen},\ and\ \citenamefont {Tomberg}}]{Figueroa:2021zah}%
  \BibitemOpen
  \bibfield  {author} {\bibinfo {author} {\bibfnamefont {D.~G.}\ \bibnamefont {Figueroa}}, \bibinfo {author} {\bibfnamefont {S.}~\bibnamefont {Raatikainen}}, \bibinfo {author} {\bibfnamefont {S.}~\bibnamefont {Rasanen}}, \ and\ \bibinfo {author} {\bibfnamefont {E.}~\bibnamefont {Tomberg}},\ }\href {\doibase 10.1088/1475-7516/2022/05/027} {\bibfield  {journal} {\bibinfo  {journal} {JCAP}\ }\textbf {\bibinfo {volume} {05}},\ \bibinfo {pages} {027} (\bibinfo {year} {2022})},\ \Eprint {http://arxiv.org/abs/2111.07437} {arXiv:2111.07437 [astro-ph.CO]} \BibitemShut {NoStop}%
\bibitem [{\citenamefont {Dom\`enech}\ \emph {et~al.}(2020)\citenamefont {Dom\`enech}, \citenamefont {Pi},\ and\ \citenamefont {Sasaki}}]{Domenech:2020kqm}%
  \BibitemOpen
  \bibfield  {author} {\bibinfo {author} {\bibfnamefont {G.}~\bibnamefont {Dom\`enech}}, \bibinfo {author} {\bibfnamefont {S.}~\bibnamefont {Pi}}, \ and\ \bibinfo {author} {\bibfnamefont {M.}~\bibnamefont {Sasaki}},\ }\href {\doibase 10.1088/1475-7516/2020/08/017} {\bibfield  {journal} {\bibinfo  {journal} {JCAP}\ }\textbf {\bibinfo {volume} {08}},\ \bibinfo {pages} {017} (\bibinfo {year} {2020})},\ \Eprint {http://arxiv.org/abs/2005.12314} {arXiv:2005.12314 [gr-qc]} \BibitemShut {NoStop}%
\bibitem [{\citenamefont {Cai}\ \emph {et~al.}(2019)\citenamefont {Cai}, \citenamefont {Pi},\ and\ \citenamefont {Sasaki}}]{Cai:2018dig}%
  \BibitemOpen
  \bibfield  {author} {\bibinfo {author} {\bibfnamefont {R.-g.}\ \bibnamefont {Cai}}, \bibinfo {author} {\bibfnamefont {S.}~\bibnamefont {Pi}}, \ and\ \bibinfo {author} {\bibfnamefont {M.}~\bibnamefont {Sasaki}},\ }\href {\doibase 10.1103/PhysRevLett.122.201101} {\bibfield  {journal} {\bibinfo  {journal} {Phys. Rev. Lett.}\ }\textbf {\bibinfo {volume} {122}},\ \bibinfo {pages} {201101} (\bibinfo {year} {2019})},\ \Eprint {http://arxiv.org/abs/1810.11000} {arXiv:1810.11000 [astro-ph.CO]} \BibitemShut {NoStop}%
\bibitem [{\citenamefont {Maiti}\ \emph {et~al.}(2025{\natexlab{a}})\citenamefont {Maiti}, \citenamefont {Maity},\ and\ \citenamefont {Srikanth}}]{Maiti:2025rkn}%
  \BibitemOpen
  \bibfield  {author} {\bibinfo {author} {\bibfnamefont {S.}~\bibnamefont {Maiti}}, \bibinfo {author} {\bibfnamefont {D.}~\bibnamefont {Maity}}, \ and\ \bibinfo {author} {\bibfnamefont {R.}~\bibnamefont {Srikanth}},\ }\href {\doibase 10.1103/xvgz-9hbn} {\bibfield  {journal} {\bibinfo  {journal} {Phys. Rev. D}\ }\textbf {\bibinfo {volume} {112}},\ \bibinfo {pages} {043535} (\bibinfo {year} {2025}{\natexlab{a}})},\ \Eprint {http://arxiv.org/abs/2504.15400} {arXiv:2504.15400 [astro-ph.CO]} \BibitemShut {NoStop}%
\bibitem [{\citenamefont {Maiti}(2025)}]{Maiti:2025awl}%
  \BibitemOpen
  \bibfield  {author} {\bibinfo {author} {\bibfnamefont {S.}~\bibnamefont {Maiti}},\ }\href {\doibase 10.1103/mjxr-pp2t} {\bibfield  {journal} {\bibinfo  {journal} {Phys. Rev. D}\ }\textbf {\bibinfo {volume} {112}},\ \bibinfo {pages} {043536} (\bibinfo {year} {2025})},\ \Eprint {http://arxiv.org/abs/2506.06183} {arXiv:2506.06183 [astro-ph.CO]} \BibitemShut {NoStop}%
\bibitem [{\citenamefont {Maiti}\ \emph {et~al.}(2025{\natexlab{b}})\citenamefont {Maiti}, \citenamefont {Maity},\ and\ \citenamefont {Srikanth}}]{Maiti:2025cbi}%
  \BibitemOpen
  \bibfield  {author} {\bibinfo {author} {\bibfnamefont {S.}~\bibnamefont {Maiti}}, \bibinfo {author} {\bibfnamefont {D.}~\bibnamefont {Maity}}, \ and\ \bibinfo {author} {\bibfnamefont {R.}~\bibnamefont {Srikanth}},\ }\href {\doibase 10.1103/4n86-9nsc} {\bibfield  {journal} {\bibinfo  {journal} {Phys. Rev. D}\ }\textbf {\bibinfo {volume} {112}},\ \bibinfo {pages} {063552} (\bibinfo {year} {2025}{\natexlab{b}})},\ \Eprint {http://arxiv.org/abs/2505.13623} {arXiv:2505.13623 [astro-ph.CO]} \BibitemShut {NoStop}%
\bibitem [{\citenamefont {Chakraborty}\ \emph {et~al.}(2025)\citenamefont {Chakraborty}, \citenamefont {Maiti},\ and\ \citenamefont {Maity}}]{Chakraborty:2024rgl}%
  \BibitemOpen
  \bibfield  {author} {\bibinfo {author} {\bibfnamefont {A.}~\bibnamefont {Chakraborty}}, \bibinfo {author} {\bibfnamefont {S.}~\bibnamefont {Maiti}}, \ and\ \bibinfo {author} {\bibfnamefont {D.}~\bibnamefont {Maity}},\ }\href {\doibase 10.1103/PhysRevD.111.083505} {\bibfield  {journal} {\bibinfo  {journal} {Phys. Rev. D}\ }\textbf {\bibinfo {volume} {111}},\ \bibinfo {pages} {083505} (\bibinfo {year} {2025})},\ \Eprint {http://arxiv.org/abs/2408.07767} {arXiv:2408.07767 [astro-ph.CO]} \BibitemShut {NoStop}%
\bibitem [{\citenamefont {Chakraborty}\ \emph {et~al.}(2023)\citenamefont {Chakraborty}, \citenamefont {Haque}, \citenamefont {Maity},\ and\ \citenamefont {Mondal}}]{Chakraborty:2023ocr}%
  \BibitemOpen
  \bibfield  {author} {\bibinfo {author} {\bibfnamefont {A.}~\bibnamefont {Chakraborty}}, \bibinfo {author} {\bibfnamefont {M.~R.}\ \bibnamefont {Haque}}, \bibinfo {author} {\bibfnamefont {D.}~\bibnamefont {Maity}}, \ and\ \bibinfo {author} {\bibfnamefont {R.}~\bibnamefont {Mondal}},\ }\href {\doibase 10.1103/PhysRevD.108.023515} {\bibfield  {journal} {\bibinfo  {journal} {Phys. Rev. D}\ }\textbf {\bibinfo {volume} {108}},\ \bibinfo {pages} {023515} (\bibinfo {year} {2023})},\ \Eprint {http://arxiv.org/abs/2304.13637} {arXiv:2304.13637 [astro-ph.CO]} \BibitemShut {NoStop}%
\bibitem [{\citenamefont {Hoory}\ \emph {et~al.}(2025)\citenamefont {Hoory}, \citenamefont {Martin}, \citenamefont {Paul},\ and\ \citenamefont {Sriramkumar}}]{Hoory:2025qgm}%
  \BibitemOpen
  \bibfield  {author} {\bibinfo {author} {\bibfnamefont {A.}~\bibnamefont {Hoory}}, \bibinfo {author} {\bibfnamefont {J.}~\bibnamefont {Martin}}, \bibinfo {author} {\bibfnamefont {A.}~\bibnamefont {Paul}}, \ and\ \bibinfo {author} {\bibfnamefont {L.}~\bibnamefont {Sriramkumar}},\ }\href@noop {} {\  (\bibinfo {year} {2025})},\ \Eprint {http://arxiv.org/abs/2512.03959} {arXiv:2512.03959 [astro-ph.CO]} \BibitemShut {NoStop}%
\bibitem [{\citenamefont {Inui}\ \emph {et~al.}(2024)\citenamefont {Inui}, \citenamefont {Jaraba}, \citenamefont {Kuroyanagi},\ and\ \citenamefont {Yokoyama}}]{Inui:2023qsd}%
  \BibitemOpen
  \bibfield  {author} {\bibinfo {author} {\bibfnamefont {R.}~\bibnamefont {Inui}}, \bibinfo {author} {\bibfnamefont {S.}~\bibnamefont {Jaraba}}, \bibinfo {author} {\bibfnamefont {S.}~\bibnamefont {Kuroyanagi}}, \ and\ \bibinfo {author} {\bibfnamefont {S.}~\bibnamefont {Yokoyama}},\ }\href {\doibase 10.1088/1475-7516/2024/05/082} {\bibfield  {journal} {\bibinfo  {journal} {JCAP}\ }\textbf {\bibinfo {volume} {05}},\ \bibinfo {pages} {082} (\bibinfo {year} {2024})},\ \Eprint {http://arxiv.org/abs/2311.05423} {arXiv:2311.05423 [astro-ph.CO]} \BibitemShut {NoStop}%
\bibitem [{\citenamefont {Abe}\ \emph {et~al.}(2023)\citenamefont {Abe}, \citenamefont {Inui}, \citenamefont {Tada},\ and\ \citenamefont {Yokoyama}}]{Abe:2022xur}%
  \BibitemOpen
  \bibfield  {author} {\bibinfo {author} {\bibfnamefont {K.~T.}\ \bibnamefont {Abe}}, \bibinfo {author} {\bibfnamefont {R.}~\bibnamefont {Inui}}, \bibinfo {author} {\bibfnamefont {Y.}~\bibnamefont {Tada}}, \ and\ \bibinfo {author} {\bibfnamefont {S.}~\bibnamefont {Yokoyama}},\ }\href {\doibase 10.1088/1475-7516/2023/05/044} {\bibfield  {journal} {\bibinfo  {journal} {JCAP}\ }\textbf {\bibinfo {volume} {05}},\ \bibinfo {pages} {044} (\bibinfo {year} {2023})},\ \Eprint {http://arxiv.org/abs/2209.13891} {arXiv:2209.13891 [astro-ph.CO]} \BibitemShut {NoStop}%
\bibitem [{\citenamefont {Okano}\ and\ \citenamefont {Fujita}(2021)}]{Okano:2020uyr}%
  \BibitemOpen
  \bibfield  {author} {\bibinfo {author} {\bibfnamefont {S.}~\bibnamefont {Okano}}\ and\ \bibinfo {author} {\bibfnamefont {T.}~\bibnamefont {Fujita}},\ }\href {\doibase 10.1088/1475-7516/2021/03/026} {\bibfield  {journal} {\bibinfo  {journal} {JCAP}\ }\textbf {\bibinfo {volume} {03}},\ \bibinfo {pages} {026} (\bibinfo {year} {2021})},\ \Eprint {http://arxiv.org/abs/2005.13833} {arXiv:2005.13833 [astro-ph.CO]} \BibitemShut {NoStop}%
\bibitem [{\citenamefont {Fletcher}(2011)}]{fletcher2011magnetic}%
  \BibitemOpen
  \bibfield  {author} {\bibinfo {author} {\bibfnamefont {A.}~\bibnamefont {Fletcher}},\ }\href@noop {} {\bibfield  {journal} {\bibinfo  {journal} {arXiv preprint arXiv:1104.2427}\ } (\bibinfo {year} {2011})}\BibitemShut {NoStop}%
\bibitem [{\citenamefont {Beck}(2016)}]{beck2016magnetic}%
  \BibitemOpen
  \bibfield  {author} {\bibinfo {author} {\bibfnamefont {R.}~\bibnamefont {Beck}},\ }\href@noop {} {\bibfield  {journal} {\bibinfo  {journal} {The Astronomy and Astrophysics Review}\ }\textbf {\bibinfo {volume} {24}},\ \bibinfo {pages} {4} (\bibinfo {year} {2016})}\BibitemShut {NoStop}%
\bibitem [{\citenamefont {Haverkorn}\ \emph {et~al.}(2008)\citenamefont {Haverkorn}, \citenamefont {Brown}, \citenamefont {Gaensler},\ and\ \citenamefont {McClure-Griffiths}}]{Haverkorn:2008tb}%
  \BibitemOpen
  \bibfield  {author} {\bibinfo {author} {\bibfnamefont {M.}~\bibnamefont {Haverkorn}}, \bibinfo {author} {\bibfnamefont {J.~C.}\ \bibnamefont {Brown}}, \bibinfo {author} {\bibfnamefont {B.~M.}\ \bibnamefont {Gaensler}}, \ and\ \bibinfo {author} {\bibfnamefont {N.~M.}\ \bibnamefont {McClure-Griffiths}},\ }\href {\doibase 10.1086/587165} {\bibfield  {journal} {\bibinfo  {journal} {Astrophys. J.}\ }\textbf {\bibinfo {volume} {680}},\ \bibinfo {pages} {362} (\bibinfo {year} {2008})},\ \Eprint {http://arxiv.org/abs/0802.2740} {arXiv:0802.2740 [astro-ph]} \BibitemShut {NoStop}%
\bibitem [{\citenamefont {Kronberg}\ \emph {et~al.}(2001)\citenamefont {Kronberg}, \citenamefont {Dufton}, \citenamefont {Li},\ and\ \citenamefont {Colgate}}]{kronberg2001magnetic}%
  \BibitemOpen
  \bibfield  {author} {\bibinfo {author} {\bibfnamefont {P.}~\bibnamefont {Kronberg}}, \bibinfo {author} {\bibfnamefont {Q.}~\bibnamefont {Dufton}}, \bibinfo {author} {\bibfnamefont {H.}~\bibnamefont {Li}}, \ and\ \bibinfo {author} {\bibfnamefont {S.}~\bibnamefont {Colgate}},\ }\href@noop {} {\bibfield  {journal} {\bibinfo  {journal} {The Astrophysical Journal}\ }\textbf {\bibinfo {volume} {560}},\ \bibinfo {pages} {178} (\bibinfo {year} {2001})}\BibitemShut {NoStop}%
\bibitem [{\citenamefont {Brandenburg}\ and\ \citenamefont {Ntormousi}(2023)}]{brandenburg:2023}%
  \BibitemOpen
  \bibfield  {author} {\bibinfo {author} {\bibfnamefont {A.}~\bibnamefont {Brandenburg}}\ and\ \bibinfo {author} {\bibfnamefont {E.}~\bibnamefont {Ntormousi}},\ }\href@noop {} {\bibfield  {journal} {\bibinfo  {journal} {Annual Review of Astronomy and Astrophysics}\ }\textbf {\bibinfo {volume} {61}},\ \bibinfo {pages} {561} (\bibinfo {year} {2023})},\ \Eprint {http://arxiv.org/abs/2211.03476} {arXiv:2211.03476 [astro-ph.GA]} \BibitemShut {NoStop}%
\bibitem [{\citenamefont {Turner}\ and\ \citenamefont {Widrow}(1988)}]{PhysRevD.37.2743}%
  \BibitemOpen
  \bibfield  {author} {\bibinfo {author} {\bibfnamefont {M.~S.}\ \bibnamefont {Turner}}\ and\ \bibinfo {author} {\bibfnamefont {L.~M.}\ \bibnamefont {Widrow}},\ }\href {\doibase 10.1103/PhysRevD.37.2743} {\bibfield  {journal} {\bibinfo  {journal} {Phys. Rev. D}\ }\textbf {\bibinfo {volume} {37}},\ \bibinfo {pages} {2743} (\bibinfo {year} {1988})}\BibitemShut {NoStop}%
\bibitem [{\citenamefont {Mazzitelli}\ and\ \citenamefont {Spedalieri}(1995)}]{PhysRevD.52.6694}%
  \BibitemOpen
  \bibfield  {author} {\bibinfo {author} {\bibfnamefont {F.~D.}\ \bibnamefont {Mazzitelli}}\ and\ \bibinfo {author} {\bibfnamefont {F.~M.}\ \bibnamefont {Spedalieri}},\ }\href {\doibase 10.1103/PhysRevD.52.6694} {\bibfield  {journal} {\bibinfo  {journal} {Phys. Rev. D}\ }\textbf {\bibinfo {volume} {52}},\ \bibinfo {pages} {6694} (\bibinfo {year} {1995})}\BibitemShut {NoStop}%
\bibitem [{\citenamefont {Lambiase}\ and\ \citenamefont {Prasanna}(2004)}]{PhysRevD.70.063502}%
  \BibitemOpen
  \bibfield  {author} {\bibinfo {author} {\bibfnamefont {G.}~\bibnamefont {Lambiase}}\ and\ \bibinfo {author} {\bibfnamefont {A.~R.}\ \bibnamefont {Prasanna}},\ }\href {\doibase 10.1103/PhysRevD.70.063502} {\bibfield  {journal} {\bibinfo  {journal} {Phys. Rev. D}\ }\textbf {\bibinfo {volume} {70}},\ \bibinfo {pages} {063502} (\bibinfo {year} {2004})}\BibitemShut {NoStop}%
\bibitem [{\citenamefont {{Yanagihara}}\ \emph {et~al.}(2023)\citenamefont {{Yanagihara}}, \citenamefont {{Uchida}}, \citenamefont {{Fujita}},\ and\ \citenamefont {{Tsujikawa}}}]{2023arXiv231207938Y}%
  \BibitemOpen
  \bibfield  {author} {\bibinfo {author} {\bibfnamefont {K.}~\bibnamefont {{Yanagihara}}}, \bibinfo {author} {\bibfnamefont {F.}~\bibnamefont {{Uchida}}}, \bibinfo {author} {\bibfnamefont {T.}~\bibnamefont {{Fujita}}}, \ and\ \bibinfo {author} {\bibfnamefont {S.}~\bibnamefont {{Tsujikawa}}},\ }\href {\doibase 10.48550/arXiv.2312.07938} {\bibfield  {journal} {\bibinfo  {journal} {arXiv e-prints}\ ,\ \bibinfo {eid} {arXiv:2312.07938}} (\bibinfo {year} {2023})},\ \Eprint {http://arxiv.org/abs/2312.07938} {arXiv:2312.07938 [astro-ph.CO]} \BibitemShut {NoStop}%
\bibitem [{\citenamefont {Caprini}\ and\ \citenamefont {Durrer}(2001)}]{PhysRevD.65.023517}%
  \BibitemOpen
  \bibfield  {author} {\bibinfo {author} {\bibfnamefont {C.}~\bibnamefont {Caprini}}\ and\ \bibinfo {author} {\bibfnamefont {R.}~\bibnamefont {Durrer}},\ }\href {\doibase 10.1103/PhysRevD.65.023517} {\bibfield  {journal} {\bibinfo  {journal} {Phys. Rev. D}\ }\textbf {\bibinfo {volume} {65}},\ \bibinfo {pages} {023517} (\bibinfo {year} {2001})}\BibitemShut {NoStop}%
\bibitem [{\citenamefont {Maiti}\ and\ \citenamefont {Maity}(2026{\natexlab{a}})}]{Maiti:2025ijr}%
  \BibitemOpen
  \bibfield  {author} {\bibinfo {author} {\bibfnamefont {S.}~\bibnamefont {Maiti}}\ and\ \bibinfo {author} {\bibfnamefont {D.}~\bibnamefont {Maity}},\ }\href {\doibase 10.1088/1475-7516/2026/04/020} {\bibfield  {journal} {\bibinfo  {journal} {JCAP}\ }\textbf {\bibinfo {volume} {04}},\ \bibinfo {pages} {020} (\bibinfo {year} {2026}{\natexlab{a}})},\ \Eprint {http://arxiv.org/abs/2508.19217} {arXiv:2508.19217 [astro-ph.CO]} \BibitemShut {NoStop}%
\bibitem [{\citenamefont {Fujita}\ and\ \citenamefont {Yokoyama}(2013)}]{Fujita:2013qxa}%
  \BibitemOpen
  \bibfield  {author} {\bibinfo {author} {\bibfnamefont {T.}~\bibnamefont {Fujita}}\ and\ \bibinfo {author} {\bibfnamefont {S.}~\bibnamefont {Yokoyama}},\ }\href {\doibase 10.1088/1475-7516/2013/09/009} {\bibfield  {journal} {\bibinfo  {journal} {JCAP}\ }\textbf {\bibinfo {volume} {09}},\ \bibinfo {pages} {009} (\bibinfo {year} {2013})},\ \Eprint {http://arxiv.org/abs/1306.2992} {arXiv:1306.2992 [astro-ph.CO]} \BibitemShut {NoStop}%
\bibitem [{\citenamefont {Fujita}\ and\ \citenamefont {Namba}(2016)}]{PhysRevD.94.043523}%
  \BibitemOpen
  \bibfield  {author} {\bibinfo {author} {\bibfnamefont {T.}~\bibnamefont {Fujita}}\ and\ \bibinfo {author} {\bibfnamefont {R.}~\bibnamefont {Namba}},\ }\href {\doibase 10.1103/PhysRevD.94.043523} {\bibfield  {journal} {\bibinfo  {journal} {Phys. Rev. D}\ }\textbf {\bibinfo {volume} {94}},\ \bibinfo {pages} {043523} (\bibinfo {year} {2016})}\BibitemShut {NoStop}%
\bibitem [{\citenamefont {Amaro-Seoane}\ \emph {et~al.}(2013)\citenamefont {Amaro-Seoane} \emph {et~al.}}]{Amaro-Seoane:2012aqc}%
  \BibitemOpen
  \bibfield  {author} {\bibinfo {author} {\bibfnamefont {P.}~\bibnamefont {Amaro-Seoane}} \emph {et~al.},\ }\href@noop {} {\bibfield  {journal} {\bibinfo  {journal} {GW Notes}\ }\textbf {\bibinfo {volume} {6}},\ \bibinfo {pages} {4} (\bibinfo {year} {2013})},\ \Eprint {http://arxiv.org/abs/1201.3621} {arXiv:1201.3621 [astro-ph.CO]} \BibitemShut {NoStop}%
\bibitem [{\citenamefont {Barausse}\ \emph {et~al.}(2020)\citenamefont {Barausse} \emph {et~al.}}]{Barausse:2020rsu}%
  \BibitemOpen
  \bibfield  {author} {\bibinfo {author} {\bibfnamefont {E.}~\bibnamefont {Barausse}} \emph {et~al.},\ }\href {\doibase 10.1007/s10714-020-02691-1} {\bibfield  {journal} {\bibinfo  {journal} {Gen. Rel. Grav.}\ }\textbf {\bibinfo {volume} {52}},\ \bibinfo {pages} {81} (\bibinfo {year} {2020})},\ \Eprint {http://arxiv.org/abs/2001.09793} {arXiv:2001.09793 [gr-qc]} \BibitemShut {NoStop}%
\bibitem [{\citenamefont {Caprini}\ \emph {et~al.}(2016)\citenamefont {Caprini} \emph {et~al.}}]{Caprini:2015zlo}%
  \BibitemOpen
  \bibfield  {author} {\bibinfo {author} {\bibfnamefont {C.}~\bibnamefont {Caprini}} \emph {et~al.},\ }\href {\doibase 10.1088/1475-7516/2016/04/001} {\bibfield  {journal} {\bibinfo  {journal} {JCAP}\ }\textbf {\bibinfo {volume} {04}},\ \bibinfo {pages} {001} (\bibinfo {year} {2016})},\ \Eprint {http://arxiv.org/abs/1512.06239} {arXiv:1512.06239 [astro-ph.CO]} \BibitemShut {NoStop}%
\bibitem [{\citenamefont {Crowder}\ and\ \citenamefont {Cornish}(2005)}]{Crowder:2005nr}%
  \BibitemOpen
  \bibfield  {author} {\bibinfo {author} {\bibfnamefont {J.}~\bibnamefont {Crowder}}\ and\ \bibinfo {author} {\bibfnamefont {N.~J.}\ \bibnamefont {Cornish}},\ }\href {\doibase 10.1103/PhysRevD.72.083005} {\bibfield  {journal} {\bibinfo  {journal} {Phys. Rev. D}\ }\textbf {\bibinfo {volume} {72}},\ \bibinfo {pages} {083005} (\bibinfo {year} {2005})},\ \Eprint {http://arxiv.org/abs/gr-qc/0506015} {arXiv:gr-qc/0506015} \BibitemShut {NoStop}%
\bibitem [{\citenamefont {Corbin}\ and\ \citenamefont {Cornish}(2006)}]{Corbin:2005ny}%
  \BibitemOpen
  \bibfield  {author} {\bibinfo {author} {\bibfnamefont {V.}~\bibnamefont {Corbin}}\ and\ \bibinfo {author} {\bibfnamefont {N.~J.}\ \bibnamefont {Cornish}},\ }\href {\doibase 10.1088/0264-9381/23/7/014} {\bibfield  {journal} {\bibinfo  {journal} {Class. Quant. Grav.}\ }\textbf {\bibinfo {volume} {23}},\ \bibinfo {pages} {2435} (\bibinfo {year} {2006})},\ \Eprint {http://arxiv.org/abs/gr-qc/0512039} {arXiv:gr-qc/0512039} \BibitemShut {NoStop}%
\bibitem [{\citenamefont {Baker}\ \emph {et~al.}(2019)\citenamefont {Baker} \emph {et~al.}}]{Baker:2019pnp}%
  \BibitemOpen
  \bibfield  {author} {\bibinfo {author} {\bibfnamefont {J.}~\bibnamefont {Baker}} \emph {et~al.},\ }\href@noop {} {\bibfield  {journal} {\bibinfo  {journal} {Bull. Am. Astron. Soc.}\ }\textbf {\bibinfo {volume} {51}},\ \bibinfo {pages} {243} (\bibinfo {year} {2019})},\ \Eprint {http://arxiv.org/abs/1907.11305} {arXiv:1907.11305 [astro-ph.IM]} \BibitemShut {NoStop}%
\bibitem [{\citenamefont {Seto}\ \emph {et~al.}(2001)\citenamefont {Seto}, \citenamefont {Kawamura},\ and\ \citenamefont {Nakamura}}]{Seto:2001qf}%
  \BibitemOpen
  \bibfield  {author} {\bibinfo {author} {\bibfnamefont {N.}~\bibnamefont {Seto}}, \bibinfo {author} {\bibfnamefont {S.}~\bibnamefont {Kawamura}}, \ and\ \bibinfo {author} {\bibfnamefont {T.}~\bibnamefont {Nakamura}},\ }\href {\doibase 10.1103/PhysRevLett.87.221103} {\bibfield  {journal} {\bibinfo  {journal} {Phys. Rev. Lett.}\ }\textbf {\bibinfo {volume} {87}},\ \bibinfo {pages} {221103} (\bibinfo {year} {2001})},\ \Eprint {http://arxiv.org/abs/astro-ph/0108011} {arXiv:astro-ph/0108011} \BibitemShut {NoStop}%
\bibitem [{\citenamefont {Kawamura}\ \emph {et~al.}(2011)\citenamefont {Kawamura} \emph {et~al.}}]{Kawamura:2011zz}%
  \BibitemOpen
  \bibfield  {author} {\bibinfo {author} {\bibfnamefont {S.}~\bibnamefont {Kawamura}} \emph {et~al.},\ }\href {\doibase 10.1088/0264-9381/28/9/094011} {\bibfield  {journal} {\bibinfo  {journal} {Class. Quant. Grav.}\ }\textbf {\bibinfo {volume} {28}},\ \bibinfo {pages} {094011} (\bibinfo {year} {2011})}\BibitemShut {NoStop}%
\bibitem [{\citenamefont {Suemasa}\ \emph {et~al.}(2017)\citenamefont {Suemasa}, \citenamefont {Nakagawa},\ and\ \citenamefont {Musha}}]{Suemasa:2017ppd}%
  \BibitemOpen
  \bibfield  {author} {\bibinfo {author} {\bibfnamefont {A.}~\bibnamefont {Suemasa}}, \bibinfo {author} {\bibfnamefont {K.}~\bibnamefont {Nakagawa}}, \ and\ \bibinfo {author} {\bibfnamefont {M.}~\bibnamefont {Musha}},\ }\href {\doibase 10.1117/12.2304209} {\bibfield  {journal} {\bibinfo  {journal} {Proc. SPIE Int. Soc. Opt. Eng.}\ }\textbf {\bibinfo {volume} {10563}},\ \bibinfo {pages} {105632V} (\bibinfo {year} {2017})}\BibitemShut {NoStop}%
\bibitem [{\citenamefont {Janssen}\ \emph {et~al.}(2015)\citenamefont {Janssen} \emph {et~al.}}]{Janssen:2014dka}%
  \BibitemOpen
  \bibfield  {author} {\bibinfo {author} {\bibfnamefont {G.}~\bibnamefont {Janssen}} \emph {et~al.},\ }\href {\doibase 10.22323/1.215.0037} {\bibfield  {journal} {\bibinfo  {journal} {PoS}\ }\textbf {\bibinfo {volume} {AASKA14}},\ \bibinfo {pages} {037} (\bibinfo {year} {2015})},\ \Eprint {http://arxiv.org/abs/1501.00127} {arXiv:1501.00127 [astro-ph.IM]} \BibitemShut {NoStop}%
\bibitem [{\citenamefont {Agazie}\ \emph {et~al.}(2023)\citenamefont {Agazie} \emph {et~al.}}]{NANOGrav:2023gor}%
  \BibitemOpen
  \bibfield  {author} {\bibinfo {author} {\bibfnamefont {G.}~\bibnamefont {Agazie}} \emph {et~al.} (\bibinfo {collaboration} {NANOGrav}),\ }\href {\doibase 10.3847/2041-8213/acdac6} {\bibfield  {journal} {\bibinfo  {journal} {Astrophys. J. Lett.}\ }\textbf {\bibinfo {volume} {951}},\ \bibinfo {pages} {L8} (\bibinfo {year} {2023})},\ \Eprint {http://arxiv.org/abs/2306.16213} {arXiv:2306.16213 [astro-ph.HE]} \BibitemShut {NoStop}%
\bibitem [{\citenamefont {et~al}(2023)}]{2023arXiv230616224A}%
  \BibitemOpen
  \bibfield  {author} {\bibinfo {author} {\bibfnamefont {A.}~\bibnamefont {et~al}},\ }\href {\doibase 10.48550/arXiv.2306.16224} {\bibfield  {journal} {\bibinfo  {journal} {arXiv e-prints}\ ,\ \bibinfo {eid} {arXiv:2306.16224}} (\bibinfo {year} {2023})},\ \Eprint {http://arxiv.org/abs/2306.16224} {arXiv:2306.16224 [astro-ph.HE]} \BibitemShut {NoStop}%
\bibitem [{\citenamefont {Reardon}\ \emph {et~al.}(2023)\citenamefont {Reardon} \emph {et~al.}}]{Reardon:2023gzh}%
  \BibitemOpen
  \bibfield  {author} {\bibinfo {author} {\bibfnamefont {D.~J.}\ \bibnamefont {Reardon}} \emph {et~al.},\ }\href {\doibase 10.3847/2041-8213/acdd02} {\bibfield  {journal} {\bibinfo  {journal} {Astrophys. J. Lett.}\ }\textbf {\bibinfo {volume} {951}},\ \bibinfo {pages} {L6} (\bibinfo {year} {2023})},\ \Eprint {http://arxiv.org/abs/2306.16215} {arXiv:2306.16215 [astro-ph.HE]} \BibitemShut {NoStop}%
\bibitem [{\citenamefont {Zic}\ \emph {et~al.}(2023)\citenamefont {Zic} \emph {et~al.}}]{Zic:2023gta}%
  \BibitemOpen
  \bibfield  {author} {\bibinfo {author} {\bibfnamefont {A.}~\bibnamefont {Zic}} \emph {et~al.},\ }\href {\doibase 10.1017/pasa.2023.36} {\bibfield  {journal} {\bibinfo  {journal} {Publ. Astron. Soc. Austral.}\ }\textbf {\bibinfo {volume} {40}},\ \bibinfo {pages} {e049} (\bibinfo {year} {2023})},\ \Eprint {http://arxiv.org/abs/2306.16230} {arXiv:2306.16230 [astro-ph.HE]} \BibitemShut {NoStop}%
\bibitem [{\citenamefont {Xu}\ \emph {et~al.}(2023)\citenamefont {Xu} \emph {et~al.}}]{Xu:2023wog}%
  \BibitemOpen
  \bibfield  {author} {\bibinfo {author} {\bibfnamefont {H.}~\bibnamefont {Xu}} \emph {et~al.},\ }\href {\doibase 10.1088/1674-4527/acdfa5} {\bibfield  {journal} {\bibinfo  {journal} {Res. Astron. Astrophys.}\ }\textbf {\bibinfo {volume} {23}},\ \bibinfo {pages} {075024} (\bibinfo {year} {2023})},\ \Eprint {http://arxiv.org/abs/2306.16216} {arXiv:2306.16216 [astro-ph.HE]} \BibitemShut {NoStop}%
\bibitem [{\citenamefont {Luo}\ \emph {et~al.}(2016)\citenamefont {Luo} \emph {et~al.}}]{TianQin:2015yph}%
  \BibitemOpen
  \bibfield  {author} {\bibinfo {author} {\bibfnamefont {J.}~\bibnamefont {Luo}} \emph {et~al.} (\bibinfo {collaboration} {TianQin}),\ }\href {\doibase 10.1088/0264-9381/33/3/035010} {\bibfield  {journal} {\bibinfo  {journal} {Class. Quant. Grav.}\ }\textbf {\bibinfo {volume} {33}},\ \bibinfo {pages} {035010} (\bibinfo {year} {2016})},\ \Eprint {http://arxiv.org/abs/1512.02076} {arXiv:1512.02076 [astro-ph.IM]} \BibitemShut {NoStop}%
\bibitem [{\citenamefont {Ferreira}\ \emph {et~al.}(2013)\citenamefont {Ferreira}, \citenamefont {Jain},\ and\ \citenamefont {Sloth}}]{Ferreira:2013sqa}%
  \BibitemOpen
  \bibfield  {author} {\bibinfo {author} {\bibfnamefont {R.~J.~Z.}\ \bibnamefont {Ferreira}}, \bibinfo {author} {\bibfnamefont {R.~K.}\ \bibnamefont {Jain}}, \ and\ \bibinfo {author} {\bibfnamefont {M.~S.}\ \bibnamefont {Sloth}},\ }\href {\doibase 10.1088/1475-7516/2013/10/004} {\bibfield  {journal} {\bibinfo  {journal} {JCAP}\ }\textbf {\bibinfo {volume} {10}},\ \bibinfo {pages} {004} (\bibinfo {year} {2013})},\ \Eprint {http://arxiv.org/abs/1305.7151} {arXiv:1305.7151 [astro-ph.CO]} \BibitemShut {NoStop}%
\bibitem [{\citenamefont {Kobayashi}(2014)}]{Kobayashi:2014sga}%
  \BibitemOpen
  \bibfield  {author} {\bibinfo {author} {\bibfnamefont {T.}~\bibnamefont {Kobayashi}},\ }\href {\doibase 10.1088/1475-7516/2014/05/040} {\bibfield  {journal} {\bibinfo  {journal} {JCAP}\ }\textbf {\bibinfo {volume} {05}},\ \bibinfo {pages} {040} (\bibinfo {year} {2014})},\ \Eprint {http://arxiv.org/abs/1403.5168} {arXiv:1403.5168 [astro-ph.CO]} \BibitemShut {NoStop}%
\bibitem [{\citenamefont {Kobayashi}\ and\ \citenamefont {Sloth}(2019{\natexlab{a}})}]{PhysRevD.100.023524}%
  \BibitemOpen
  \bibfield  {author} {\bibinfo {author} {\bibfnamefont {T.}~\bibnamefont {Kobayashi}}\ and\ \bibinfo {author} {\bibfnamefont {M.~S.}\ \bibnamefont {Sloth}},\ }\href {\doibase 10.1103/PhysRevD.100.023524} {\bibfield  {journal} {\bibinfo  {journal} {Phys. Rev. D}\ }\textbf {\bibinfo {volume} {100}},\ \bibinfo {pages} {023524} (\bibinfo {year} {2019}{\natexlab{a}})}\BibitemShut {NoStop}%
\bibitem [{\citenamefont {Haque}\ \emph {et~al.}(2021{\natexlab{b}})\citenamefont {Haque}, \citenamefont {Maity},\ and\ \citenamefont {Pal}}]{Haque:2020bip}%
  \BibitemOpen
  \bibfield  {author} {\bibinfo {author} {\bibfnamefont {M.~R.}\ \bibnamefont {Haque}}, \bibinfo {author} {\bibfnamefont {D.}~\bibnamefont {Maity}}, \ and\ \bibinfo {author} {\bibfnamefont {S.}~\bibnamefont {Pal}},\ }\href {\doibase 10.1103/PhysRevD.103.103540} {\bibfield  {journal} {\bibinfo  {journal} {Phys. Rev. D}\ }\textbf {\bibinfo {volume} {103}},\ \bibinfo {pages} {103540} (\bibinfo {year} {2021}{\natexlab{b}})},\ \Eprint {http://arxiv.org/abs/2012.10859} {arXiv:2012.10859 [hep-th]} \BibitemShut {NoStop}%
\bibitem [{\citenamefont {Maity}\ \emph {et~al.}(2021)\citenamefont {Maity}, \citenamefont {Pal},\ and\ \citenamefont {Paul}}]{Maity:2021qps}%
  \BibitemOpen
  \bibfield  {author} {\bibinfo {author} {\bibfnamefont {D.}~\bibnamefont {Maity}}, \bibinfo {author} {\bibfnamefont {S.}~\bibnamefont {Pal}}, \ and\ \bibinfo {author} {\bibfnamefont {T.}~\bibnamefont {Paul}},\ }\href {\doibase 10.1088/1475-7516/2021/05/045} {\bibfield  {journal} {\bibinfo  {journal} {JCAP}\ }\textbf {\bibinfo {volume} {05}},\ \bibinfo {pages} {045} (\bibinfo {year} {2021})},\ \Eprint {http://arxiv.org/abs/2103.02411} {arXiv:2103.02411 [hep-th]} \BibitemShut {NoStop}%
\bibitem [{\citenamefont {Tripathy}\ \emph {et~al.}(2022)\citenamefont {Tripathy}, \citenamefont {Chowdhury}, \citenamefont {Jain},\ and\ \citenamefont {Sriramkumar}}]{Tripathy:2021sfb}%
  \BibitemOpen
  \bibfield  {author} {\bibinfo {author} {\bibfnamefont {S.}~\bibnamefont {Tripathy}}, \bibinfo {author} {\bibfnamefont {D.}~\bibnamefont {Chowdhury}}, \bibinfo {author} {\bibfnamefont {R.~K.}\ \bibnamefont {Jain}}, \ and\ \bibinfo {author} {\bibfnamefont {L.}~\bibnamefont {Sriramkumar}},\ }\href {\doibase 10.1103/PhysRevD.105.063519} {\bibfield  {journal} {\bibinfo  {journal} {Phys. Rev. D}\ }\textbf {\bibinfo {volume} {105}},\ \bibinfo {pages} {063519} (\bibinfo {year} {2022})},\ \Eprint {http://arxiv.org/abs/2111.01478} {arXiv:2111.01478 [astro-ph.CO]} \BibitemShut {NoStop}%
\bibitem [{\citenamefont {Li}\ and\ \citenamefont {Zhang}(2022)}]{Li:2022yqb}%
  \BibitemOpen
  \bibfield  {author} {\bibinfo {author} {\bibfnamefont {Y.}~\bibnamefont {Li}}\ and\ \bibinfo {author} {\bibfnamefont {L.-Y.}\ \bibnamefont {Zhang}},\ }\href {\doibase 10.1142/S0217732322500699} {\bibfield  {journal} {\bibinfo  {journal} {Mod. Phys. Lett. A}\ }\textbf {\bibinfo {volume} {37}},\ \bibinfo {pages} {2250069} (\bibinfo {year} {2022})}\BibitemShut {NoStop}%
\bibitem [{\citenamefont {Benevides}\ \emph {et~al.}(2018)\citenamefont {Benevides}, \citenamefont {Dabholkar},\ and\ \citenamefont {Kobayashi}}]{Benevides:2018mwx}%
  \BibitemOpen
  \bibfield  {author} {\bibinfo {author} {\bibfnamefont {A.}~\bibnamefont {Benevides}}, \bibinfo {author} {\bibfnamefont {A.}~\bibnamefont {Dabholkar}}, \ and\ \bibinfo {author} {\bibfnamefont {T.}~\bibnamefont {Kobayashi}},\ }\href {\doibase 10.1007/JHEP11(2018)039} {\bibfield  {journal} {\bibinfo  {journal} {JHEP}\ }\textbf {\bibinfo {volume} {11}},\ \bibinfo {pages} {039} (\bibinfo {year} {2018})},\ \Eprint {http://arxiv.org/abs/1808.08237} {arXiv:1808.08237 [hep-th]} \BibitemShut {NoStop}%
\bibitem [{\citenamefont {Kobayashi}\ and\ \citenamefont {Sloth}(2019{\natexlab{b}})}]{Kobayashi:2019uqs}%
  \BibitemOpen
  \bibfield  {author} {\bibinfo {author} {\bibfnamefont {T.}~\bibnamefont {Kobayashi}}\ and\ \bibinfo {author} {\bibfnamefont {M.~S.}\ \bibnamefont {Sloth}},\ }\href {\doibase 10.1103/PhysRevD.100.023524} {\bibfield  {journal} {\bibinfo  {journal} {Phys. Rev. D}\ }\textbf {\bibinfo {volume} {100}},\ \bibinfo {pages} {023524} (\bibinfo {year} {2019}{\natexlab{b}})},\ \Eprint {http://arxiv.org/abs/1903.02561} {arXiv:1903.02561 [astro-ph.CO]} \BibitemShut {NoStop}%
\bibitem [{\citenamefont {Hort\'ua}\ and\ \citenamefont {Casta\~neda}(2014)}]{Hortua:2014wna}%
  \BibitemOpen
  \bibfield  {author} {\bibinfo {author} {\bibfnamefont {H.~J.}\ \bibnamefont {Hort\'ua}}\ and\ \bibinfo {author} {\bibfnamefont {L.}~\bibnamefont {Casta\~neda}},\ }\href {\doibase 10.1103/physrevd.90.123520} {\bibfield  {journal} {\bibinfo  {journal} {Phys. Rev. D}\ }\textbf {\bibinfo {volume} {90}},\ \bibinfo {pages} {123520} (\bibinfo {year} {2014})},\ \Eprint {http://arxiv.org/abs/1405.1786} {arXiv:1405.1786 [gr-qc]} \BibitemShut {NoStop}%
\bibitem [{\citenamefont {Campanelli}(2009)}]{Campanelli:2008kh}%
  \BibitemOpen
  \bibfield  {author} {\bibinfo {author} {\bibfnamefont {L.}~\bibnamefont {Campanelli}},\ }\href {\doibase 10.1142/S0218271809015175} {\bibfield  {journal} {\bibinfo  {journal} {Int. J. Mod. Phys. D}\ }\textbf {\bibinfo {volume} {18}},\ \bibinfo {pages} {1395} (\bibinfo {year} {2009})},\ \Eprint {http://arxiv.org/abs/0805.0575} {arXiv:0805.0575 [astro-ph]} \BibitemShut {NoStop}%
\bibitem [{\citenamefont {Jain}\ \emph {et~al.}(2014)\citenamefont {Jain}, \citenamefont {Durrer},\ and\ \citenamefont {Hollenstein}}]{Jain:2012jy}%
  \BibitemOpen
  \bibfield  {author} {\bibinfo {author} {\bibfnamefont {R.~K.}\ \bibnamefont {Jain}}, \bibinfo {author} {\bibfnamefont {R.}~\bibnamefont {Durrer}}, \ and\ \bibinfo {author} {\bibfnamefont {L.}~\bibnamefont {Hollenstein}},\ }\href {\doibase 10.1088/1742-6596/484/1/012062} {\bibfield  {journal} {\bibinfo  {journal} {J. Phys. Conf. Ser.}\ }\textbf {\bibinfo {volume} {484}},\ \bibinfo {pages} {012062} (\bibinfo {year} {2014})},\ \Eprint {http://arxiv.org/abs/1204.2409} {arXiv:1204.2409 [astro-ph.CO]} \BibitemShut {NoStop}%
\bibitem [{\citenamefont {Sharma}\ \emph {et~al.}(2018)\citenamefont {Sharma}, \citenamefont {Subramanian},\ and\ \citenamefont {Seshadri}}]{Sharma:2018kgs}%
  \BibitemOpen
  \bibfield  {author} {\bibinfo {author} {\bibfnamefont {R.}~\bibnamefont {Sharma}}, \bibinfo {author} {\bibfnamefont {K.}~\bibnamefont {Subramanian}}, \ and\ \bibinfo {author} {\bibfnamefont {T.~R.}\ \bibnamefont {Seshadri}},\ }\href {\doibase 10.1103/PhysRevD.97.083503} {\bibfield  {journal} {\bibinfo  {journal} {Phys. Rev. D}\ }\textbf {\bibinfo {volume} {97}},\ \bibinfo {pages} {083503} (\bibinfo {year} {2018})},\ \Eprint {http://arxiv.org/abs/1802.04847} {arXiv:1802.04847 [astro-ph.CO]} \BibitemShut {NoStop}%
\bibitem [{\citenamefont {Bamba}\ \emph {et~al.}(2022)\citenamefont {Bamba}, \citenamefont {Odintsov}, \citenamefont {Paul},\ and\ \citenamefont {Maity}}]{Bamba:2021wyx}%
  \BibitemOpen
  \bibfield  {author} {\bibinfo {author} {\bibfnamefont {K.}~\bibnamefont {Bamba}}, \bibinfo {author} {\bibfnamefont {S.~D.}\ \bibnamefont {Odintsov}}, \bibinfo {author} {\bibfnamefont {T.}~\bibnamefont {Paul}}, \ and\ \bibinfo {author} {\bibfnamefont {D.}~\bibnamefont {Maity}},\ }\href {\doibase 10.1016/j.dark.2022.101025} {\bibfield  {journal} {\bibinfo  {journal} {Phys. Dark Univ.}\ }\textbf {\bibinfo {volume} {36}},\ \bibinfo {pages} {101025} (\bibinfo {year} {2022})},\ \Eprint {http://arxiv.org/abs/2107.11524} {arXiv:2107.11524 [gr-qc]} \BibitemShut {NoStop}%
\bibitem [{\citenamefont {Sharma}\ \emph {et~al.}(2017)\citenamefont {Sharma}, \citenamefont {Jagannathan}, \citenamefont {Seshadri},\ and\ \citenamefont {Subramanian}}]{Sharma:2017eps}%
  \BibitemOpen
  \bibfield  {author} {\bibinfo {author} {\bibfnamefont {R.}~\bibnamefont {Sharma}}, \bibinfo {author} {\bibfnamefont {S.}~\bibnamefont {Jagannathan}}, \bibinfo {author} {\bibfnamefont {T.~R.}\ \bibnamefont {Seshadri}}, \ and\ \bibinfo {author} {\bibfnamefont {K.}~\bibnamefont {Subramanian}},\ }\href {\doibase 10.1103/PhysRevD.96.083511} {\bibfield  {journal} {\bibinfo  {journal} {Phys. Rev. D}\ }\textbf {\bibinfo {volume} {96}},\ \bibinfo {pages} {083511} (\bibinfo {year} {2017})},\ \Eprint {http://arxiv.org/abs/1708.08119} {arXiv:1708.08119 [astro-ph.CO]} \BibitemShut {NoStop}%
\bibitem [{\citenamefont {Papanikolaou}\ \emph {et~al.}(2025)\citenamefont {Papanikolaou}, \citenamefont {Tzerefos}, \citenamefont {Capozziello},\ and\ \citenamefont {Lambiase}}]{Papanikolaou:2024cwr}%
  \BibitemOpen
  \bibfield  {author} {\bibinfo {author} {\bibfnamefont {T.}~\bibnamefont {Papanikolaou}}, \bibinfo {author} {\bibfnamefont {C.}~\bibnamefont {Tzerefos}}, \bibinfo {author} {\bibfnamefont {S.}~\bibnamefont {Capozziello}}, \ and\ \bibinfo {author} {\bibfnamefont {G.}~\bibnamefont {Lambiase}},\ }\href {\doibase 10.1088/1475-7516/2025/01/051} {\bibfield  {journal} {\bibinfo  {journal} {JCAP}\ }\textbf {\bibinfo {volume} {01}},\ \bibinfo {pages} {051} (\bibinfo {year} {2025})},\ \Eprint {http://arxiv.org/abs/2408.17259} {arXiv:2408.17259 [astro-ph.CO]} \BibitemShut {NoStop}%
\bibitem [{\citenamefont {Dai}\ \emph {et~al.}(2014)\citenamefont {Dai}, \citenamefont {Kamionkowski},\ and\ \citenamefont {Wang}}]{Kamionkowski-2014}%
  \BibitemOpen
  \bibfield  {author} {\bibinfo {author} {\bibfnamefont {L.}~\bibnamefont {Dai}}, \bibinfo {author} {\bibfnamefont {M.}~\bibnamefont {Kamionkowski}}, \ and\ \bibinfo {author} {\bibfnamefont {J.}~\bibnamefont {Wang}},\ }\href {\doibase 10.1103/PhysRevLett.113.041302} {\bibfield  {journal} {\bibinfo  {journal} {Phys. Rev. Lett.}\ }\textbf {\bibinfo {volume} {113}},\ \bibinfo {pages} {041302} (\bibinfo {year} {2014})}\BibitemShut {NoStop}%
\bibitem [{\citenamefont {Subramanian}(2016)}]{Subramanian:2015lua}%
  \BibitemOpen
  \bibfield  {author} {\bibinfo {author} {\bibfnamefont {K.}~\bibnamefont {Subramanian}},\ }\href {\doibase 10.1088/0034-4885/79/7/076901} {\bibfield  {journal} {\bibinfo  {journal} {Rept. Prog. Phys.}\ }\textbf {\bibinfo {volume} {79}},\ \bibinfo {pages} {076901} (\bibinfo {year} {2016})},\ \Eprint {http://arxiv.org/abs/1504.02311} {arXiv:1504.02311 [astro-ph.CO]} \BibitemShut {NoStop}%
\bibitem [{\citenamefont {Paoletti}\ \emph {et~al.}(2009)\citenamefont {Paoletti}, \citenamefont {Finelli},\ and\ \citenamefont {Paci}}]{Paoletti:2008ck}%
  \BibitemOpen
  \bibfield  {author} {\bibinfo {author} {\bibfnamefont {D.}~\bibnamefont {Paoletti}}, \bibinfo {author} {\bibfnamefont {F.}~\bibnamefont {Finelli}}, \ and\ \bibinfo {author} {\bibfnamefont {F.}~\bibnamefont {Paci}},\ }\href {\doibase 10.1111/j.1365-2966.2009.14727.x} {\bibfield  {journal} {\bibinfo  {journal} {Mon. Not. Roy. Astron. Soc.}\ }\textbf {\bibinfo {volume} {396}},\ \bibinfo {pages} {523} (\bibinfo {year} {2009})},\ \Eprint {http://arxiv.org/abs/0811.0230} {arXiv:0811.0230 [astro-ph]} \BibitemShut {NoStop}%
\bibitem [{\citenamefont {Ade}\ \emph {et~al.}(2016)\citenamefont {Ade} \emph {et~al.}}]{Planck:2015zrl}%
  \BibitemOpen
  \bibfield  {author} {\bibinfo {author} {\bibfnamefont {P.~A.~R.}\ \bibnamefont {Ade}} \emph {et~al.} (\bibinfo {collaboration} {Planck}),\ }\href {\doibase 10.1051/0004-6361/201525821} {\bibfield  {journal} {\bibinfo  {journal} {Astron. Astrophys.}\ }\textbf {\bibinfo {volume} {594}},\ \bibinfo {pages} {A19} (\bibinfo {year} {2016})},\ \Eprint {http://arxiv.org/abs/1502.01594} {arXiv:1502.01594 [astro-ph.CO]} \BibitemShut {NoStop}%
\bibitem [{\citenamefont {Zucca}\ \emph {et~al.}(2017)\citenamefont {Zucca}, \citenamefont {Li},\ and\ \citenamefont {Pogosian}}]{Zucca:2016iur}%
  \BibitemOpen
  \bibfield  {author} {\bibinfo {author} {\bibfnamefont {A.}~\bibnamefont {Zucca}}, \bibinfo {author} {\bibfnamefont {Y.}~\bibnamefont {Li}}, \ and\ \bibinfo {author} {\bibfnamefont {L.}~\bibnamefont {Pogosian}},\ }\href {\doibase 10.1103/PhysRevD.95.063506} {\bibfield  {journal} {\bibinfo  {journal} {Phys. Rev. D}\ }\textbf {\bibinfo {volume} {95}},\ \bibinfo {pages} {063506} (\bibinfo {year} {2017})},\ \Eprint {http://arxiv.org/abs/1611.00757} {arXiv:1611.00757 [astro-ph.CO]} \BibitemShut {NoStop}%
\bibitem [{\citenamefont {Paoletti}\ \emph {et~al.}(2022)\citenamefont {Paoletti}, \citenamefont {Chluba}, \citenamefont {Finelli},\ and\ \citenamefont {Rubi{\~n}o-Mart{\'\i}n}}]{paoletti2022constraints}%
  \BibitemOpen
  \bibfield  {author} {\bibinfo {author} {\bibfnamefont {D.}~\bibnamefont {Paoletti}}, \bibinfo {author} {\bibfnamefont {J.}~\bibnamefont {Chluba}}, \bibinfo {author} {\bibfnamefont {F.}~\bibnamefont {Finelli}}, \ and\ \bibinfo {author} {\bibfnamefont {J.}~\bibnamefont {Rubi{\~n}o-Mart{\'\i}n}},\ }\href@noop {} {\bibfield  {journal} {\bibinfo  {journal} {Monthly Notices of the Royal Astronomical Society}\ }\textbf {\bibinfo {volume} {517}},\ \bibinfo {pages} {3916} (\bibinfo {year} {2022})}\BibitemShut {NoStop}%
\bibitem [{\citenamefont {Ade}\ \emph {et~al.}(2017)\citenamefont {Ade} \emph {et~al.}}]{BICEP2:2017lpa}%
  \BibitemOpen
  \bibfield  {author} {\bibinfo {author} {\bibfnamefont {P.~A.~R.}\ \bibnamefont {Ade}} \emph {et~al.} (\bibinfo {collaboration} {BICEP2, Keck Arrary}),\ }\href {\doibase 10.1103/PhysRevD.96.102003} {\bibfield  {journal} {\bibinfo  {journal} {Phys. Rev. D}\ }\textbf {\bibinfo {volume} {96}},\ \bibinfo {pages} {102003} (\bibinfo {year} {2017})},\ \Eprint {http://arxiv.org/abs/1705.02523} {arXiv:1705.02523 [astro-ph.CO]} \BibitemShut {NoStop}%
\bibitem [{\citenamefont {Acciari}\ \emph {et~al.}(2023)\citenamefont {Acciari} \emph {et~al.}}]{MAGIC:2022piy}%
  \BibitemOpen
  \bibfield  {author} {\bibinfo {author} {\bibfnamefont {V.~A.}\ \bibnamefont {Acciari}} \emph {et~al.} (\bibinfo {collaboration} {MAGIC}),\ }\href {\doibase 10.1051/0004-6361/202244126} {\bibfield  {journal} {\bibinfo  {journal} {Astron. Astrophys.}\ }\textbf {\bibinfo {volume} {670}},\ \bibinfo {pages} {A145} (\bibinfo {year} {2023})},\ \Eprint {http://arxiv.org/abs/2210.03321} {arXiv:2210.03321 [astro-ph.HE]} \BibitemShut {NoStop}%
\bibitem [{\citenamefont {Takahashi}\ \emph {et~al.}(2011)\citenamefont {Takahashi}, \citenamefont {Mori}, \citenamefont {Ichiki},\ and\ \citenamefont {Inoue}}]{takahashi2011lower}%
  \BibitemOpen
  \bibfield  {author} {\bibinfo {author} {\bibfnamefont {K.}~\bibnamefont {Takahashi}}, \bibinfo {author} {\bibfnamefont {M.}~\bibnamefont {Mori}}, \bibinfo {author} {\bibfnamefont {K.}~\bibnamefont {Ichiki}}, \ and\ \bibinfo {author} {\bibfnamefont {S.}~\bibnamefont {Inoue}},\ }\href@noop {} {\bibfield  {journal} {\bibinfo  {journal} {The Astrophysical Journal Letters}\ }\textbf {\bibinfo {volume} {744}},\ \bibinfo {pages} {L7} (\bibinfo {year} {2011})}\BibitemShut {NoStop}%
\bibitem [{\citenamefont {Arlen}\ \emph {et~al.}(2014{\natexlab{a}})\citenamefont {Arlen}, \citenamefont {Vassilev}, \citenamefont {Weisgarber}, \citenamefont {Wakely},\ and\ \citenamefont {Shafi}}]{Arlen_2014}%
  \BibitemOpen
  \bibfield  {author} {\bibinfo {author} {\bibfnamefont {T.~C.}\ \bibnamefont {Arlen}}, \bibinfo {author} {\bibfnamefont {V.~V.}\ \bibnamefont {Vassilev}}, \bibinfo {author} {\bibfnamefont {T.}~\bibnamefont {Weisgarber}}, \bibinfo {author} {\bibfnamefont {S.~P.}\ \bibnamefont {Wakely}}, \ and\ \bibinfo {author} {\bibfnamefont {S.~Y.}\ \bibnamefont {Shafi}},\ }\href {\doibase 10.1088/0004-637X/796/1/18} {\bibfield  {journal} {\bibinfo  {journal} {The Astrophysical Journal}\ }\textbf {\bibinfo {volume} {796}},\ \bibinfo {pages} {18} (\bibinfo {year} {2014}{\natexlab{a}})}\BibitemShut {NoStop}%
\bibitem [{\citenamefont {Arlen}\ \emph {et~al.}(2014{\natexlab{b}})\citenamefont {Arlen}, \citenamefont {Vassiliev}, \citenamefont {Weisgarber}, \citenamefont {Wakely},\ and\ \citenamefont {Shafi}}]{Arlen:2012iy}%
  \BibitemOpen
  \bibfield  {author} {\bibinfo {author} {\bibfnamefont {T.~C.}\ \bibnamefont {Arlen}}, \bibinfo {author} {\bibfnamefont {V.~V.}\ \bibnamefont {Vassiliev}}, \bibinfo {author} {\bibfnamefont {T.}~\bibnamefont {Weisgarber}}, \bibinfo {author} {\bibfnamefont {S.~P.}\ \bibnamefont {Wakely}}, \ and\ \bibinfo {author} {\bibfnamefont {S.~Y.}\ \bibnamefont {Shafi}},\ }\href {\doibase 10.1088/0004-637X/796/1/18} {\bibfield  {journal} {\bibinfo  {journal} {Astrophys. J.}\ }\textbf {\bibinfo {volume} {796}},\ \bibinfo {pages} {18} (\bibinfo {year} {2014}{\natexlab{b}})},\ \Eprint {http://arxiv.org/abs/1210.2802} {arXiv:1210.2802 [astro-ph.HE]} \BibitemShut {NoStop}%
\bibitem [{\citenamefont {{Taylor}}\ \emph {et~al.}(2011)\citenamefont {{Taylor}}, \citenamefont {{Vovk}},\ and\ \citenamefont {{Neronov}}}]{2011A&A...529A.144T}%
  \BibitemOpen
  \bibfield  {author} {\bibinfo {author} {\bibfnamefont {A.~M.}\ \bibnamefont {{Taylor}}}, \bibinfo {author} {\bibfnamefont {I.}~\bibnamefont {{Vovk}}}, \ and\ \bibinfo {author} {\bibfnamefont {A.}~\bibnamefont {{Neronov}}},\ }\href {\doibase 10.1051/0004-6361/201116441} {\bibfield  {journal} {\bibinfo  {journal} {aap}\ }\textbf {\bibinfo {volume} {529}},\ \bibinfo {eid} {A144} (\bibinfo {year} {2011})},\ \Eprint {http://arxiv.org/abs/1101.0932} {arXiv:1101.0932 [astro-ph.HE]} \BibitemShut {NoStop}%
\bibitem [{\citenamefont {{Neronov}}\ and\ \citenamefont {{Vovk}}(2010)}]{2010Sci...328...73N}%
  \BibitemOpen
  \bibfield  {author} {\bibinfo {author} {\bibfnamefont {A.}~\bibnamefont {{Neronov}}}\ and\ \bibinfo {author} {\bibfnamefont {I.}~\bibnamefont {{Vovk}}},\ }\href {\doibase 10.1126/science.1184192} {\bibfield  {journal} {\bibinfo  {journal} {Science}\ }\textbf {\bibinfo {volume} {328}},\ \bibinfo {pages} {73} (\bibinfo {year} {2010})},\ \Eprint {http://arxiv.org/abs/1006.3504} {arXiv:1006.3504 [astro-ph.HE]} \BibitemShut {NoStop}%
\bibitem [{\citenamefont {Pshirkov}\ \emph {et~al.}(2016)\citenamefont {Pshirkov}, \citenamefont {Tinyakov},\ and\ \citenamefont {Urban}}]{PhysRevLett.116.191302}%
  \BibitemOpen
  \bibfield  {author} {\bibinfo {author} {\bibfnamefont {M.~S.}\ \bibnamefont {Pshirkov}}, \bibinfo {author} {\bibfnamefont {P.~G.}\ \bibnamefont {Tinyakov}}, \ and\ \bibinfo {author} {\bibfnamefont {F.~R.}\ \bibnamefont {Urban}},\ }\href {\doibase 10.1103/PhysRevLett.116.191302} {\bibfield  {journal} {\bibinfo  {journal} {Phys. Rev. Lett.}\ }\textbf {\bibinfo {volume} {116}},\ \bibinfo {pages} {191302} (\bibinfo {year} {2016})}\BibitemShut {NoStop}%
\bibitem [{\citenamefont {Kamada}\ and\ \citenamefont {Long}(2016{\natexlab{a}})}]{Kamada:2016cnb}%
  \BibitemOpen
  \bibfield  {author} {\bibinfo {author} {\bibfnamefont {K.}~\bibnamefont {Kamada}}\ and\ \bibinfo {author} {\bibfnamefont {A.~J.}\ \bibnamefont {Long}},\ }\href {\doibase 10.1103/PhysRevD.94.123509} {\bibfield  {journal} {\bibinfo  {journal} {Phys. Rev. D}\ }\textbf {\bibinfo {volume} {94}},\ \bibinfo {pages} {123509} (\bibinfo {year} {2016}{\natexlab{a}})},\ \Eprint {http://arxiv.org/abs/1610.03074} {arXiv:1610.03074 [hep-ph]} \BibitemShut {NoStop}%
\bibitem [{\citenamefont {Kamada}\ and\ \citenamefont {Long}(2016{\natexlab{b}})}]{Kamada:2016eeb}%
  \BibitemOpen
  \bibfield  {author} {\bibinfo {author} {\bibfnamefont {K.}~\bibnamefont {Kamada}}\ and\ \bibinfo {author} {\bibfnamefont {A.~J.}\ \bibnamefont {Long}},\ }\href {\doibase 10.1103/PhysRevD.94.063501} {\bibfield  {journal} {\bibinfo  {journal} {Phys. Rev. D}\ }\textbf {\bibinfo {volume} {94}},\ \bibinfo {pages} {063501} (\bibinfo {year} {2016}{\natexlab{b}})},\ \Eprint {http://arxiv.org/abs/1606.08891} {arXiv:1606.08891 [astro-ph.CO]} \BibitemShut {NoStop}%
\bibitem [{\citenamefont {Fujita}\ and\ \citenamefont {Kamada}(2016)}]{Fujita:2016igl}%
  \BibitemOpen
  \bibfield  {author} {\bibinfo {author} {\bibfnamefont {T.}~\bibnamefont {Fujita}}\ and\ \bibinfo {author} {\bibfnamefont {K.}~\bibnamefont {Kamada}},\ }\href {\doibase 10.1103/PhysRevD.93.083520} {\bibfield  {journal} {\bibinfo  {journal} {Phys. Rev. D}\ }\textbf {\bibinfo {volume} {93}},\ \bibinfo {pages} {083520} (\bibinfo {year} {2016})},\ \Eprint {http://arxiv.org/abs/1602.02109} {arXiv:1602.02109 [hep-ph]} \BibitemShut {NoStop}%
\bibitem [{\citenamefont {Kamada}\ \emph {et~al.}(2021)\citenamefont {Kamada}, \citenamefont {Uchida},\ and\ \citenamefont {Yokoyama}}]{Kamada:2020bmb}%
  \BibitemOpen
  \bibfield  {author} {\bibinfo {author} {\bibfnamefont {K.}~\bibnamefont {Kamada}}, \bibinfo {author} {\bibfnamefont {F.}~\bibnamefont {Uchida}}, \ and\ \bibinfo {author} {\bibfnamefont {J.}~\bibnamefont {Yokoyama}},\ }\href {\doibase 10.1088/1475-7516/2021/04/034} {\bibfield  {journal} {\bibinfo  {journal} {JCAP}\ }\textbf {\bibinfo {volume} {04}},\ \bibinfo {pages} {034} (\bibinfo {year} {2021})},\ \Eprint {http://arxiv.org/abs/2012.14435} {arXiv:2012.14435 [astro-ph.CO]} \BibitemShut {NoStop}%
\bibitem [{\citenamefont {Bassett}\ \emph {et~al.}(2006)\citenamefont {Bassett}, \citenamefont {Tsujikawa},\ and\ \citenamefont {Wands}}]{Bassett:2005xm}%
  \BibitemOpen
  \bibfield  {author} {\bibinfo {author} {\bibfnamefont {B.~A.}\ \bibnamefont {Bassett}}, \bibinfo {author} {\bibfnamefont {S.}~\bibnamefont {Tsujikawa}}, \ and\ \bibinfo {author} {\bibfnamefont {D.}~\bibnamefont {Wands}},\ }\href {\doibase 10.1103/RevModPhys.78.537} {\bibfield  {journal} {\bibinfo  {journal} {Rev. Mod. Phys.}\ }\textbf {\bibinfo {volume} {78}},\ \bibinfo {pages} {537} (\bibinfo {year} {2006})},\ \Eprint {http://arxiv.org/abs/astro-ph/0507632} {arXiv:astro-ph/0507632} \BibitemShut {NoStop}%
\bibitem [{\citenamefont {Wands}\ \emph {et~al.}(2000)\citenamefont {Wands}, \citenamefont {Malik}, \citenamefont {Lyth},\ and\ \citenamefont {Liddle}}]{Wands:2000dp}%
  \BibitemOpen
  \bibfield  {author} {\bibinfo {author} {\bibfnamefont {D.}~\bibnamefont {Wands}}, \bibinfo {author} {\bibfnamefont {K.~A.}\ \bibnamefont {Malik}}, \bibinfo {author} {\bibfnamefont {D.~H.}\ \bibnamefont {Lyth}}, \ and\ \bibinfo {author} {\bibfnamefont {A.~R.}\ \bibnamefont {Liddle}},\ }\href {\doibase 10.1103/PhysRevD.62.043527} {\bibfield  {journal} {\bibinfo  {journal} {Phys. Rev. D}\ }\textbf {\bibinfo {volume} {62}},\ \bibinfo {pages} {043527} (\bibinfo {year} {2000})},\ \Eprint {http://arxiv.org/abs/astro-ph/0003278} {arXiv:astro-ph/0003278} \BibitemShut {NoStop}%
\bibitem [{\citenamefont {Baumann}(2009)}]{Baumann:2009ds}%
  \BibitemOpen
  \bibfield  {author} {\bibinfo {author} {\bibfnamefont {D.}~\bibnamefont {Baumann}},\ }\href@noop {} {\bibfield  {journal} {\bibinfo  {journal} {arXiv:0907.5424}\ } (\bibinfo {year} {2009})}\BibitemShut {NoStop}%
\bibitem [{\citenamefont {Malik}\ and\ \citenamefont {Wands}(2009)}]{Malik:2008im}%
  \BibitemOpen
  \bibfield  {author} {\bibinfo {author} {\bibfnamefont {K.~A.}\ \bibnamefont {Malik}}\ and\ \bibinfo {author} {\bibfnamefont {D.}~\bibnamefont {Wands}},\ }\href {\doibase 10.1016/j.physrep.2009.03.001} {\bibfield  {journal} {\bibinfo  {journal} {Phys. Rept.}\ }\textbf {\bibinfo {volume} {475}},\ \bibinfo {pages} {1} (\bibinfo {year} {2009})},\ \Eprint {http://arxiv.org/abs/0809.4944} {arXiv:0809.4944 [astro-ph]} \BibitemShut {NoStop}%
\bibitem [{\citenamefont {Bonvin}\ \emph {et~al.}(2013)\citenamefont {Bonvin}, \citenamefont {Caprini},\ and\ \citenamefont {Durrer}}]{PhysRevD.88.083515}%
  \BibitemOpen
  \bibfield  {author} {\bibinfo {author} {\bibfnamefont {C.}~\bibnamefont {Bonvin}}, \bibinfo {author} {\bibfnamefont {C.}~\bibnamefont {Caprini}}, \ and\ \bibinfo {author} {\bibfnamefont {R.}~\bibnamefont {Durrer}},\ }\href {\doibase 10.1103/PhysRevD.88.083515} {\bibfield  {journal} {\bibinfo  {journal} {Phys. Rev. D}\ }\textbf {\bibinfo {volume} {88}},\ \bibinfo {pages} {083515} (\bibinfo {year} {2013})}\BibitemShut {NoStop}%
\bibitem [{\citenamefont {Shaw}\ and\ \citenamefont {Lewis}(2010)}]{PhysRevD.81.043517}%
  \BibitemOpen
  \bibfield  {author} {\bibinfo {author} {\bibfnamefont {J.~R.}\ \bibnamefont {Shaw}}\ and\ \bibinfo {author} {\bibfnamefont {A.}~\bibnamefont {Lewis}},\ }\href {\doibase 10.1103/PhysRevD.81.043517} {\bibfield  {journal} {\bibinfo  {journal} {Phys. Rev. D}\ }\textbf {\bibinfo {volume} {81}},\ \bibinfo {pages} {043517} (\bibinfo {year} {2010})}\BibitemShut {NoStop}%
\bibitem [{\citenamefont {Saga}\ \emph {et~al.}(2020)\citenamefont {Saga}, \citenamefont {Tashiro},\ and\ \citenamefont {Yokoyama}}]{Saga_2020}%
  \BibitemOpen
  \bibfield  {author} {\bibinfo {author} {\bibfnamefont {S.}~\bibnamefont {Saga}}, \bibinfo {author} {\bibfnamefont {H.}~\bibnamefont {Tashiro}}, \ and\ \bibinfo {author} {\bibfnamefont {S.}~\bibnamefont {Yokoyama}},\ }\href {\doibase 10.1088/1475-7516/2020/05/039} {\bibfield  {journal} {\bibinfo  {journal} {Journal of Cosmology and Astroparticle Physics}\ }\textbf {\bibinfo {volume} {2020}},\ \bibinfo {pages} {039} (\bibinfo {year} {2020})}\BibitemShut {NoStop}%
\bibitem [{\citenamefont {Maiti}\ and\ \citenamefont {Maity}(2026{\natexlab{b}})}]{Maiti:2026lvx}%
  \BibitemOpen
  \bibfield  {author} {\bibinfo {author} {\bibfnamefont {S.}~\bibnamefont {Maiti}}\ and\ \bibinfo {author} {\bibfnamefont {D.}~\bibnamefont {Maity}},\ }\href@noop {} {\  (\bibinfo {year} {2026}{\natexlab{b}})},\ \Eprint {http://arxiv.org/abs/2605.14044} {arXiv:2605.14044 [astro-ph.CO]} \BibitemShut {NoStop}%
\bibitem [{\citenamefont {Dom{\`e}nech}\ and\ \citenamefont {Tr{\"a}nkle}(2025)}]{Domenech:2025ffb}%
  \BibitemOpen
  \bibfield  {author} {\bibinfo {author} {\bibfnamefont {G.}~\bibnamefont {Dom{\`e}nech}}\ and\ \bibinfo {author} {\bibfnamefont {J.}~\bibnamefont {Tr{\"a}nkle}},\ }\href {\doibase 10.1088/1475-7516/2025/12/017} {\bibfield  {journal} {\bibinfo  {journal} {JCAP}\ }\textbf {\bibinfo {volume} {12}},\ \bibinfo {pages} {017} (\bibinfo {year} {2025})},\ \Eprint {http://arxiv.org/abs/2509.02122} {arXiv:2509.02122 [gr-qc]} \BibitemShut {NoStop}%
\bibitem [{\citenamefont {Dom{\`e}nech}\ \emph {et~al.}(2024)\citenamefont {Dom{\`e}nech}, \citenamefont {Pi}, \citenamefont {Wang},\ and\ \citenamefont {Wang}}]{Domenech:2024rks}%
  \BibitemOpen
  \bibfield  {author} {\bibinfo {author} {\bibfnamefont {G.}~\bibnamefont {Dom{\`e}nech}}, \bibinfo {author} {\bibfnamefont {S.}~\bibnamefont {Pi}}, \bibinfo {author} {\bibfnamefont {A.}~\bibnamefont {Wang}}, \ and\ \bibinfo {author} {\bibfnamefont {J.}~\bibnamefont {Wang}},\ }\href {\doibase 10.1088/1475-7516/2024/08/054} {\bibfield  {journal} {\bibinfo  {journal} {JCAP}\ }\textbf {\bibinfo {volume} {08}},\ \bibinfo {pages} {054} (\bibinfo {year} {2024})},\ \Eprint {http://arxiv.org/abs/2402.18965} {arXiv:2402.18965 [astro-ph.CO]} \BibitemShut {NoStop}%
\bibitem [{\citenamefont {Dom\`enech}(2021)}]{Domenech:2021ztg}%
  \BibitemOpen
  \bibfield  {author} {\bibinfo {author} {\bibfnamefont {G.}~\bibnamefont {Dom\`enech}},\ }\href {\doibase 10.3390/universe7110398} {\bibfield  {journal} {\bibinfo  {journal} {Universe}\ }\textbf {\bibinfo {volume} {7}},\ \bibinfo {pages} {398} (\bibinfo {year} {2021})},\ \Eprint {http://arxiv.org/abs/2109.01398} {arXiv:2109.01398 [gr-qc]} \BibitemShut {NoStop}%
\bibitem [{\citenamefont {Dom{\`e}nech}\ and\ \citenamefont {Sasaki}(2021)}]{Domenech:2020xin}%
  \BibitemOpen
  \bibfield  {author} {\bibinfo {author} {\bibfnamefont {G.}~\bibnamefont {Dom{\`e}nech}}\ and\ \bibinfo {author} {\bibfnamefont {M.}~\bibnamefont {Sasaki}},\ }\href {\doibase 10.1103/PhysRevD.103.063531} {\bibfield  {journal} {\bibinfo  {journal} {Phys. Rev. D}\ }\textbf {\bibinfo {volume} {103}},\ \bibinfo {pages} {063531} (\bibinfo {year} {2021})},\ \Eprint {http://arxiv.org/abs/2012.14016} {arXiv:2012.14016 [gr-qc]} \BibitemShut {NoStop}%
\bibitem [{\citenamefont {Dom{\`e}nech}(2020)}]{Domenech:2019quo}%
  \BibitemOpen
  \bibfield  {author} {\bibinfo {author} {\bibfnamefont {G.}~\bibnamefont {Dom{\`e}nech}},\ }\href {\doibase 10.1142/S0218271820500285} {\bibfield  {journal} {\bibinfo  {journal} {Int. J. Mod. Phys. D}\ }\textbf {\bibinfo {volume} {29}},\ \bibinfo {pages} {2050028} (\bibinfo {year} {2020})},\ \Eprint {http://arxiv.org/abs/1912.05583} {arXiv:1912.05583 [gr-qc]} \BibitemShut {NoStop}%
\bibitem [{\citenamefont {Balaji}\ \emph {et~al.}(2023)\citenamefont {Balaji}, \citenamefont {Dom{\`e}nech},\ and\ \citenamefont {Franciolini}}]{Balaji:2023ehk}%
  \BibitemOpen
  \bibfield  {author} {\bibinfo {author} {\bibfnamefont {S.}~\bibnamefont {Balaji}}, \bibinfo {author} {\bibfnamefont {G.}~\bibnamefont {Dom{\`e}nech}}, \ and\ \bibinfo {author} {\bibfnamefont {G.}~\bibnamefont {Franciolini}},\ }\href {\doibase 10.1088/1475-7516/2023/10/041} {\bibfield  {journal} {\bibinfo  {journal} {JCAP}\ }\textbf {\bibinfo {volume} {10}},\ \bibinfo {pages} {041} (\bibinfo {year} {2023})},\ \Eprint {http://arxiv.org/abs/2307.08552} {arXiv:2307.08552 [gr-qc]} \BibitemShut {NoStop}%
\bibitem [{\citenamefont {Balaji}\ \emph {et~al.}(2022)\citenamefont {Balaji}, \citenamefont {Domenech},\ and\ \citenamefont {Silk}}]{Balaji:2022dbi}%
  \BibitemOpen
  \bibfield  {author} {\bibinfo {author} {\bibfnamefont {S.}~\bibnamefont {Balaji}}, \bibinfo {author} {\bibfnamefont {G.}~\bibnamefont {Domenech}}, \ and\ \bibinfo {author} {\bibfnamefont {J.}~\bibnamefont {Silk}},\ }\href {\doibase 10.1088/1475-7516/2022/09/016} {\bibfield  {journal} {\bibinfo  {journal} {JCAP}\ }\textbf {\bibinfo {volume} {09}},\ \bibinfo {pages} {016} (\bibinfo {year} {2022})},\ \Eprint {http://arxiv.org/abs/2205.01696} {arXiv:2205.01696 [astro-ph.CO]} \BibitemShut {NoStop}%
\bibitem [{\citenamefont {Inui}\ \emph {et~al.}(2025)\citenamefont {Inui}, \citenamefont {Joana}, \citenamefont {Motohashi}, \citenamefont {Pi}, \citenamefont {Tada},\ and\ \citenamefont {Yokoyama}}]{Inui:2024fgk}%
  \BibitemOpen
  \bibfield  {author} {\bibinfo {author} {\bibfnamefont {R.}~\bibnamefont {Inui}}, \bibinfo {author} {\bibfnamefont {C.}~\bibnamefont {Joana}}, \bibinfo {author} {\bibfnamefont {H.}~\bibnamefont {Motohashi}}, \bibinfo {author} {\bibfnamefont {S.}~\bibnamefont {Pi}}, \bibinfo {author} {\bibfnamefont {Y.}~\bibnamefont {Tada}}, \ and\ \bibinfo {author} {\bibfnamefont {S.}~\bibnamefont {Yokoyama}},\ }\href {\doibase 10.1088/1475-7516/2025/03/021} {\bibfield  {journal} {\bibinfo  {journal} {JCAP}\ }\textbf {\bibinfo {volume} {03}},\ \bibinfo {pages} {021} (\bibinfo {year} {2025})},\ \Eprint {http://arxiv.org/abs/2411.07647} {arXiv:2411.07647 [astro-ph.CO]} \BibitemShut {NoStop}%
\bibitem [{\citenamefont {Baumann}\ \emph {et~al.}(2007)\citenamefont {Baumann}, \citenamefont {Steinhardt}, \citenamefont {Takahashi},\ and\ \citenamefont {Ichiki}}]{Baumann:2007zm}%
  \BibitemOpen
  \bibfield  {author} {\bibinfo {author} {\bibfnamefont {D.}~\bibnamefont {Baumann}}, \bibinfo {author} {\bibfnamefont {P.~J.}\ \bibnamefont {Steinhardt}}, \bibinfo {author} {\bibfnamefont {K.}~\bibnamefont {Takahashi}}, \ and\ \bibinfo {author} {\bibfnamefont {K.}~\bibnamefont {Ichiki}},\ }\href {\doibase 10.1103/PhysRevD.76.084019} {\bibfield  {journal} {\bibinfo  {journal} {Phys. Rev. D}\ }\textbf {\bibinfo {volume} {76}},\ \bibinfo {pages} {084019} (\bibinfo {year} {2007})},\ \Eprint {http://arxiv.org/abs/hep-th/0703290} {arXiv:hep-th/0703290} \BibitemShut {NoStop}%
\bibitem [{\citenamefont {Ragavendra}\ \emph {et~al.}(2021{\natexlab{b}})\citenamefont {Ragavendra}, \citenamefont {Saha}, \citenamefont {Sriramkumar},\ and\ \citenamefont {Silk}}]{Ragavendra:2020sop}%
  \BibitemOpen
  \bibfield  {author} {\bibinfo {author} {\bibfnamefont {H.~V.}\ \bibnamefont {Ragavendra}}, \bibinfo {author} {\bibfnamefont {P.}~\bibnamefont {Saha}}, \bibinfo {author} {\bibfnamefont {L.}~\bibnamefont {Sriramkumar}}, \ and\ \bibinfo {author} {\bibfnamefont {J.}~\bibnamefont {Silk}},\ }\href {\doibase 10.1103/PhysRevD.103.083510} {\bibfield  {journal} {\bibinfo  {journal} {Phys. Rev. D}\ }\textbf {\bibinfo {volume} {103}},\ \bibinfo {pages} {083510} (\bibinfo {year} {2021}{\natexlab{b}})},\ \Eprint {http://arxiv.org/abs/2008.12202} {arXiv:2008.12202 [astro-ph.CO]} \BibitemShut {NoStop}%
\bibitem [{\citenamefont {Ragavendra}\ and\ \citenamefont {Sriramkumar}(2023)}]{Ragavendra:2023ret}%
  \BibitemOpen
  \bibfield  {author} {\bibinfo {author} {\bibfnamefont {H.~V.}\ \bibnamefont {Ragavendra}}\ and\ \bibinfo {author} {\bibfnamefont {L.}~\bibnamefont {Sriramkumar}},\ }\href {\doibase 10.3390/galaxies11010034} {\bibfield  {journal} {\bibinfo  {journal} {Galaxies}\ }\textbf {\bibinfo {volume} {11}},\ \bibinfo {pages} {34} (\bibinfo {year} {2023})},\ \Eprint {http://arxiv.org/abs/2301.08887} {arXiv:2301.08887 [astro-ph.CO]} \BibitemShut {NoStop}%
\bibitem [{\citenamefont {Dimastrogiovanni}\ \emph {et~al.}(2023)\citenamefont {Dimastrogiovanni}, \citenamefont {Fasiello}, \citenamefont {Malhotra},\ and\ \citenamefont {Tasinato}}]{Dimastrogiovanni:2022eir}%
  \BibitemOpen
  \bibfield  {author} {\bibinfo {author} {\bibfnamefont {E.}~\bibnamefont {Dimastrogiovanni}}, \bibinfo {author} {\bibfnamefont {M.}~\bibnamefont {Fasiello}}, \bibinfo {author} {\bibfnamefont {A.}~\bibnamefont {Malhotra}}, \ and\ \bibinfo {author} {\bibfnamefont {G.}~\bibnamefont {Tasinato}},\ }\href {\doibase 10.1088/1475-7516/2023/01/018} {\bibfield  {journal} {\bibinfo  {journal} {JCAP}\ }\textbf {\bibinfo {volume} {01}},\ \bibinfo {pages} {018} (\bibinfo {year} {2023})},\ \Eprint {http://arxiv.org/abs/2205.05644} {arXiv:2205.05644 [astro-ph.CO]} \BibitemShut {NoStop}%
\bibitem [{\citenamefont {Cai}\ \emph {et~al.}(2020)\citenamefont {Cai}, \citenamefont {Pi},\ and\ \citenamefont {Sasaki}}]{Cai:2019cdl}%
  \BibitemOpen
  \bibfield  {author} {\bibinfo {author} {\bibfnamefont {R.-G.}\ \bibnamefont {Cai}}, \bibinfo {author} {\bibfnamefont {S.}~\bibnamefont {Pi}}, \ and\ \bibinfo {author} {\bibfnamefont {M.}~\bibnamefont {Sasaki}},\ }\href {\doibase 10.1103/PhysRevD.102.083528} {\bibfield  {journal} {\bibinfo  {journal} {Phys. Rev. D}\ }\textbf {\bibinfo {volume} {102}},\ \bibinfo {pages} {083528} (\bibinfo {year} {2020})},\ \Eprint {http://arxiv.org/abs/1909.13728} {arXiv:1909.13728 [astro-ph.CO]} \BibitemShut {NoStop}%
\bibitem [{\citenamefont {Papanikolaou}(2022)}]{Papanikolaou:2022chm}%
  \BibitemOpen
  \bibfield  {author} {\bibinfo {author} {\bibfnamefont {T.}~\bibnamefont {Papanikolaou}},\ }\href {\doibase 10.1088/1475-7516/2022/10/089} {\bibfield  {journal} {\bibinfo  {journal} {JCAP}\ }\textbf {\bibinfo {volume} {10}},\ \bibinfo {pages} {089} (\bibinfo {year} {2022})},\ \Eprint {http://arxiv.org/abs/2207.11041} {arXiv:2207.11041 [astro-ph.CO]} \BibitemShut {NoStop}%
\bibitem [{\citenamefont {Maity}\ \emph {et~al.}(2025)\citenamefont {Maity}, \citenamefont {Bhaumik}, \citenamefont {Haque}, \citenamefont {Maity},\ and\ \citenamefont {Sriramkumar}}]{Maity:2024odg}%
  \BibitemOpen
  \bibfield  {author} {\bibinfo {author} {\bibfnamefont {S.}~\bibnamefont {Maity}}, \bibinfo {author} {\bibfnamefont {N.}~\bibnamefont {Bhaumik}}, \bibinfo {author} {\bibfnamefont {M.~R.}\ \bibnamefont {Haque}}, \bibinfo {author} {\bibfnamefont {D.}~\bibnamefont {Maity}}, \ and\ \bibinfo {author} {\bibfnamefont {L.}~\bibnamefont {Sriramkumar}},\ }\href {\doibase 10.1088/1475-7516/2025/01/118} {\bibfield  {journal} {\bibinfo  {journal} {JCAP}\ }\textbf {\bibinfo {volume} {01}},\ \bibinfo {pages} {118} (\bibinfo {year} {2025})},\ \Eprint {http://arxiv.org/abs/2403.16963} {arXiv:2403.16963 [astro-ph.CO]} \BibitemShut {NoStop}%
\bibitem [{\citenamefont {Kohri}\ and\ \citenamefont {Terada}(2018)}]{Kohri:2018awv}%
  \BibitemOpen
  \bibfield  {author} {\bibinfo {author} {\bibfnamefont {K.}~\bibnamefont {Kohri}}\ and\ \bibinfo {author} {\bibfnamefont {T.}~\bibnamefont {Terada}},\ }\href {\doibase 10.1103/PhysRevD.97.123532} {\bibfield  {journal} {\bibinfo  {journal} {Phys. Rev. D}\ }\textbf {\bibinfo {volume} {97}},\ \bibinfo {pages} {123532} (\bibinfo {year} {2018})},\ \Eprint {http://arxiv.org/abs/1804.08577} {arXiv:1804.08577 [gr-qc]} \BibitemShut {NoStop}%
\bibitem [{\citenamefont {Inomata}\ \emph {et~al.}(2017)\citenamefont {Inomata}, \citenamefont {Kawasaki}, \citenamefont {Mukaida}, \citenamefont {Tada},\ and\ \citenamefont {Yanagida}}]{Inomata:2016rbd}%
  \BibitemOpen
  \bibfield  {author} {\bibinfo {author} {\bibfnamefont {K.}~\bibnamefont {Inomata}}, \bibinfo {author} {\bibfnamefont {M.}~\bibnamefont {Kawasaki}}, \bibinfo {author} {\bibfnamefont {K.}~\bibnamefont {Mukaida}}, \bibinfo {author} {\bibfnamefont {Y.}~\bibnamefont {Tada}}, \ and\ \bibinfo {author} {\bibfnamefont {T.~T.}\ \bibnamefont {Yanagida}},\ }\href {\doibase 10.1103/PhysRevD.95.123510} {\bibfield  {journal} {\bibinfo  {journal} {Phys. Rev. D}\ }\textbf {\bibinfo {volume} {95}},\ \bibinfo {pages} {123510} (\bibinfo {year} {2017})},\ \Eprint {http://arxiv.org/abs/1611.06130} {arXiv:1611.06130 [astro-ph.CO]} \BibitemShut {NoStop}%
\bibitem [{\citenamefont {Maiti}\ \emph {et~al.}(2024)\citenamefont {Maiti}, \citenamefont {Maity},\ and\ \citenamefont {Sriramkumar}}]{Maiti:2024nhv}%
  \BibitemOpen
  \bibfield  {author} {\bibinfo {author} {\bibfnamefont {S.}~\bibnamefont {Maiti}}, \bibinfo {author} {\bibfnamefont {D.}~\bibnamefont {Maity}}, \ and\ \bibinfo {author} {\bibfnamefont {L.}~\bibnamefont {Sriramkumar}},\ }\href@noop {} {\  (\bibinfo {year} {2024})},\ \Eprint {http://arxiv.org/abs/2401.01864} {arXiv:2401.01864 [gr-qc]} \BibitemShut {NoStop}%
\bibitem [{\citenamefont {Bhaumik}\ \emph {et~al.}(2025)\citenamefont {Bhaumik}, \citenamefont {Papanikolaou},\ and\ \citenamefont {Ghoshal}}]{Bhaumik:2025kuj}%
  \BibitemOpen
  \bibfield  {author} {\bibinfo {author} {\bibfnamefont {A.}~\bibnamefont {Bhaumik}}, \bibinfo {author} {\bibfnamefont {T.}~\bibnamefont {Papanikolaou}}, \ and\ \bibinfo {author} {\bibfnamefont {A.}~\bibnamefont {Ghoshal}},\ }\href {\doibase 10.1088/1475-7516/2025/08/054} {\bibfield  {journal} {\bibinfo  {journal} {JCAP}\ }\textbf {\bibinfo {volume} {08}},\ \bibinfo {pages} {054} (\bibinfo {year} {2025})},\ \Eprint {http://arxiv.org/abs/2504.10477} {arXiv:2504.10477 [astro-ph.CO]} \BibitemShut {NoStop}%
\bibitem [{\citenamefont {Ragavendra}\ \emph {et~al.}(2026)\citenamefont {Ragavendra}, \citenamefont {Tasinato},\ and\ \citenamefont {Sriramkumar}}]{Ragavendra:2026fgs}%
  \BibitemOpen
  \bibfield  {author} {\bibinfo {author} {\bibfnamefont {H.~V.}\ \bibnamefont {Ragavendra}}, \bibinfo {author} {\bibfnamefont {G.}~\bibnamefont {Tasinato}}, \ and\ \bibinfo {author} {\bibfnamefont {L.}~\bibnamefont {Sriramkumar}},\ }\href@noop {} {\  (\bibinfo {year} {2026})},\ \Eprint {http://arxiv.org/abs/2602.16575} {arXiv:2602.16575 [astro-ph.CO]} \BibitemShut {NoStop}%
\bibitem [{\citenamefont {Boyle}\ and\ \citenamefont {Steinhardt}(2008)}]{Boyle:2005se}%
  \BibitemOpen
  \bibfield  {author} {\bibinfo {author} {\bibfnamefont {L.~A.}\ \bibnamefont {Boyle}}\ and\ \bibinfo {author} {\bibfnamefont {P.~J.}\ \bibnamefont {Steinhardt}},\ }\href {\doibase 10.1103/PhysRevD.77.063504} {\bibfield  {journal} {\bibinfo  {journal} {Phys. Rev. D}\ }\textbf {\bibinfo {volume} {77}},\ \bibinfo {pages} {063504} (\bibinfo {year} {2008})},\ \Eprint {http://arxiv.org/abs/astro-ph/0512014} {arXiv:astro-ph/0512014} \BibitemShut {NoStop}%
\bibitem [{\citenamefont {Atkins}\ \emph {et~al.}(2025)\citenamefont {Atkins}, \citenamefont {Chowdhury}, \citenamefont {Marriott-Best},\ and\ \citenamefont {Tasinato}}]{Atkins:2025pvg}%
  \BibitemOpen
  \bibfield  {author} {\bibinfo {author} {\bibfnamefont {B.}~\bibnamefont {Atkins}}, \bibinfo {author} {\bibfnamefont {D.}~\bibnamefont {Chowdhury}}, \bibinfo {author} {\bibfnamefont {A.}~\bibnamefont {Marriott-Best}}, \ and\ \bibinfo {author} {\bibfnamefont {G.}~\bibnamefont {Tasinato}},\ }\href {\doibase 10.1103/btjz-pqgv} {\bibfield  {journal} {\bibinfo  {journal} {Phys. Rev. D}\ }\textbf {\bibinfo {volume} {112}},\ \bibinfo {pages} {063534} (\bibinfo {year} {2025})},\ \Eprint {http://arxiv.org/abs/2507.01772} {arXiv:2507.01772 [astro-ph.CO]} \BibitemShut {NoStop}%
\bibitem [{\citenamefont {Papanikolaou}\ \emph {et~al.}(2021)\citenamefont {Papanikolaou}, \citenamefont {Vennin},\ and\ \citenamefont {Langlois}}]{Papanikolaou:2020qtd}%
  \BibitemOpen
  \bibfield  {author} {\bibinfo {author} {\bibfnamefont {T.}~\bibnamefont {Papanikolaou}}, \bibinfo {author} {\bibfnamefont {V.}~\bibnamefont {Vennin}}, \ and\ \bibinfo {author} {\bibfnamefont {D.}~\bibnamefont {Langlois}},\ }\href {\doibase 10.1088/1475-7516/2021/03/053} {\bibfield  {journal} {\bibinfo  {journal} {JCAP}\ }\textbf {\bibinfo {volume} {03}},\ \bibinfo {pages} {053} (\bibinfo {year} {2021})},\ \Eprint {http://arxiv.org/abs/2010.11573} {arXiv:2010.11573 [astro-ph.CO]} \BibitemShut {NoStop}%
\bibitem [{\citenamefont {Punturo}\ \emph {et~al.}(2010)\citenamefont {Punturo} \emph {et~al.}}]{Punturo:2010zz}%
  \BibitemOpen
  \bibfield  {author} {\bibinfo {author} {\bibfnamefont {M.}~\bibnamefont {Punturo}} \emph {et~al.},\ }\href {\doibase 10.1088/0264-9381/27/19/194002} {\bibfield  {journal} {\bibinfo  {journal} {Class. Quant. Grav.}\ }\textbf {\bibinfo {volume} {27}},\ \bibinfo {pages} {194002} (\bibinfo {year} {2010})}\BibitemShut {NoStop}%
\bibitem [{\citenamefont {Sathyaprakash}\ \emph {et~al.}(2012)\citenamefont {Sathyaprakash} \emph {et~al.}}]{Sathyaprakash:2012jk}%
  \BibitemOpen
  \bibfield  {author} {\bibinfo {author} {\bibfnamefont {B.}~\bibnamefont {Sathyaprakash}} \emph {et~al.},\ }\href {\doibase 10.1088/0264-9381/29/12/124013} {\bibfield  {journal} {\bibinfo  {journal} {Class. Quant. Grav.}\ }\textbf {\bibinfo {volume} {29}},\ \bibinfo {pages} {124013} (\bibinfo {year} {2012})},\ \bibinfo {note} {[Erratum: Class.Quant.Grav. 30, 079501 (2013)]},\ \Eprint {http://arxiv.org/abs/1206.0331} {arXiv:1206.0331 [gr-qc]} \BibitemShut {NoStop}%
\bibitem [{\citenamefont {Clarke}\ \emph {et~al.}(2020{\natexlab{b}})\citenamefont {Clarke}, \citenamefont {Copeland},\ and\ \citenamefont {Moss}}]{clarke2020constraints}%
  \BibitemOpen
  \bibfield  {author} {\bibinfo {author} {\bibfnamefont {T.~J.}\ \bibnamefont {Clarke}}, \bibinfo {author} {\bibfnamefont {E.~J.}\ \bibnamefont {Copeland}}, \ and\ \bibinfo {author} {\bibfnamefont {A.}~\bibnamefont {Moss}},\ }\href@noop {} {\bibfield  {journal} {\bibinfo  {journal} {Journal of Cosmology and Astroparticle Physics}\ }\textbf {\bibinfo {volume} {2020}},\ \bibinfo {pages} {002} (\bibinfo {year} {2020}{\natexlab{b}})}\BibitemShut {NoStop}%
\bibitem [{\citenamefont {{Rajagopal}}\ and\ \citenamefont {{Romani}}(1995)}]{1995ApJ...446..543R}%
  \BibitemOpen
  \bibfield  {author} {\bibinfo {author} {\bibfnamefont {M.}~\bibnamefont {{Rajagopal}}}\ and\ \bibinfo {author} {\bibfnamefont {R.~W.}\ \bibnamefont {{Romani}}},\ }\href {\doibase 10.1086/175813} {\bibfield  {journal} {\bibinfo  {journal} {\apj}\ }\textbf {\bibinfo {volume} {446}},\ \bibinfo {pages} {543} (\bibinfo {year} {1995})},\ \Eprint {http://arxiv.org/abs/astro-ph/9412038} {arXiv:astro-ph/9412038 [astro-ph]} \BibitemShut {NoStop}%
\bibitem [{\citenamefont {Jaffe}\ and\ \citenamefont {Backer}(2003)}]{Jaffe:2002rt}%
  \BibitemOpen
  \bibfield  {author} {\bibinfo {author} {\bibfnamefont {A.~H.}\ \bibnamefont {Jaffe}}\ and\ \bibinfo {author} {\bibfnamefont {D.~C.}\ \bibnamefont {Backer}},\ }\href {\doibase 10.1086/345443} {\bibfield  {journal} {\bibinfo  {journal} {Astrophys. J.}\ }\textbf {\bibinfo {volume} {583}},\ \bibinfo {pages} {616} (\bibinfo {year} {2003})},\ \Eprint {http://arxiv.org/abs/astro-ph/0210148} {arXiv:astro-ph/0210148} \BibitemShut {NoStop}%
\bibitem [{\citenamefont {Wyithe}\ and\ \citenamefont {Loeb}(2003)}]{Wyithe:2002ep}%
  \BibitemOpen
  \bibfield  {author} {\bibinfo {author} {\bibfnamefont {J.~S.~B.}\ \bibnamefont {Wyithe}}\ and\ \bibinfo {author} {\bibfnamefont {A.}~\bibnamefont {Loeb}},\ }\href {\doibase 10.1086/375187} {\bibfield  {journal} {\bibinfo  {journal} {Astrophys. J.}\ }\textbf {\bibinfo {volume} {590}},\ \bibinfo {pages} {691} (\bibinfo {year} {2003})},\ \Eprint {http://arxiv.org/abs/astro-ph/0211556} {arXiv:astro-ph/0211556} \BibitemShut {NoStop}%
\bibitem [{\citenamefont {Enoki}\ and\ \citenamefont {Nagashima}(2007)}]{Enoki:2006kj}%
  \BibitemOpen
  \bibfield  {author} {\bibinfo {author} {\bibfnamefont {M.}~\bibnamefont {Enoki}}\ and\ \bibinfo {author} {\bibfnamefont {M.}~\bibnamefont {Nagashima}},\ }\href {\doibase 10.1143/PTP.117.241} {\bibfield  {journal} {\bibinfo  {journal} {Prog. Theor. Phys.}\ }\textbf {\bibinfo {volume} {117}},\ \bibinfo {pages} {241} (\bibinfo {year} {2007})},\ \Eprint {http://arxiv.org/abs/astro-ph/0609377} {arXiv:astro-ph/0609377} \BibitemShut {NoStop}%
\bibitem [{\citenamefont {Sesana}\ \emph {et~al.}(2008)\citenamefont {Sesana}, \citenamefont {Vecchio},\ and\ \citenamefont {Colacino}}]{Sesana:2008mz}%
  \BibitemOpen
  \bibfield  {author} {\bibinfo {author} {\bibfnamefont {A.}~\bibnamefont {Sesana}}, \bibinfo {author} {\bibfnamefont {A.}~\bibnamefont {Vecchio}}, \ and\ \bibinfo {author} {\bibfnamefont {C.~N.}\ \bibnamefont {Colacino}},\ }\href {\doibase 10.1111/j.1365-2966.2008.13682.x} {\bibfield  {journal} {\bibinfo  {journal} {Mon. Not. Roy. Astron. Soc.}\ }\textbf {\bibinfo {volume} {390}},\ \bibinfo {pages} {192} (\bibinfo {year} {2008})},\ \Eprint {http://arxiv.org/abs/0804.4476} {arXiv:0804.4476 [astro-ph]} \BibitemShut {NoStop}%
\bibitem [{\citenamefont {Kelley}\ \emph {et~al.}(2017)\citenamefont {Kelley}, \citenamefont {Blecha}, \citenamefont {Hernquist}, \citenamefont {Sesana},\ and\ \citenamefont {Taylor}}]{Kelley:2017lek}%
  \BibitemOpen
  \bibfield  {author} {\bibinfo {author} {\bibfnamefont {L.~Z.}\ \bibnamefont {Kelley}}, \bibinfo {author} {\bibfnamefont {L.}~\bibnamefont {Blecha}}, \bibinfo {author} {\bibfnamefont {L.}~\bibnamefont {Hernquist}}, \bibinfo {author} {\bibfnamefont {A.}~\bibnamefont {Sesana}}, \ and\ \bibinfo {author} {\bibfnamefont {S.~R.}\ \bibnamefont {Taylor}},\ }\href {\doibase 10.1093/mnras/stx1638} {\bibfield  {journal} {\bibinfo  {journal} {Mon. Not. Roy. Astron. Soc.}\ }\textbf {\bibinfo {volume} {471}},\ \bibinfo {pages} {4508} (\bibinfo {year} {2017})},\ \Eprint {http://arxiv.org/abs/1702.02180} {arXiv:1702.02180 [astro-ph.HE]} \BibitemShut {NoStop}%
\bibitem [{\citenamefont {Benetti}\ \emph {et~al.}(2022)\citenamefont {Benetti}, \citenamefont {Graef},\ and\ \citenamefont {Vagnozzi}}]{Benetti:2021uea}%
  \BibitemOpen
  \bibfield  {author} {\bibinfo {author} {\bibfnamefont {M.}~\bibnamefont {Benetti}}, \bibinfo {author} {\bibfnamefont {L.~L.}\ \bibnamefont {Graef}}, \ and\ \bibinfo {author} {\bibfnamefont {S.}~\bibnamefont {Vagnozzi}},\ }\href {\doibase 10.1103/PhysRevD.105.043520} {\bibfield  {journal} {\bibinfo  {journal} {Phys. Rev. D}\ }\textbf {\bibinfo {volume} {105}},\ \bibinfo {pages} {043520} (\bibinfo {year} {2022})},\ \Eprint {http://arxiv.org/abs/2111.04758} {arXiv:2111.04758 [astro-ph.CO]} \BibitemShut {NoStop}%
\bibitem [{\citenamefont {Vagnozzi}(2021)}]{Vagnozzi:2020gtf}%
  \BibitemOpen
  \bibfield  {author} {\bibinfo {author} {\bibfnamefont {S.}~\bibnamefont {Vagnozzi}},\ }\href {\doibase 10.1093/mnrasl/slaa203} {\bibfield  {journal} {\bibinfo  {journal} {Mon. Not. Roy. Astron. Soc.}\ }\textbf {\bibinfo {volume} {502}},\ \bibinfo {pages} {L11} (\bibinfo {year} {2021})},\ \Eprint {http://arxiv.org/abs/2009.13432} {arXiv:2009.13432 [astro-ph.CO]} \BibitemShut {NoStop}%
\bibitem [{\citenamefont {Vagnozzi}(2023)}]{Vagnozzi:2023lwo}%
  \BibitemOpen
  \bibfield  {author} {\bibinfo {author} {\bibfnamefont {S.}~\bibnamefont {Vagnozzi}},\ }\href {\doibase 10.1016/j.jheap.2023.07.001} {\bibfield  {journal} {\bibinfo  {journal} {JHEAp}\ }\textbf {\bibinfo {volume} {39}},\ \bibinfo {pages} {81} (\bibinfo {year} {2023})},\ \Eprint {http://arxiv.org/abs/2306.16912} {arXiv:2306.16912 [astro-ph.CO]} \BibitemShut {NoStop}%
\bibitem [{\citenamefont {Zhu}\ \emph {et~al.}(2024)\citenamefont {Zhu}, \citenamefont {Zhao}, \citenamefont {Wang},\ and\ \citenamefont {Zhang}}]{Zhu:2023gmx}%
  \BibitemOpen
  \bibfield  {author} {\bibinfo {author} {\bibfnamefont {Q.-H.}\ \bibnamefont {Zhu}}, \bibinfo {author} {\bibfnamefont {Z.-C.}\ \bibnamefont {Zhao}}, \bibinfo {author} {\bibfnamefont {S.}~\bibnamefont {Wang}}, \ and\ \bibinfo {author} {\bibfnamefont {X.}~\bibnamefont {Zhang}},\ }\href {\doibase 10.1088/1674-1137/ad79d5} {\bibfield  {journal} {\bibinfo  {journal} {Chin. Phys. C}\ }\textbf {\bibinfo {volume} {48}},\ \bibinfo {pages} {125105} (\bibinfo {year} {2024})},\ \Eprint {http://arxiv.org/abs/2307.13574} {arXiv:2307.13574 [astro-ph.CO]} \BibitemShut {NoStop}%
\bibitem [{\citenamefont {Mitridate}\ \emph {et~al.}(2023)\citenamefont {Mitridate}, \citenamefont {Wright}, \citenamefont {von Eckardstein}, \citenamefont {Schr{\"o}der}, \citenamefont {Nay}, \citenamefont {Olum}, \citenamefont {Schmitz},\ and\ \citenamefont {Trickle}}]{mitridate2023ptarcade}%
  \BibitemOpen
  \bibfield  {author} {\bibinfo {author} {\bibfnamefont {A.}~\bibnamefont {Mitridate}}, \bibinfo {author} {\bibfnamefont {D.}~\bibnamefont {Wright}}, \bibinfo {author} {\bibfnamefont {R.}~\bibnamefont {von Eckardstein}}, \bibinfo {author} {\bibfnamefont {T.}~\bibnamefont {Schr{\"o}der}}, \bibinfo {author} {\bibfnamefont {J.}~\bibnamefont {Nay}}, \bibinfo {author} {\bibfnamefont {K.}~\bibnamefont {Olum}}, \bibinfo {author} {\bibfnamefont {K.}~\bibnamefont {Schmitz}}, \ and\ \bibinfo {author} {\bibfnamefont {T.}~\bibnamefont {Trickle}},\ }\href@noop {} {\bibfield  {journal} {\bibinfo  {journal} {arXiv preprint arXiv:2306.16377}\ } (\bibinfo {year} {2023})}\BibitemShut {NoStop}%
\bibitem [{\citenamefont {{Ellis}}\ \emph {et~al.}(2019)\citenamefont {{Ellis}}, \citenamefont {{Vallisneri}}, \citenamefont {{Taylor}},\ and\ \citenamefont {{Baker}}}]{2019ascl.soft12015E}%
  \BibitemOpen
  \bibfield  {author} {\bibinfo {author} {\bibfnamefont {J.~A.}\ \bibnamefont {{Ellis}}}, \bibinfo {author} {\bibfnamefont {M.}~\bibnamefont {{Vallisneri}}}, \bibinfo {author} {\bibfnamefont {S.~R.}\ \bibnamefont {{Taylor}}}, \ and\ \bibinfo {author} {\bibfnamefont {P.~T.}\ \bibnamefont {{Baker}}},\ }\href@noop {} {\enquote {\bibinfo {title} {{ENTERPRISE: Enhanced Numerical Toolbox Enabling a Robust PulsaR Inference SuitE}},}\ }\bibinfo {howpublished} {Astrophysics Source Code Library, record ascl:1912.015} (\bibinfo {year} {2019}),\ \Eprint {http://arxiv.org/abs/1912.015} {ascl:1912.015} \BibitemShut {NoStop}%
\bibitem [{\citenamefont {Lamb}\ \emph {et~al.}(2023)\citenamefont {Lamb}, \citenamefont {Taylor},\ and\ \citenamefont {van Haasteren}}]{lamb2023need}%
  \BibitemOpen
  \bibfield  {author} {\bibinfo {author} {\bibfnamefont {W.~G.}\ \bibnamefont {Lamb}}, \bibinfo {author} {\bibfnamefont {S.~R.}\ \bibnamefont {Taylor}}, \ and\ \bibinfo {author} {\bibfnamefont {R.}~\bibnamefont {van Haasteren}},\ }\href@noop {} {\enquote {\bibinfo {title} {The need for speed: Rapid refitting techniques for bayesian spectral characterization of the gravitational wave background using ptas},}\ } (\bibinfo {year} {2023}),\ \Eprint {http://arxiv.org/abs/2303.15442} {arXiv:2303.15442 [astro-ph. HE]} \BibitemShut {NoStop}%
\end{thebibliography}%

\end{document}